%% file: thesis.tex
\pdfoutput=1 
\documentclass[a4paper,11pt]{memoir}
\usepackage[T1]{fontenc}
\usepackage[utf8]{inputenc}
\usepackage[sc]{mathpazo}
\usepackage{xcolor}
\usepackage{graphicx}
\usepackage[final,tracking=true,kerning=true]{microtype}

\input{style/thesis_style}

\colorlet{cite}{mpired}
\colorlet{ref}{black}
\colorlet{url}{black}

\input{style/packages}
\input{style/macros}

\thesistype{Bachelor thesis}
\discipline{Physics}
\title{Electrostatic interaction between colloids with constant surface potentials at fluid interfaces}
\author{Rick Bebon}
\institution{Theory of Inhomogeneous Condensed Matter \\ Max Planck Institute for Intelligent Systems}
\anotherinstitution{Institute for Theoretical Physics IV \\ University of Stuttgart}
\supervisor{Dr.~Arghya Majee}
\examiner{Priv.-Doz.~Dr.~Markus Bier}

\begin{document}

\frontmatter
\include{frontmatter/title}
\include{frontmatter/abstract}
\include{frontmatter/abstract_de}

\cleardoublepage
\tableofcontents
\clearpage


\mainmatter
\include{chapters/01_Introduction}
\include{chapters/02_Formalism}
\include{chapters/03_Electrostatic_Potential}
\include{chapters/04_Interaction_Energy}
\include{chapters/05_Discussion}
\include{chapters/06_Conclusion}

\backmatter

\bibliographystyle{unsrt}
\bibliography{bibliography.bib}

\include{backmatter/acknowledgements}
\include{backmatter/declaration}

\end{document}

%% file: style/thesis_style.tex
\definecolor{mpiblue}{HTML}{008B8B}
\definecolor{darkgrey}{HTML}{4d4d4d}
\definecolor{lightgrey}{HTML}{e3e3e3}
\definecolor{mpired}{HTML}{8b0000}
\definecolor{mpigreen}{HTML}{008b46}
\definecolor{mpidarkblue}{HTML}{00468b}
\definecolor{shadecolor}{HTML}{eeeeee}

\DeclareRobustCommand{\spacedallcaps}[1]{\textls[160]{{\oldstylenums{\textsc{\MakeTextUppercase{#1}}}}}}%
\DeclareRobustCommand{\spacedlowsmallcaps}[1]{\textls[80]{{\oldstylenums{\textsc{\MakeTextLowercase{#1}}}}}}%

\setsecnumdepth{subsection}

\makeatletter
\makechapterstyle{noether}{%
	\chapterstyle{thatcher}

	\setsecheadstyle{\spacedlowsmallcaps}
	\setsecindent{-3.5ex}
	\setbeforesecskip{-4ex plus -1ex minus -.2ex}
	\setaftersecskip{4ex plus 1ex minus .2ex}

	\setsubsecheadstyle{\itshape}
	\setbeforesubsecskip{-4ex plus -1ex minus -.2ex}
	\setaftersubsecskip{4ex plus 1ex minus .2ex}
	\setsecnumformat{\spacedlowsmallcaps{\csname the##1\endcsname\quad}}

	\setparaheadstyle{\color{mpidarkblue}$\triangleright \ $ \color{black} \normalsize\itshape}

}
\makeatother
\chapterstyle{noether}

\makeatletter
\newcommand\thesistype[1]{\renewcommand\@thesistype{#1}}
\newcommand\@thesistype{}

\newcommand\discipline[1]{\renewcommand\@discipline{#1}}
\newcommand\@discipline{}

\newcommand\institution[1]{\renewcommand\@institution{#1}}
\newcommand\@institution{}

\newcommand\anotherinstitution[1]{\renewcommand\@anotherinstitution{#1}}
\newcommand\@anotherinstitution{}

\newcommand\supervisor[1]{\renewcommand\@supervisor{#1}}
\newcommand\@supervisor{}

\newcommand\examiner[1]{\renewcommand\@examiner{#1}}
\newcommand\@examiner{}

\newcommand{\applytolist}[3][{,\\\vspace{0.1cm}}]{%
	\def\nextitem{\def\nextitem{#1}}%
	\@for \el:=#3\do{\nextitem{#2{\el}}}%
}

\newcommand{\smarttitle}{
{\begingroup
\thispagestyle{empty}
\makeatletter
\raggedright
\vfill
{\LARGE \spacedallcaps{\@thesistype} \\}
{\Large \spacedlowsmallcaps{\@discipline}\\}
\vfill
{\Huge \textcolor{mpidarkblue}{\spacedallcaps{\@title}}\\}
\vfill
{\LARGE \applytolist{\spacedallcaps}{\@author}\\}
\vfill
{\Large \applytolist{\itshape}{\@institution}}
\vfill
{\Large \applytolist{\itshape}{\@anotherinstitution}}
\vfill
{\Large \spacedlowsmallcaps{Supervised by}\\}
\vspace{0.1cm}
{\large \applytolist{\spacedallcaps}{\@supervisor}}
\vfill
{\Large \spacedlowsmallcaps{Examined by}\\}
\vspace{0.1cm}
{\large \applytolist{\spacedallcaps}{\@examiner}}
\vfill
\vfill
{\Large \spacedlowsmallcaps{August 18, 2018}}
\vfill
{\Large \spacedlowsmallcaps{Corrected Version}}
\vfill
\makeatother
\endgroup}}

\makeatletter

\makepagestyle{standard} 

\makeatletter                 
\makeevenfoot{standard}{\thepage}{}{} %
\makeoddfoot{standard}{}{}{\thepage}  %
\makeevenhead{standard}{\color{mpidarkblue}\spacedlowsmallcaps\leftmark}{}{}
\makeoddhead{standard}{}{}{\color{mpidarkblue}\spacedlowsmallcaps\rightmark}
\makeatother  
 
\makepagestyle{backmatter} 

\makeatletter                 
\makeevenfoot{backmatter}{\thepage}{}{} %
\makeoddfoot{backmatter}{}{}{\thepage}  %
\makeevenhead{backmatter}{}{}{}
\makeoddhead{backmatter}{}{}{}
\makeatother

\makepagestyle{chap} 

\makeatletter
\makeevenfoot{chap}{\thepage}{}{} 
\makeoddfoot{chap}{}{}{\thepage}  %
\makeevenhead{chap}{}{}{}   %
\makeoddhead{chap}{}{}{}    %
\makeatother

\nouppercaseheads
\aliaspagestyle{chapter}{chap}

\captionstyle{\small}
\captionnamefont{\slshape\color{mpidarkblue}}

%% file: style/packages.tex
\usepackage{csquotes} 

\usepackage[ngerman,english]{babel}

\usepackage{latexsym,amsmath,amssymb,amsthm,amscd}

\usepackage{mathtools} 

\usepackage[pdfencoding=auto,pdfborderstyle={/S/U/W 1},linkbordercolor=ref,colorlinks,linkcolor=ref,citecolor=cite,urlcolor=url]{hyperref}

\usepackage[all]{hypcap}

\usepackage{physics}

\usepackage{enumitem}

\usepackage{tikz}

\usetikzlibrary{decorations.pathreplacing} 

\usepackage{framed}

\usepackage{booktabs}

\usepackage{xcolor}

\usepackage{tikz}

\usepackage{float}

\usepackage{textcase}

\usepackage{graphicx}

\usepackage{siunitx} 

%% file: style/macros.tex
\renewcommand{\epsilon}{\varepsilon}
\renewcommand{\rho}{\varrho}
\renewcommand{\tilde}{\widetilde}
\newcommand{\kone}{\kappa_{1}}
\newcommand{\ktwo}{\kappa_{2}}
\newcommand{\eone}{\varepsilon_{1}}
\newcommand{\etwo}{\varepsilon_{2}}
\newcommand{\PsiD}{\Psi_{D}}
\newcommand{\PsiP}{\Psi_{P}}
\newcommand{\tpsi}{\tilde{\psi}}

\newcommand{\V}{\mathcal{V}}
\newcommand{\kb}{k_{\textrm{B}}}
\newcommand{\n}{\vb{n}(\vb{r})}
\newcommand{\D}{\vb{D}(\vb{r}, \qty[\phi_\pm])}
\newcommand{\Psii}{\Psi(\vb{r},\qty[\phi_\pm])}
\newcommand{\TO}{\beta \tilde{\Omega} \qty[\phi_{\pm}]}

\newcommand{\obar}{\overline{\omega}}

%% file: frontmatter/title.tex

\smarttitle

%% file: frontmatter/abstract.tex

{\begingroup
\cleardoublepage
\begin{abstract}
In this thesis, the electrostatic interaction between two chemically identical colloids, both carrying 
constant surface potential is studied in the limit of short inter-particle separation at the interface 
of two immiscible fluids.
Using an appropriate model system, the problem is solved analytically within the framework 
of linearized Poisson-Boltzmann theory and classical density functional theory.
The governing equation of the electrostatic problem is derived by minimization of the corresponding 
density functional, which leads to the Debye-Hückel equation.
Subsequently, the Debye-Hückel equation is solved by exact calculations as well as by applying
the widely used superposition approximation, each providing expressions for the electrostatic
potential distribution inside the system.
Furthermore, the obtained results of the electrostatic problem are used to calculate surface and 
line interaction energy densities between the colloidal particles.
In all cases, the superposition approximation fails to predict the interaction 
energies correctly for both small and large separations.
Additionally, analytic expressions for the surface tensions, line tension, and interfacial tension, 
which are all independent of the inter-particle separation, are obtained.
The results of this thesis are expected to enrich the description of electrostatic interaction between 
colloids at fluid interfaces.
\end{abstract}
\endgroup}

%% file: frontmatter/abstract_de.tex

{\begingroup
\cleardoublepage
\begin{otherlanguage}{ngerman}
\begin{abstract}
Die vorliegende Arbeit befasst sich mit der elektrostatischen Wechselwirkung zweier identischer Kolloide, 
die in einer Grenzfläche zwischen zwei nicht mischbaren Fluiden energetisch gefangen sind. 
Die Kolloide werden mit konstantem elektrostatischem Potential auf ihrer Oberfläche modelliert 
und haben einen kleinen inter-partikulären Abstand zueinander.
Das Problem wird mit Hilfe eines geeigneten Modellsystems im Rahmen der linearisierten Poisson-Boltzmann Theorie, 
sowohl der klassischen Dichtefunktionaltheorie analytisch gelöst.
Die Debye-Hückel Gleichung des elektrostatischen Potentials innerhalb des Systems wird durch die Minimierung 
des systemspezifischen Dichtefunktionals hergeleitet und anschließend, unter Anwendung exakter Berechnung 
sowohl als auch der weitverbreiteten Superpositionnäherung, gelöst.
Die somit gewonnene Beschreibung des elektrostatischen Potentials innerhalb des betrachteten Systems 
wird verwendet um Oberflächen- und Linienwechselwirkungsenergiedichten zwischen den Kolloiden zu bestimmen.
Als Resultat wird dabei festgehalten, dass die Superpositionnährung sowohl für kleine, 
als auch für große Teilchenabstände das Wechselwirkungsverhalten nicht korrekt vorhersagt.
Zusätzlich werden analytische Ausdrücke für die Oberflächen-, Linien- und Grenzflächenspannung hergeleitet.
Die Ergebnisse dieser Arbeit sollen dazu beitragen die Modellierung von Kolloiden an Flüssigkeitsgrenzflächen zu verbessern. 
\end{abstract}
\end{otherlanguage}
\endgroup}

%% file: chapters/01_Introduction.tex

\chapter{Introduction} 

The first chapter provides a brief overview of the theoretical background required for the present work.
Section~\ref{sec:1_1} introduces colloidal particles, their behavior in systems featuring fluid interfaces
and presents the problem which serves as the motivation for this thesis.
Section~\ref{sec:1_2} provides an introduction to the formalism of classical density functional theory,
commonly used mathematical conventions are presented in Section~\ref{sec:1_3} and the overall structure of 
this thesis is given in Section~\ref{sec:1_4}.

\section{Theoretical background and motivation}
\label{sec:1_1}

\subsection{Colloidal particles}

Colloidal particles are usually sized in the range of approximately 10 to \SI{1000}{\nm} and are an integral part 
of systems featuring dispersion of microscopic matter in a liquid or gas.
Suspensions of such particles are surface-active, meaning they are able to spontaneously self-assemble at
the interface of two immiscible fluids.
Although the surface activity might have its origin in the amphiphilic nature of the particle, it is not necessary 
for the spontaneous accumulation at the interface.
In general, colloidal particles have either homogeneous, heterogeneous or amphiphilic chemical composition 
and properties on their surface~\cite{binks2006colloidal}.
In this thesis, suspensions of colloidal particles are considered to be
characterized by their homogeneous chemical composition 
and spherical shape.

\subsection{Colloidal particles at fluid interfaces}

Systems featuring a fluid interface between two immiscible fluids, such as oil and water, 
try to reduce their free energy by minimizing the interfacial area since fluid interfaces
are generally energetically expensive and characterized by high interfacial tension.
Colloidal particles suspended in either fluid can therefore spontaneously accumulate at the fluid interface, 
if the reduction in free energy exceeds the thermal energy and thus form a two-dimensional 
monolayer~\cite{ramsden1904separation}.

\begin{figure}[htbp]
	\centering	
	\input{figure/tikz/contact_angle.tikz}
	\caption{
		A spherical particle trapped at an oil-water interface at $x=0$. 
		The contact angle is given by $\theta$ and the three-phase contact 
		line with radius $r_{\textrm{c}}$ is depressed with $x_{\textrm{c}}$. 
		Adapted from~\cite{binks2006colloidal}.
		}
	\label{fig:angle}
\end{figure}
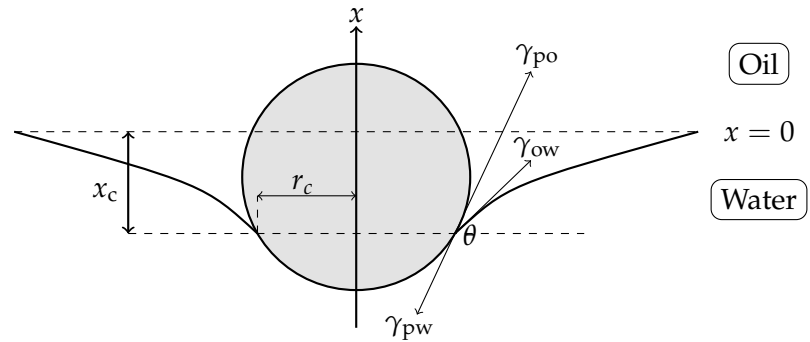 

When dealing with colloidal particles at a fluid interface, the three-phase contact angle $\theta$
(formed between the tangents to the particle surface and fluid-fluid interface at the three-phase contact line 
as seen in Fig.~\ref{fig:angle}) is a characteristic key parameter.
Standardly, the contact angle $\theta$ is measured inside the more polar fluid phase.
Within this thesis the particles are assumed to be equally wetted, resulting in a contact angle of $90^\circ$.
The structure and stability of colloidal monolayer systems have been heavily studied during the last two decades 
and serve as theoretical models to study phase behavior, structure and, dynamics of condensed matter, as well as
the stabilisation of emulsions and foams~\cite{dickinson1989food, tambe1994effect}.
Furthermore, self-assembled monolayers of colloids find application in a vast number of different fields 
including purification of water and oil recovery, pharmacy, bio-medicine, food and, the cosmetics 
industry~\cite{binks2006colloidal, ray2009submicrometer, isa2017two}.

The movement of colloidal particles trapped at the fluid interface is restricted to lateral directions
and depends on particle-particle interaction. 
Generally speaking, this interaction consists of van der Waals, steric, capillary, magnetic or electrostatic contributions. 
Within this thesis, the focus lies on the electrostatic interaction between metal colloids 
or non-metallic colloids with a metal coating.

\subsection{Electrostatic interaction of charged colloids}

In 1980 Pieranski~\cite{pieranski1980two} has shown that the electrostatic interaction
between charged colloidal particles, which are trapped at a fluid interface, is described by a long-range dipole-dipole interaction in a system featuring a low concentration of charged colloids.
The electric dipole, which is perpendicular to the fluid-fluid interface, stems from an asymmetric charge distribution around the colloid.
Subsequently assuming point-like particles, Hurd was able to study the exponential and power-law contributions for such system within 
linearized Poisson-Boltzmann theory~\cite{hurd1985electrostatic}.
In the last two decades, the work of Pieranski and Hurd has been extended in numerous directions~\cite{aveyard2002measurement, aveyard2000compression},
however, the vast majority of studies exclusively discuss the case of large inter-particle separation. 
Within this limit, the linear superposition approximation is commonly used, in which the electrostatic
potential between two colloids is approximated by superposing the electrostatic potential of a single colloidal particle. 

Meanwhile, in systems with high colloid number densities or during processes like aggregation the separation between colloidal particles 
decreases significantly~\cite{reincke2004, toor2016self}.
As presented in Ref.~\cite{majee2014}, the electrostatic interaction between two identically charged colloids
is analytically solvable within linear Poisson-Boltzmann theory
by neglecting the particle curvature and treating them as flat walls. 
This simplification is in the spirit of the Derjaguin approximation and can be justified since the colloids are in their immediate vicinity.
As it turns out, the superposition approximation fails to predict
the correct behavior for this model system, even by taking the limit of large inter-particle separation into account.
Later on, the above-mentioned approach was generalized for systems featuring non-identical charged particles in Ref.~\cite{majee2018} as well.

\subsection{Metallic colloids}

The spontaneous formation of reflective monolayer films, consisting of metallic colloidal particles
at fluid interfaces, has interesting optical properties and
has been a topic in numerous studies. 
Such reflective monolayers find application as 
liquid mirrors~\cite{collier1997, farbman1992optical, brousseau2008, yogev1988novel, schaming2011photocurrents} 
or surface-enhanced Raman scattering subtrates~\cite{talley2005surface, ko2008nanostructured}.
Moreover, colloidal gold is also relevant in the context of medical studies, where it is used to improve targeted drug delivery~\cite{han2007}.
Staying in the realm of medical studies, gold nanoparticles also find application in 
cancer research, where they are used to improve the detection of tumors
by using the above-mentioned application in surface-enhanced Raman
spectroscopy~\cite{qian2008vivo}.
Within all these monolayers of metallic particles, the inter-particle separation can easily be very small 
compared to their radii~\cite{reincke2004, toor2016self}.
However, an appropriate description of the electrostatic interaction between metallic colloids situated close to each other is missing so far.

The results of the above-mentioned studies for short inter-particle separations are suitable for non-metallic (e.g. polystyrene or silica) colloids, which can be described by a constant charge density assumption at their surfaces.
Metallic colloids, on the contrary, are characterized by constant surface potentials due to mobile electrons
which form a different boundary condition for the electrostatic problem.
Therefore, the goal of this thesis is to study
the electrostatic interaction between a pair of metallic colloids
situated in close vicinity of each other at a fluid interface.

\section{Classical density functional theory}
\label{sec:1_2}

This section provides a brief introduction to the classical density functional theory
as presented in Ref.~\cite{evans1979nature}.
In the framework of statistical physics, the equilibrium probability density $p_0$ for $N$ particles at temperature $T$,
enclosed in system volume $\V \subseteq \mathbb{R}^d$, is given by 
\begin{equation}
	p_0 = Z^{-1} \exp(-\beta(\mathcal{H} - \mu N)),
\end{equation}
in a grand canonical ensemble, where $\beta^{-1} = \kb T$ is the thermal energy and $\mu$ the chemical potential. 
The grand partition function $Z$ reads  
\begin{equation}
	Z = \Tr \exp(-\beta(\mathcal{H} - \mu N)),
\end{equation}
with the classical trace defined as
\begin{equation}
	\Tr \coloneqq 1 + \sum\limits_{N=1}^\infty \frac{1}{h^{dN}N!} \prod\limits_{i=1}^N 
	\qty( \int_\V \dd[d]{r_i} \int_{\mathbb{R}^d} \dd[d]{p_i}).
\end{equation}
The position of a particle is denoted by $\vb{r} \in \V$, the momentum by $\vb{p} \in \mathbb{R}^d$ 
and the Planck constant $h$.
The Hamiltonian $\mathcal{H}$ of this system reads  
\begin{equation}
	\mathcal{H} = 
	\frac{1}{2m}\sum\limits_{i=1}^N \vb{p}_i^2 
	+ \sum\limits_{i=1}^N V(\vb{r}_i) 
	+  \sum\limits_{\substack{i,j =1 \\ i < j}}^N U(\vb{r}_i,\vb{r}_j),
\end{equation} 
where $U(\vb{r},\vb{r}')$ is the interaction potential between the particles, $V(\vb{r})$ is an arbitrary
external potential and $m$ the particle mass. 
The Mermin functional~\cite{mermin1965nd} is defined as
\begin{equation}
	\mathcal{M} \qty[p] := \Tr p \qty[\beta\mathcal{H} - \beta\mu N + \ln p],
\end{equation}
and leads to the grand potential $ \beta\Omega$ for the equilibrium probability density $p_0$ 
and probability densities with $\Tr p = 1$
\begin{equation}
	\beta \mathcal{M}\qty[p_0] = \beta \Omega = - \ln Z.
\end{equation}
Note that $\mathcal{M}\qty[p] > \mathcal{M}\qty[p_0]$ for $p \neq p_0$. 
The single particle density observable is defined as
\begin{equation}
	\tilde{\rho}^{(1)}(\vb{r}) \coloneqq \sum\limits_{i=1}^N \delta(\vb{r}- \vb{r}_i),
\end{equation}
and therefore the single particle density reads  
\begin{equation}
	\rho^{(1)}(\vb{r}) \coloneqq \rho(\vb{r}) \coloneqq \expval{\tilde{\rho}^{(1)}(\vb{r})} 
	= \Tr \qty[p \tilde{\rho}^{(1)}(\vb{r})].
\end{equation}
In the same fashion the two particle density observable reads  
\begin{equation}
	\tilde{\rho}^{(2)}(\vb{r},\vb{r'}) \coloneqq 
	\sum\limits_{\substack{i,j =1 \\ i \neq j}}^N \delta(\vb{r}- \vb{r}_i)\delta(\vb{r'}- \vb{r}_j),
\end{equation}
and the two particle density is given by
\begin{equation}
	\rho^{(2)}(\vb{r},\vb{r'}) \coloneqq \expval{\tilde{\rho}^{(2)}(\vb{r},\vb{r'})} 
	= \Tr \qty[p \tilde{\rho}^{(2)}(\vb{r},\vb{r'})].
\end{equation}
Additionally, the pair distribution function is defined as 
\begin{equation}
	g(\vb{r},\vb{r'}) \coloneqq 
	\frac{\tilde{\rho}^{(2)}(\vb{r},\vb{r'})}{\tilde{\rho}^{(1)}(\vb{r})\tilde{\rho}^{(1)}(\vb{r'})}.
\end{equation}
Suppose an arbitrary density function $\rho(\vb{r}): \V \to \mathbb{R}^+$, which fulfills the relation
\begin{equation}
	p|\rho \equiv \Tr \qty[p \tilde{\rho}(\vb{r})] = \rho(\vb{r}).
\end{equation}
The grand potential in equilibrium now reads  
\begin{equation}
	\beta \Omega_0 = \min_{p} \mathcal{M}\qty[p] = \min_{\rho} \min_{\substack{p \\ p|\rho}} \mathcal{M}\qty[p].
\end{equation}
Finally, the density functional can be defined as 
\begin{equation}
	\beta \Omega\qty[\rho] \coloneqq \min_{\substack{p \\ p|\rho}} \mathcal{M}\qty[p],
\end{equation}
which is minimized by the single particle density $\rho^{(1)}_0(\vb{r})$ in equilibrium
\begin{equation}
	\beta \Omega\qty[\rho_0] = \beta \Omega_0.
\end{equation}
The Euler-Lagrange equation 
\begin{equation}
	\fdv{\beta\Omega\qty[\rho]}{\rho} = 0,
\end{equation}
is a necessary condition for the single particle density in equilibrium.
As an example, the density functional of the ideal gas (where $U(\vb{r},\vb{r}')=0$) is given by
\begin{equation}
	\beta \Omega^{\textrm{id}}\qty[\rho]  = 
	\int\limits_\V \dd[d]{\vb{r}} \rho(\vb{r}) \qty[\ln\rho(\vb{r})\Lambda^d - 1 - \beta \mu + \beta V(\vb{r})],
\end{equation}
with the thermal de Broglie wavelength $\Lambda \coloneqq \sqrt{(2\pi\hbar^2\beta)/m}$.
In general, the exact density functional for interacting particles with $U (\vb{r},\vb{r'}) \neq 0$ is unknown. 
In the present work the mean-field like random phase approximation (RPA)
($g(\vb{r},\vb{r'})= 1 \ \forall \ \vb{r},\vb{r'} \in \V$) is appropriate, providing
the density functional
\begin{equation}
	\beta \Omega\qty[\rho] 
	= \beta F^{\textrm{ex}}\qty[\rho] 
	+ \beta \Omega^{\textrm{id}}\qty[\rho],
\end{equation}
with the excess functional defined as 
\begin{equation}
	\beta F^{\textrm{ex}}\qty[\rho] \coloneqq 
	\frac{1}{2} \int\limits_\V \dd[d]{\vb{r}} \int\limits_\V \dd[d]{\vb{r'}} \beta U(\vb{r},\vb{r'}) \rho(\vb{r})\rho(\vb{r'}).
	\label{eq:excess}
\end{equation}
\section{Mathematical conventions}
\label{sec:1_3}

\paragraph{Fourier Sine Transformation}

The Fourier sine transform of a piece-wise continuous function $f(z)$, which is absolute integrable
over the interval $[0,\infty)$, is defined as 
\begin{equation}
	\mathcal{F}\qty{f(z)} = \hat{f}(q) = \int\limits_0^\infty \dd{z} f(z) \sin(qz) \qquad q > 0.
	\label{eq:sine_trafo}
\end{equation}
In similar fashion the inverse Fourier sine transformation reads
\begin{equation}
	\mathcal{F}^{-1} \qty{\hat{f}(q)} 
	= f(z) 
	= \frac{2}{\pi}\int\limits_0^\infty \dd{q} \hat{f}(q) \sin(qz) \qquad z\geq 0.
	\label{eq:inv_sine_trafo}
\end{equation}
\paragraph{Kronecker delta}

The Kronecker delta, as it appears in many areas of mathematics and physics, is defined as
\begin{equation}
	\delta_{\alpha,\beta} = 
	\begin{cases}
		0 \qquad \textrm{if} \ \alpha \neq \beta
		\\ 
		1 \qquad \textrm{if} \ \alpha = \beta. 
	\end{cases}
\label{eq:kronecker_delta}
\end{equation}
\paragraph{Heaviside step function}

The discontinuous Heaviside step function is defined as 
\begin{equation}
	\Theta \coloneqq \mathbb{R} \to \qty{0,1}\qquad x \mapsto 
	\begin{cases}
		0: \qquad x<0
		\\ 
		1: \qquad x \geq 0. 
	\end{cases}
\label{eq:heaviside}
\end{equation}
\section{Structure of the thesis}
\label{sec:1_4}

The thesis is organized in the following way:

\paragraph{Chapter 2}

The second chapter provides a mathematical description of the model system.
Furthermore, the density functional of said system is derived and minimized, yielding the linearized
Poisson-Boltzmann equation (also known as Debye-Hückel equation). 
Additionally, an expression for the effective interactions in terms of the electrostatic potential is presented.

\paragraph{Chapter 3}

The third chapter solves the Debye-Hückel equation, both within exact calculations and superposition approximation, 
to obtain the electrostatic potential distribution inside the system.

\paragraph{Chapter 4}

In the fourth chapter, the results from Chapter~\ref{ch:2} and Chapter~\ref{ch:3} are used to obtain analytic
expressions for the surface interaction energy densities, line interaction energy density, 
as well as the surface, line and interfacial tension.

\paragraph{Chapter 5}

In the fifth chapter, results for the surface interaction energy densities, line interaction energy density and
electrostatic potential within exact and superposition calculations are compared and discussed for different system parameters.

\paragraph{Chapter 6}

The sixth chapter summarizes the results and provides an outlook for further research.

%% file: figure/tikz/contact_angle.tikz
\begin{tikzpicture}
		\filldraw[lightgrey] (0,0) circle[radius=1.5];
		\draw[thick] (0,0) circle[radius=1.5];
		\draw[thick] (-4.5,0.6) .. controls (-2,-0.1) .. (-1.2990,-0.75);
		\draw[thick] (4.5,0.6) .. controls (2,-0.1) .. (1.2990,-0.75);
		\draw[dashed] (-4.5,0.6) -- (4.5,0.6);
		\draw[dashed] (-3,-0.75) -- (3,-0.75);
		\draw[<->,thick] (-3,-0.75) -- (-3,0.6);
		\draw[->,thick] (0,-2) -- (0,2);
		\draw[dashed] (-1.2990,-0.75) -- (-1.2990,-0.25);
		\draw[<->] (-1.2990,-0.25) -- (0,-0.25);
		\draw[->] (1.2990,-0.75) -- (2.2990,0.22);
		\draw[->] (1.2990,-0.75) -- (2.3,1.4);
		\draw[->] (1.2990,-0.75) -- (0.8,-1.825);
		\draw (-0.7,-0.1) node{$r_c$};
		\draw (2.38, 0.39) node{$\gamma_{\textrm{ow}}$};
		\draw (2.38,1.6) node{$\gamma_{\textrm{po}}$};
		\draw (0.7, -2) node{$\gamma_{\textrm{pw}}$};
		\draw (5.3,0.6) node{$x=0$};
		\draw (0,2.15) node{$x$};
		\draw (-3.3,-0.2) node{$x_{\textrm{c}}$};
		\draw (1.5,-0.77) node{$\theta$};
		\draw (5.3, 1.5) node[draw,rounded corners]{Oil};
		\draw (5.3, -0.3) node[draw,rounded corners]{Water};
\end{tikzpicture}

%% file: chapters/02_Formalism.tex

\chapter{Formalism}
\label{ch:2}

The following chapter provides a theoretical description of the system treated in this thesis.
Within the framework of classical density functional theory the governing equation for the 
electrostatic potential, the corresponding boundary conditions and an expression for the effective 
interactions in terms of the electrostatic potential are obtained in a self-consistent manner.

\section{System}
\label{sec:system}

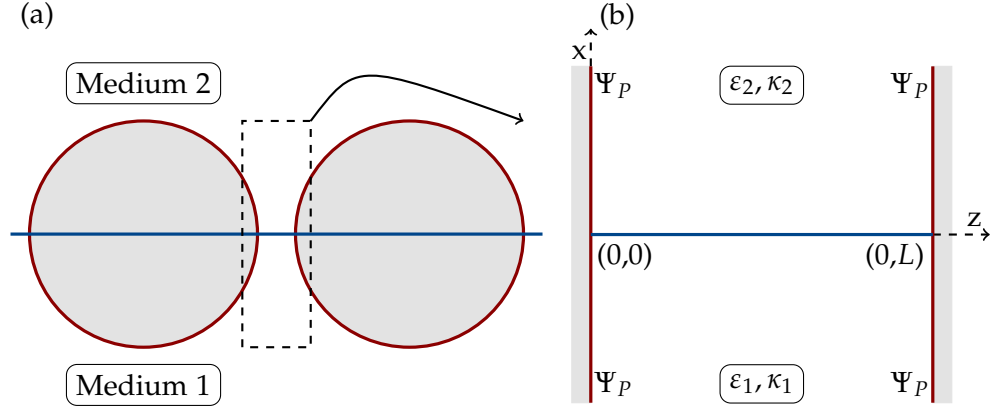
\begin{figure}[htbp]
	\centering	
	\hfil
	\input{figure/tikz/colloids_interface.tikz}
	\hfil
	\input{figure/tikz/walls_interface.tikz}
	\hfil
	\caption{
			(a) Cross-section of two identical spherical colloids, trapped at a fluid-fluid interface 
			(indicated by the horizontal blue line).
			The particles are in close vicinity with a colloid-fluid contact angle of $90^\circ$. 
			(b) Sketch of the simplified system (ignoring the curvature of the sphere) of the boxed region in panel (a).
			Due to the small surface-surface separation, the particle curvatures are approximated as two flat walls at $z=0$ 
			and $z=L$ which carry the constant surface potential $\PsiP$. 
			For the region $z \in \qty[0,L]$, the space is filled by two immiscible fluids forming 
			a flat fluid interface at $x=0$. 
			The media are characterized by their permittivities $\eone$, $\etwo$ and 
			inverse Debye lengths $\kone$, $\ktwo$, respectively.
			}
	\label{fig:system}
\end{figure}

Within the scope of this thesis, the focuses lies on the electrostatic interaction between two colloidal particles at a fluid-fluid interface.
Both particles carry the constant surface potential $\PsiP$ and are separated from each other by distance $L$,
which is considered to be small compared to their radii (See Fig.~\ref{fig:system}a).
Any deformities of the fluid-fluid interface are disregarded, resulting in a planar interface.
Additionally, a colloid-fluid contact angle of $90^\circ$ is assumed and 
due to the small inter-particle separation, each particle curvature is approximated as a flat wall.

For the mathematical description a confined space, which is bound by two parallel planar walls located at 
$z=0$ and $z=L$ in Cartesian coordinates $x,y$ and $z$, is considered.
The space between the walls is filled by two immiscible fluids, resulting in a fluid-fluid interface 
at $x=0$ which is perpendicular to both walls. 
The upper fluid (in half-space $x>0$) is denoted as medium 2 and the lower fluid (in half-space $x<0$) as medium 1
(See Fig \ref{fig:system}b).

In this setup, binary monovalent ionic species of opposing signs (e.g. $\textrm{Na}^+$ and $\textrm{Cl}^-$) are present. 
Their bulk ionic strength therefore reads  
\begin{align}
	I(\vb{r})=
	\begin{cases}
		I_1\qquad x<0
		\\
		I_2 \qquad x>0.
	\end{cases}
\end{align}
Typically, molecule and ion interaction lead to variation of the ion number density on the scale of the 
bulk correlation length, which in this case is much smaller than the length scale of the studied system 
and can, therefore, be neglected.
As a result, both media are modeled as structure-less linear dielectric fluids where the characteristic 
dielectric constant is given by
\begin{equation}
	\epsilon_i = \epsilon_0 \epsilon_{r,i}  \qquad i \in \qty{1,2},
\end{equation} 
for each medium $i$.
Here, $\epsilon_ {r,i}$ is the dimensionless relative permittivity 
and $\epsilon_0$ the vacuum permittivity. 
Thus, the overall dielectric profile of the system varies step-like at the fluid interface
\begin{align}
	\epsilon(\vb{r})=
	\begin{cases}
		\eone \qquad x<0 \qquad \textrm{medium 1}
		\\
	\etwo \qquad x>0 \qquad \textrm{medium 2}.
	\end{cases}
\end{align}
However, the charge density varies on the length scale of the Debye length (which exceeds the bulk correlation length) 
and thus, sets the length scale of interest.
The respective Debye lengths of medium $i \in \qty{1,2}$ are defined as 
\begin{align}
	\kappa(\vb{r})^{-1} =
	\begin{cases}
		\kone^{-1} \qquad x<0 \qquad \textrm{medium 1}
		\\
		\ktwo^{-1} \qquad x>0 \qquad \textrm{medium 2},
	\end{cases}
	\qquad
	\kappa_i^{-1} := \sqrt{\frac{\epsilon_{r,i}}{8\pi l_B I_i}},
	\label{eq:debye_length}
\end{align}
where the vacuum Bjerrum length defined as
\begin{equation}
	l_B := \frac{e^2}{4\pi \epsilon_0 \kb T},
\end{equation}
with Boltzmann constant $\kb$, temperature $T$ and elementary charge $e>0$.
In the grand canonical description, the ion reservoirs are provided by the bulk of both media. 
The ion solvent interaction is characterized by an external potential $V_\pm(\vb{r})$ acting on the ions,
by definition
\begin{align}
	V_\pm(\vb{r}) =
	\begin{cases}
		0 \qquad &x<0
		\\
		f_\pm \qquad &x>0.
	\end{cases}
\end{align}
Hence, $f_\pm$ is the solvation free energy difference for the ions in the two media.

\section{Density functional}

The potential distribution $U$ of two different species of ionic particles $i$ and $j$ 
with respective valencies $Z_i$ and $Z_j$ is given by 
\begin{equation}
	\beta U_{i,j}(\vb{r},\vb{r'}) = Z_i Z_j\frac{l_B}{\abs{\vb{r}-\vb{r'}}}. 
\label{eq:betaU}
\end{equation}
Meaning, two particles with single elementary charge (i.e. $\textrm{Na}^+,\textrm{Cl}^-$), which are separated by their 
Bjerrum length $l_B$ have identical electrostatic  and thermal energy $\kb T$.
The excess-functional in Eq.~\eqref{eq:excess} within the mean-field like random phase approximation 
(see Section~\ref{sec:1_2}) is given by
\begin{align}
	\beta F^{\textrm{ex}}(\rho) &= \frac{1}{2} \int\limits_\V \dd[3]{r} \int\limits_\V \dd[3]{r'}\sum_{i,j} Z_i Z_j 
	\frac{l_B}{\abs{\vb{r}-\vb{r'}}} \rho(\vb{r})\rho(\vb{r'})
	\nonumber	
	\\
	&= \frac{\beta}{2} \int\limits_\V \dd[3]{r} \rho_{\textrm{int}}(\vb{r}) 
	 \frac{1}{4 \pi \epsilon_0} \int\limits_\V \dd[3]{r'}\frac{\rho_{\textrm{int}}(\vb{r'})}{\abs{\vb{r}-\vb{r'}}}
	\nonumber	
	\\
	&= \frac{\beta}{2} \int\limits_\V \dd[3]{r} \rho_{\textrm{int}}(\vb{r}) \phi_{\textrm{int}}(\rho,\vb{r}),
	\label{eq:excess_rpa}
\end{align}
where the internal charge density $\rho_{\textrm{int}}$ and the internal potential $\phi_{\textrm{int}}$ are given by
\begin{align}
&\rho_{\textrm{int}} = e \sum_i Z_i \rho_i,
\\
&\phi_{\textrm{int}} = \frac{1}{4 \pi \epsilon_0} \int\limits_\V \dd[3]{r'} \frac{\rho_{\textrm{int}}(\vb{r'})}{\abs{\vb{r}-\vb{r'}}}.
\end{align}
In general, the external potential $V_i(\vb{r})$ consists of electrostatic contributions stemming from the 
system boundary $\partial \V$ and other contributing factors
\begin{equation}
	V_i(\vb{r}) = Z_i e \phi_{\textrm{ext}}(\vb{r}) + \overline{V}_i(\vb{r}).
\end{equation}
By using the excess-functional in Eq.~\eqref{eq:excess_rpa}, the density functional of the free energy 
within the random phase approximation now reads
\begin{align}
	\beta\Omega\qty[\rho] = &\int\limits_\V\dd[3]{r}\sum_i\rho_i(\vb{r})\qty{\ln(\rho_i(\vb{r})\Lambda^3_i)-1-\beta\mu_i+\beta V_i(\vb{r})}
	\nonumber
	\\
	&+ \int\limits_\V \dd[3]{r} \frac{\beta}{2} \rho_{\textrm{int}}(\vb{r})\phi_{\textrm{int}}(\rho,\vb{r})
	\nonumber
	\\
	= &\int\limits_\V \dd[3]{r} \sum_i \rho_i(\vb{r}) \qty{\ln(\rho_i(\vb{r}) \Lambda^3_i) -1 
	-\beta \mu_i + \beta \overline{V}_i(\vb{r})}
	\nonumber
	\\
	&+ \int\limits_\V \dd[3]{r} \frac{\beta}{2} \rho_{\textrm{int}}(\vb{r})\phi_{\textrm{int}}(\rho,\vb{r}) 
	+ \beta e \sum_i Z_i \rho_i(\vb{r})\phi_{\textrm{ext}}(\vb{r})
	\nonumber
	\\
	= &\int\limits_\V \dd[3]{r} \sum_i \rho_i(\vb{r}) \qty{\ln(\frac{\rho_i(\vb{r})}{\zeta_i})  
	-1 + \beta V_i(\vb{r})}
	\nonumber
	\\
	&+ \int\limits_\V \dd[3]{r} \frac{\beta}{2} \rho_{\textrm{int}}(\vb{r})\phi_{\textrm{int}}(\rho,\vb{r}) 
	+ \beta \rho_{\textrm{int}}\phi_{\textrm{ext}}(\vb{r}),
	\label{eq:DFT_herleitung}
\end{align}
with the fugacity defined as
\begin{equation}
	\zeta_i := \frac{\exp(\beta \mu_i)}{\Lambda^3_i}.
\end{equation}
The last term of Eq.~\eqref{eq:DFT_herleitung} can be further reduced by using the relation 
$\rho_{\textrm{int}} = \div{\vb{D}_{\textrm{int}}(\vb{r},\qty[\rho])}$, applying the divergence 
theorem and using the product rule for the divergence.
It yields the contribution for the energy density of the electric displacement field 
\begin{equation}
	\int\limits_\V \dd[3]{r} \frac{\beta \vb{D}(\vb{r},\qty[\rho])^2}{2\epsilon(\vb{r})},
\end{equation}
and a surface contribution which is the work the system has to provide to keep the potentials 
at constant value $\PsiP$ at each surface
\begin{equation}
	\PsiP\int\limits_{\partial\V} \dd[2]{r}\vb{n}(\vb{r})\vdot\vb{D}(\vb{r},\qty[\rho]).
\end{equation}
The final expression of the functional is therefore given by
\begin{align}
	\beta \Omega \qty[\rho] =& \int\limits_\V \dd[3]{r} \qty[\sum_i \rho_i(\vb{r}) 
	\qty{\ln(\frac{\rho_i(\vb{r})}{\zeta_i}) -1 
	+ \beta V_i(\vb{r})} + \frac{\beta \vb{D}(\vb{r},\qty[\rho])^2}{2\epsilon(\vb{r})} ]
	\nonumber
	\\
	&+ \beta\PsiP\int\limits_{\partial\V} \dd[2]{r}\vb{n}(\vb{r})\vdot \vb{D}(\vb{r},\qty[\rho]).
\label{eq:DF}
\end{align}

\section{Linearized Poisson-Boltzmann equation}
\label{sec:lin_PB}

After the more general grand potential functional within random phase approximation has been derived in Eq.~\eqref{eq:DF}, the 
functional of the system described in Sec.~\ref{sec:system} now reads
\begin{align}
	\beta \Omega \qty[\rho_\pm] =& \int\limits_\V \dd[3]{r} \qty[\sum_{i=\pm} \rho_i(\vb{r}) 
	\qty{\ln(\frac{\rho_i(\vb{r})}{\zeta_i}) -1 + \beta V_i(\vb{r})} 
	+ \frac{\beta \vb{D}(\vb{r},\qty[\rho_\pm])^2}{2\epsilon(\vb{r})} ]
	\nonumber
	\\
	&+ \beta\PsiP\int\limits_{\partial\V} \dd[2]{r} \vb{n}(\vb{r})\vdot \vb{D}(\vb{r},\qty[\rho_\pm]),
\label{eq:DF_system}
\end{align}
where the indices $+$ and $-$ indicate positive and negative ions with $Z_\pm = \pm 1$.
Additionally, the deviations of the ion number density from the bulk ionic strength are defined 
as $\phi_\pm := \rho_\pm(\vb{r}) - I(\vb{r})$.
In the next step, the grand potential functional in Eq.~\eqref{eq:DF_system} is expanded up to quadratic order in terms of
$\phi_{\pm}$, which provides the following expression
\begin{align}
	\beta \Omega \qty[\rho_\pm] &= \int\limits_\V \dd[3]{r}  \sum_{i=\pm} I(\vb{r}) 
	\qty[\ln(\frac{I(\vb{r})}{\zeta_i}) -1 + \beta V_i(\vb{r})]
	\nonumber	
	\\
	\nonumber
	&+ \int\limits_\V \dd[3]{r} \qty[ \sum_{i=\pm} \phi_i(\vb{r}) \qty{\ln(\frac{I(\vb{r})}{\zeta_i})+ \beta V_i(\vb{r}) 
	+ \frac{\phi_i(\vb{r})}{2I(\vb{r})}} + \frac{\beta \vb{D}(\vb{r},\qty[\rho_\pm])^2}{2\epsilon(\vb{r})}]
	\nonumber
	\\
	&+ \beta\PsiP \int\limits_{\partial\V} \dd[2]{r}  \vb{n}(\vb{r})\vdot \vb{D(\vb{r}},\qty[\rho_\pm]) + \mathcal{O}(\phi^3) .
	\label{eq:exp_DF}
\end{align}
The first line in Eq.~\eqref{eq:exp_DF} of order $\mathcal{O}(\phi_i^0)$ describes the bulk contribution
to the free energy, the second and third line of order $\mathcal{O}(\phi_i^n)$ with $n\geq 1$ denote 
the surface and line contributions. 
For further calculations, the surface and line contributions to the free energy up to second order are defined as
\begin{align}
	\beta \tilde{\Omega} \qty[\phi_{\pm}] :=& 
	\int\limits_\V \dd[3]{r} \qty[\sum_{i=\pm} \phi_i(\vb{r}) \qty{\ln(\frac{I(\vb{r})}{\zeta_i})+ \beta V_i(\vb{r}) 
	+ \frac{\phi_i(\vb{r})}{2I(\vb{r})}} + \frac{\beta \vb{D}(\vb{r},\qty[\phi_\pm])^2}{2\epsilon(\vb{r})}] 
	\nonumber
	\\
	&+ \beta\PsiP\int\limits_{\partial\V} \dd[2]{r}\vb{n}(\vb{r})\vdot \vb{D(\vb{r}},\qty[\phi_\pm]).
\label{eq:Surface_Line_DF}
\end{align}
The equilibrium profile of the ionic number density is obtained by minimizing the functional in
Eq.~\eqref{eq:Surface_Line_DF}, which leads to the Euler-Lagrange equation
\begin{equation}
	\delta \qty(\TO) = 0.
	\label{eq:ELG}
\end{equation}
According to Eq.~\eqref{eq:Surface_Line_DF}, one can write
\begin{align}
	\delta \qty(\beta \tilde{\Omega}\qty[\phi_\pm]) =& \nonumber
	\int\limits_\V \dd[3]{r} \sum_{i=\pm} \delta I(\vb{r}) \qty{\ln(\frac{I(\vb{r})}{\zeta_i})
	+ \beta V_i(\vb{r}) + \frac{\phi_i(\vb{r})}{2I(\vb{r})}}
	\nonumber
	\\
	&+ \int\limits_\V \dd[3]{r} \frac{\beta \vb{D}(\vb{r},\qty[\phi_\pm])}{\epsilon(\vb{r})}
	\delta \vb{D}(\vb{r},\qty[\phi_{\pm}]) 
	\nonumber	
	\\
	&+ \beta\PsiP \int\limits_{\partial\V} \dd[2]{r} \vb{n}(\vb{r})\vdot \delta\vb{D(\vb{r}},\qty[\phi_\pm]).
\label{eq:ELG1}
\end{align}
Furthermore, the electric displacement field $\D$ can be expressed in terms of the electrostatic potential as 
$\D = -\epsilon(\vb{r})\grad{\Psii}$. 
Using this relation and applying the divergence theorem, the second line of Eq.~\eqref{eq:ELG1} now reads
\begin{align}
	&\int\limits_\V \dd[3]{r} \frac{\beta\vb{D}(\vb{r},\qty[\phi_{\pm}])}{\varepsilon(\vb{r})} 
	\delta \vb{D}(\vb{r},\qty[\phi_{\pm}])
	\\
	&= \int\limits_\V \dd[3]{r} \qty[-\grad{\Psi(\vb{r},\qty[\phi_\pm])}] \delta \vb{D}(\vb{r},
	\qty[\phi_{\pm}])
	\\
	&= \beta e \sum_{i=\pm} Z_i \int\limits_\V \dd[3]{r} \Psi(\vb{r},\qty[\phi_\pm]) \delta \phi_i(\vb{r}) 
	- \beta \int\limits_{\partial\V} \dd[2]{r} \Psi(\vb{r},\qty[\phi_\pm]) \qty[\vb{n}(\vb{r}) 
	\vdot \delta \vb{D}(\vb{r},\qty[\phi_{\pm}])].
\end{align}
As a result, the Euler-Lagrange equation in Eq.~\eqref{eq:ELG1} now yields
\begin{align}
	0=&\int\limits_\V \dd[3]{r} \sum_{i=\pm} \delta \phi_i(\vb{r}) \qty[\ln(\frac{I(\vb{r})}{\zeta_i})+ \beta V_i(\vb{r}) 
	+ \frac{\phi_i(\vb{r})}{I(\vb{r})} + \beta e Z_i \Psi(\vb{r},\qty[\phi_\pm])]
	\nonumber
	\\
	&- \beta \int\limits_\V \dd[2]{r} \qty[\Psi(\vb{r},\qty[\phi_\pm]) - \PsiP] \vb{n(\vb{r})} 
	\vdot \delta \vb{D}(\vb{r},\qty[\rho]).
\label{eq:ELG2}
\end{align}
The first line of Eq.~\eqref{eq:ELG2} directly implies
\begin{equation}
	\ln(\frac{I(\vb{r})}{\zeta_i})+ \beta V_i(\vb{r}) + \frac{\phi_i(\vb{r})}{I(\vb{r})} 
	+ \beta e Z_i \Psi(\vb{r},\qty[\phi_\pm]) = 0.
\label{eq:LN}
\end{equation}
Additionally, the second line of Eq.~\eqref{eq:ELG2} provides the Dirichlet boundary condition for the 
electrostatic potential at the surface $z=0$ and $z=L$ of system volume $\V$
\begin{equation}
	\Psi(x,0) = \Psi(x,L) = \PsiP.
\label{eq:PsiP}
\end{equation}
\subsection{Bulk of medium 1}

In the bulk of medium 1 the quantities read $I(\vb{r}) = I_1$, $\beta V_\pm(\vb{r}) = 0$, $\phi_\pm(\vb{r}) = 0$ and $\Psii=0$.
Hence, Eq.~\eqref{eq:LN} yields
\begin{equation}
	\ln (\frac{I_1}{\zeta_\pm}) = 0 \qquad \Rightarrow \qquad I_1 = \zeta_\pm.
\label{eq:Bulk1}
\end{equation}
\subsection{Bulk of medium 2}

In the bulk of medium 2 the quantities read $I(\vb{r}) = I_2$, 
$\beta V_\pm(\vb{r}) = \beta f_\pm$, $\phi_\pm(\vb{r}) = 0$ and $\Psii=\PsiD$ with the Donnan-Potential $\PsiD$ 
originating due to a difference in solubilties of the ions in the two media \cite{bagotsky2005}. 
Therefore, Eq.~\eqref{eq:LN} gives
\begin{equation}
	\ln(\frac{I_2}{\zeta_\pm}) + \beta f_\pm \pm \beta e \PsiD = 0.
\label{eq:Bulk2}
\end{equation}
Applying $I_1 = \zeta_\pm$, Eq.~\eqref{eq:Bulk2} can be re-written as
\begin{equation}
	\ln(\frac{I_2}{I_1}) + \beta f_\pm \pm \beta e \PsiD = 0.
\end{equation}
Adding the two expressions in Eq.~\eqref{eq:Bulk2} one obtains
\begin{align}
	&2\ln(\frac{I_2}{I_1}) + \beta (f_+ + f_-) = 0,
	\\
	\Leftrightarrow \ &\frac{I_2}{I_1} = \exp(-\frac{\beta}{2	}(f_+ + f_-)).
\end{align}
Whereas by subtracting the two equations in Eq.~\eqref{eq:Bulk2}, one arrives at the expression for the Donnan-potential as
\begin{equation}
	\PsiD = -\frac{1}{2e}(f_+ + f_-).
\end{equation}
\subsection{Non-bulk} 

Introducing  the bulk potential as
\begin{align}
	\Psi_{b}(\vb{r})=
	\begin{cases}
		\Psi_{b,1} = 0 \qquad &x<0 \qquad \textrm{medium 1}
		\\
		\Psi_{b,2} = \PsiD\qquad &x>0 \qquad \textrm{medium 2},
	\end{cases}
	\label{eq:bulkpotential}
\end{align}
Eqs.~\eqref{eq:Bulk1} and~\eqref{eq:Bulk2} can be summarized as
\begin{equation}
	\ln(\frac{I_2}{\zeta_\pm}) + \beta V_\pm(\vb{r}) + \beta e \Psi_{b}(\vb{r})= 0.
\label{eq:final_bulk}
\end{equation}
Subtracting the bulk contribution in Eq.~\eqref{eq:final_bulk} from Eq.~\eqref{eq:LN},
one obtains the equilibrium profile of the ionic number density profile
\begin{equation}
	\phi_i(\vb{r}) = -\beta e Z_i I (\vb{r}) \qty[\Psii - \Psi_{b}(\vb{r})],
\label{eq:ion_density}
\end{equation}
in medium $i \in \qty{1,2}$.
According to the electrostatic Gauss' law and with $e \sum_{i=\pm} Z_i I(\vb{r})= 0$ 
one obtains the following expression
\begin{align}
	\div{\D} &= - \div\qty[{\epsilon(\vb{r})\grad{\Psii}}]
	\nonumber
	\\
	&= e \sum_{i = \pm} Z_i \rho_i(\vb{r}) 
	\nonumber
	\\
	&= e \sum_{i = \pm} Z_i (\phi_i(\vb{r})+ I(\vb{r}))
	\nonumber
	\\
	&= -2\beta e I (\vb{r}) \qty[\Psii - \Psi_{b}(\vb{r})].
\label{eq:divD}
\end{align}
The permittivity  varies step-like at the fluid interface at $x=0$, therefore it can be expressed 
by using the Heaviside step function as introduced in Eq.~\eqref{eq:heaviside}
\begin{equation}
	\epsilon(\vb{r}) = \eone\Theta(-x) + \etwo \Theta(x).
\end{equation}
Subsequently, Eq.~\eqref{eq:divD} now reads
\begin{align}
	-\delta(x) \eone  \partial_x \Psii + \delta(x) \etwo \partial_x \Psii 
	+ \epsilon(\vb{r}) \laplacian \Psii
	\nonumber
	\\
	= 2 \beta e^2 I(\vb{r})\qty[\Psii - \Psi_{b}(\vb{r})].
\label{eq:delta}
\end{align}
As a result of integrating Eq.~\eqref{eq:delta} over the interval $x \in \qty[-\xi,\xi]$ and taking 
the limit of $\xi \rightarrow 0$, one obtains the boundary condition of the electric displacement
field $\D$ at the fluid interface 
\begin{equation}
	\eone \partial_x \Psi(\vb{r})\rvert_{x=0} - \etwo \partial_x \Psi(\vb{r})\rvert_{x=0} = 0.
\end{equation}
Finally, for $x\neq0$, the linearized Poisson-Boltzmann equation, also known as the Debye-Hückel equation,
is obtained as
\begin{align}
	\laplacian [\Psii -\Psi_{b}(\vb{r})] = \kappa^2(\vb{r})[\Psii -\Psi_{b}(\vb{r})],
\label{eq:DH_EQ}
\end{align}
with the inverse Debye length
\begin{align}
	\kappa(\vb{r}):= \sqrt{\frac{2\beta e^2I(\vb{r})}{\epsilon(\vb{r})}}.
\label{eq:inv_debye}
\end{align}
\paragraph{Remark} 

The non-linear Poisson-Boltzmann equation is derived within
the framework of density functional theory in Ref.~\cite{majee2016} to study
the electrostatic interaction between two non-metallic colloids within a numerical approach.
To obtain the non-linear Poisson Boltzmann equation, one has to minimize the density functional
given in Eq.~\eqref{eq:DF_system} without expanding it up to quadratic order in $\phi_i$.
In similar fashion to the calculations in Section~\ref{sec:lin_PB}, the Euler-Lagrange 
equation leads to the corresponding equilibrium profile of the ionic number density
\begin{equation}
	\phi_i(\vb{r}) =  I(\vb{r}) \qty[\exp(-\beta e Z_i(\Psi(\vb{r},\qty[\phi_\pm]) -  \Psi_{b}(\vb{r})))-1].
\label{eq:nonlinear_density}
\end{equation}
And therefore, the non-linear Poisson-Boltzmann equation is given by
\begin{align}
	\laplacian (\beta e \Psi(\vb{r},\qty[\phi_\pm])) 
	= \kappa^2(\vb{r}) \sinh(\beta e (\Psi(\vb{r},\qty[\phi_\pm]) -  \Psi_{b}(\vb{r}))),
\end{align}
with the inverse Debye length as defined in Eq.~\eqref{eq:inv_debye}.

\section{Interaction potential}

The interaction contributions of the surface and line parts to the free energy functional are given by 
\begin{align}
	\beta \tilde{\Omega} \qty[\phi_{\pm}] =& 
	\int\limits_\V \dd[3]{r} \sum_{i=\pm} \phi_i(\vb{r}) \qty{\ln(\frac{I(\vb{r})}{\zeta_i})+ \beta V_i(\vb{r}) 
	+ \frac{\phi_i(\vb{r})}{2I(\vb{r})}} + \frac{\beta \vb{D}(\vb{r},\qty[\phi_\pm])^2}{2\epsilon(\vb{r})} 
	\nonumber
	\\
	&+ \beta\PsiP \int\limits_{\partial\V} \dd[2]{r}  \vb{n}(\vb{r})\vdot \vb{D(\vb{r}}).
\end{align}
Using Eqs.~\eqref{eq:final_bulk} and~\eqref{eq:ion_density}, the surface and line contribution can be re-written as
\begin{align}
 \beta \tilde{\Omega} \qty[\phi_{\pm}] =& - \frac{\beta}{2} \int\limits_\V \dd[3]{r} 
 \div{\D} \qty{ \Psi_{b}(\vb{r}) + \Psii} + \frac{\beta\D^2}{2\epsilon(\vb{r})}
 \nonumber
 \\
 &+ \beta \PsiP \int\limits_{\partial\V} \dd[2]{r}\n \vdot \D,
\end{align}
which can further be reduced to
\begin{align}
	\TO =& - \frac{\beta}{2} \int\limits_\V \dd[3]{r} \qty(\div \D) \Psii + \qty(\div\D) \Psi_{b}(\vb{r}) 
	\nonumber
	\\
	&+ \D \vdot (\grad{\Psii})
	\nonumber
	\\
	&+ \beta \PsiP \int\limits_{\partial\V} \dd[2]{r} \n \vdot \D,
\label{eq:Sur_line}
\end{align}
by using 
\begin{equation}
	\frac{\beta \D^2}{2\epsilon(\vb{r})} = -\frac{\beta}{2} \D \grad{\Psii}.
\end{equation}
Applying the product rule, Eq.~\eqref{eq:Sur_line} yields
\begin{align}
	\TO =& -\frac{\beta}{2} \int\limits_\V \dd[3]{r} \div{\qty[\D(\Psii + \Psi_{b}(\vb{r}))]} 
	\nonumber
	\\	
	&- \D(\grad{\Psi_{b}(\vb{r})})
	+ \beta \PsiP \int\limits_{\partial\V} \dd[2]{r} \n \vdot \D,
\end{align}
which results in 
\begin{align}
	\tilde{\Omega}\qty[\phi_\pm] =& \int\limits_{\partial\V} \dd[2]{r} (\n \vdot \D)\qty[\Psii + \Psi_{b}(\vb{r}) - 2\PsiP]
	\nonumber
	\\
	&+ \frac{\PsiD}{2} \int\limits_{x=0} \dd[2]{r} \D \vdot \vb{e}_x,
\label{eq:omega}
\end{align}
by applying the divergence theorem and using $\grad{\Psi_{b}(\vb{r})} = \delta(x) \PsiD \vb{e}_x$.
Here, $\delta(x)$ denotes the Dirac delta distribution.
The system volume $\V$ between the two planar walls is set to 
$\V = \qty[-L_x,L_x] \times \qty[0,L_y] \times \qty[0,L]$.
Hence, Eq.~\eqref{eq:omega} yields
\begin{align}
	\tilde{\Omega}\qty[\phi_\pm] =&- \frac{L_y}{2} \int\limits_{-L_x}^0 \dd{x} \eone \partial_z \Psi_1(x,z)\rvert_{z=0} \qty[\Psi_1(x,0)-2 \PsiP] 
	\nonumber
	\\
	&+\frac{L_y}{2} \int\limits_{-L_x}^0 \dd{x} \eone \partial_z \Psi_1(x,z)\rvert_{z=L} \qty[\Psi_1(x,L)-2 \PsiP]
	\nonumber
	\\
	&-\frac{L_y}{2} \int\limits_{0}^{L_x} \dd{x} \etwo \partial_z \Psi_2(x,z)\rvert_{z=0}\qty[\Psi_2(x,0)+\PsiD - 2\PsiP]
	\nonumber
	\\
	&+ \frac{L_y}{2} \int\limits_{0}^{L_x} \dd{x} \etwo \partial_z \Psi_2(x,z)\rvert_{z=L}\qty[\Psi_2(x,L)+\PsiD - 2\PsiP]
	\nonumber
	\\
	&- \frac{\PsiD L_y}{2} \int\limits_0^L \dd{z} \eone\partial_x \Psi_1(x,z)\rvert_{x=0}.
	\label{eq:omega_tilde_long}
\end{align}
As derived in Eq.~\eqref{eq:PsiP}, the electrostatic potential in medium $i \in \qty{1,2}$ at $z=0$ and $z=L$ is 
given by the boundary condition
\begin{align}
	\Psi_i(x,0) = \Psi_i(x,L) = \PsiP. 
\end{align}
Furthermore, the symmetry of the problem provides the following relations, 
which are used to simplify Eq.~\eqref{eq:omega_tilde_long}
\begin{align}
	\epsilon_i \partial_z \Psi_i(x,z)\rvert_{z=0} &= - \epsilon_i \partial_z \Psi_i(x,z)\rvert_{z=L},
	\\
	\eone \partial_x \Psi_1(x,z)\rvert_{x=0} &=  \etwo \partial_x \Psi_2(x,z)\rvert_{x=0}.
\end{align}
Finally, the expression for the surface and line contributions to the free energy functional
in terms of the electrostatic potential is given by 

\begin{shaded*}
\begin{align}
	\tilde{\Omega}\qty[\phi_\pm] =~&\eone L_y \PsiP \int\limits_{-L_x}^0 \dd{x} \partial_z \Psi_1(x,z)\rvert_{z=0} 
	\nonumber
	\\
	&+ (\PsiP-\PsiD) \etwo L_y \int\limits_{0}^{L_x} \dd{x} \partial_z \Psi_2(x,z)\rvert_{z=0}
	\nonumber
	\\
	&- \frac{\eone \PsiD L_y}{2} \int\limits_0^L \dd{z} \partial_x \Psi_1(x,z)\rvert_{x=0}.
\label{eq:INTERACTION_POTENTIAL}
\end{align}
\end{shaded*}
%

%% file: figure/tikz/colloids_interface.tikz
\begin{tikzpicture}
		\filldraw[lightgrey] (0,0) circle[radius=1.5];
		\draw[mpired,very thick] (0,0) circle[radius=1.5];
		\filldraw[lightgrey] (3.5,0) circle[radius=1.5];
		\draw[mpired,very thick] (3.5,0) circle[radius=1.5];
		\draw[mpidarkblue,very thick] (-1.75,0) -- (5.25,0);
		\draw[dashed,thick] (1.3,-1.5) -- (1.3,1.5) -- (2.2,1.5) -- (2.2,-1.5) -- cycle;
		\draw[->,thick] (2.2,1.5) .. controls (2.8,2.3) .. (5,1.5);
		\draw (0,2) node[draw,rounded corners]{Medium 2};
		\draw (0,-2) node[draw,rounded corners]{Medium 1};
		\draw (-1.4,2.9) node{(a)};
\end{tikzpicture}

%% file: figure/tikz/walls_interface.tikz
\begin{tikzpicture}
		\filldraw[lightgrey] (0,0) -- (0.25,0) -- (0.25,4.46) -- (0,4.46) -- cycle;
		\filldraw[lightgrey] (4.75,0) -- (5,0) -- (5,4.46) -- (4.75,4.46) -- cycle;
		\draw[->,thick,dashed] (0.25,4.46) -- (0.25,4.96);
		\draw[->,thick,dashed] (4.75,2.23) -- (5.5,2.23);
		\draw[mpired,very thick] (0.25,0) -- (0.25,4.46);
		\draw[mpired,very thick] (4.75,0) -- (4.75,4.46);
		\draw[mpidarkblue,very thick] (0.25,2.23) -- (4.75,2.23);
		\draw (0.1,4.65) node {x};
		\draw (5.3,2.43) node {z};
		\draw (0.6,5.12) node {(b)};
		\draw (0.7,1.93) node{(0,0)};
		\draw (2.5,4.21) node[draw,rounded corners]{$\etwo, \ktwo$};
		\draw (2.5,0.25) node[draw,rounded corners]{$\eone, \kone$};
		\draw (4.25,1.93)	node{(0,$L$)};
		\draw (0.55,0.25) node{$\PsiP$};
		\draw (0.55,4.21) node{$\PsiP$};
		\draw (4.45,0.25) node{$\PsiP$};
		\draw (4.45,4.21) node{$\PsiP$};
\end{tikzpicture}

%% file: chapters/03_Electrostatic_Potential.tex

\chapter{Electrostatic Potential}
\label{ch:3}

This chapter aims to find the electrostatic potential $\Psi(x,z)$, which solves
the Debye-Hückel equation of the system described in Section~\ref{sec:system}. 
Section~\ref{sec:exact_potential} presents the electrostatic potential $\Psi^e(x,z)$ obtained 
within exact calculations and Section~\ref{sec:superposition} provides the electrostatic 
potential $\Psi^s(x,z)$ within the superposition approximation.

\section{Electrostatic potential}

The electrostatic potential due to two chemically identical colloidal particles, which carry the surface potential $\Psi_P$ 
and are situated at a fluid interface between medium 1 and 2 at $x=0$ reads
\begin{equation}
	\Psi(x,z) =
	\begin{cases}
		\Psi_1(x,z) \qquad x<0 \qquad \textrm{medium 1}
		\\
		\Psi_2(x,z) \qquad x>0 \qquad \textrm{medium 2}.
	\end{cases}
\end{equation}
In order to determine the electrostatic potential distribution for medium $i\in\qty{1,2}$, of the system described in Section~\ref{sec:system},
one has to solve each corresponding Debye-Hückle equation given by
\begin{equation}
	\laplacian \qty(\Psi_i(x,z)-\Psi_{b,i}) = \kappa^2_i \qty(\Psi_i(x,z) - \Psi_{b,i}).
\label{eq:total_dh}
\end{equation}
Here, potential in the bulk is given by $\Psi_{b,i}$ as specified in Eq.~\eqref{eq:bulkpotential} and the inverse
Debye length $\kappa_i$ is given by Eq.~\eqref{eq:debye_length}.

\section{Exact electrostatic potential}
\label{sec:exact_potential}

The electrostatic potential $\Psi_i^e(x,z)$ in medium $i\in\qty{1,2}$ within exact calculations (denoted with superscript $e$) 
is obtained by applying a procedure similiar to Refs.~\cite{majee2014, majee2018}, which already has been proven to be successful.
In order to obtain a solution, the actual problem is split into two different sub-problems as depicted in Fig.~\ref{fig:e_problems}.
The electrostatic potentials for each of the two sub-problems are then obtained by solving their corresponding Debye-Hückel equations.
As mentioned above, as a first step the system is split into the following two sub-problems.

\begin{figure}[htbp]
	\centering	
	\hfil
	\input{figure/tikz/exact_interface.tikz}
	\hfil
	\input{figure/tikz/exact_walls.tikz}
	\hfil
	\caption{
		(a) Sketch of sub-problem \ref{enum:E_P1}: Two fluids separated by a fluid-fluid interface at $x=0$ in absence of any walls. 
		Medium 1 $x<0$ and medium 2 $x>0$.
		(b) Sketch of sub-problem \ref{enum:E_P2}: Two walls located,one at $z=0$ the other at $z=L$, filled with medium $i \in \qty{1,2}$.
		Both walls carry the constant surface potential $\PsiP$. The fluids for each sub-problem are characterized by their dielectric permittivity
		$\eone$, $\etwo$, and inverse Debye length $\kone$, $\ktwo$, respectively.
		}
	\label{fig:e_problems}
\end{figure}
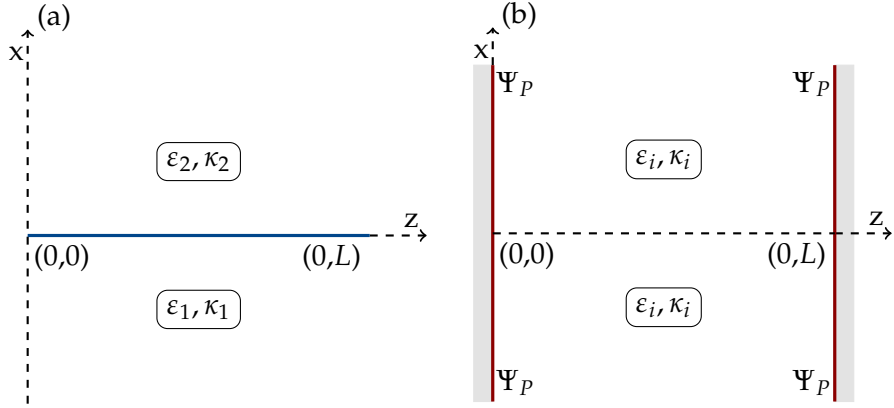
\begin{enumerate}[wide, labelindent=0pt,label={{\color{mpidarkblue}$\vartriangleright$} \textit{Sub-problem \arabic*}},ref=\arabic*]
	\item \label{enum:E_P1} 
		The first sub-problem only deals with the flat fluid interface at $x=0$ between the two fluids in the absence of any walls 
		(See Fig.~\ref{fig:e_problems}a).
		The potential $\tpsi_i(x)$ obtained by solving the Debye-Hückel equation of this system 
		is, due to the symmetry of the problem, only dependent on the $x$-coordinate. 
	\item \label{enum:E_P2}
		The second sub-problem deals with two flat walls, one located at $z=0$ and one at $z=L$, each carrying constant surface potential $\PsiP$. 
		The space in between those walls is filled with either medium 1 or 2 (See Fig.~\ref{fig:e_problems}b).
		Once again, due to the symmetry of this problem, the potential $\phi_i(z)$ is now only dependent on the $z$-coordinate and is obtained
		by solving the Debye-Hückel equation.
\end{enumerate}
Since the Debye-Hückel equation is a linear differential equation, the solution of the complete system, in principle,
can be obtained by adding the solutions for sub-problems~\ref{enum:E_P1} and~\ref{enum:E_P2} as $\Psi_i^e(x,z) = \tpsi_i(x) + \phi_i(z)$. 
As it turns out, this solution satisfies all boundary conditions but the continuity of the electrostatic potential 
and electric displacement field at the fluid interface. 
However by introducing a correction function $c_i(x,z)$, which itself is a solution of the Debye-Hückel equation, 
the continuity at the fluid interface can be restored. 
Thus, the final expression for the exact electrostatic potential now reads
\begin{equation}
	\Psi_i^e(x,z) = \tpsi_i(x) + \phi_i(z) + c_i(x,z).
	\label{eq:E:TOTAL_POT}
\end{equation}
The boundary value problem for the electrostatic potential $\Psi_i^e(x,z)$ in medium $i \in \qty{1,2}$, which fulfills
Eq.~\eqref{eq:total_dh} requires the following set of conditions to be satisfied:

\begin{enumerate}[wide, labelindent=0pt,label={{\color{mpidarkblue}$\vartriangleright$} \textit{Condition \arabic*}},ref=\arabic*]
	\item \label{E_C1} 
		The electrostatic potential in medium 1 of sub-problem~\ref{enum:E_P1} 
		has to be equal to the bulk potential $\Psi_{b,1} = 0$ far away from the fluid interface. Therefore,
		\begin{equation}
			\tpsi_1(x \rightarrow -\infty) = 0.
		\end{equation} 
	\item \label{E_C2} 
		The electrostatic potential in medium 2 of sub-problem~\ref{enum:E_P2} has to be equal to the bulk potential $\Psi_{b,2} = \Psi_D$ far away
		from the fluid interface. Therefore,
		\begin{equation}
			\tpsi_2(x \rightarrow \infty) = \PsiD.
		\end{equation}
	\item \label{E_C3} 
		The (total) electrostatic potential of the system has to be continuous at the fluid interface, meaning
		\begin{equation}
			\Psi_1^e(0^-,z) = \Psi_2^e(0^+,z).
		\end{equation}
	\item \label{E_C4} 
		The normal component of the electric displacement field has to be continuous at the fluid interface
		\begin{equation}		
			\eone \partial_x \Psi^e_1(0^-,z) = \etwo \partial_x \Psi^e_2(0^+,z).
		\end{equation}
	\item \label{E_C5}
		The electrostatic potential at the walls, located at $z=0$ and $z=L$, 
		has to be equal to the constant surface potential $\PsiP$ in both media 
		$i\in\qty{1,2}$
		\begin{equation}
				\phi_i(z=0) = \phi_i(z=L) = \PsiP .
		\end{equation}
\end{enumerate}

\subsection{Fluid interface in the absence of any walls}
\label{subsec:E_interface}

The electrostatic potential of sub-problem~\ref{enum:E_P1} (see Fig. \ref{fig:e_problems}a) is denoted as
\begin{align}
	\tpsi(x) = 
	\begin{cases}
		\tpsi_1(x) \qquad x < 0  \qquad \textrm{medium 1}
		\\
		\tpsi_2(x) \qquad x > 0 \qquad \textrm{medium 2}, 
	\end{cases}
\end{align}
where $\tpsi_i(x)$ is obtained by solving the Debye-Hückel equation
\begin{equation}
	\laplacian(\tpsi_i(x) - \Psi_{b,i}) = \kappa_{i}^{2} \qty(\tpsi_i (x) - \Psi_{b,i})
\label{eq:E_DH_P1}
\end{equation}
in medium $i \in \qty{1,2}$. 
The bulk potential is set to $\Psi_{b,1} = 0$ and $\Psi_{b,2} = \PsiD$ (Donnan-Potential),
therefore Eq. \eqref{eq:E_DH_P1} leads to the following equations to be solved
\begin{align}
	\laplacian \tilde{\psi}_{1}(x) &= \kappa_{1}^{2} \tilde{\psi}_1(x), 
	\label{eq:dh_tpsi_1}
	\\
	\laplacian(\tilde{\psi}_{2}(x) - \Psi_{D}) &= \kappa_{2}^{2} \qty(\tilde{\psi}_2(x) - \Psi_{D}) 
	\label{eq:dh_tpsi_2}.
\end{align}
Since the potential $\tpsi_i(x)$ is a function of $x$ only, the Laplacian $\laplacian$ in Eqs.~\eqref{eq:dh_tpsi_1} 
and~\eqref{eq:dh_tpsi_2} can be written as  $\laplacian \equiv \dv*[2]{x}$. 
These differential equations are solved by using an exponential ansatz
\begin{align}
	\tilde{\psi}_{1}(x) &= A e^{- \kappa_{1}x} + B e^{\kappa_{1}x},
	\label{eq:E_tpsi1_AB}
	\\
	\tilde{\psi}_{2}(x) &= C e^{- \kappa_{2}x} + D e^{\kappa_{2}x} + \Psi_{D},
	\label{eq:E_tpsi2_CD}
\end{align}
where the integration constants $A,B,C$ and $D$ are calculated by using the above-mentioned boundary conditions.
As a consequence of condition~\ref{E_C1} and~\ref{E_C2}, one can easily conclude that $A=D=0$. 
Subsequently condition \ref{E_C3} then leads to 
\begin{equation}
B = C + \PsiD.
\label{eq:B=C+PsiD}
\end{equation} 
Using $A=D=0$, one can write from Eqs.~\eqref{eq:E_tpsi1_AB} and~\eqref{eq:E_tpsi2_CD}
\begin{align}
	\partial_x \tilde{\psi}_{1}(x)\rvert_{x=0} &= \eone \kappa_{1} B,
	\\
	\partial_x \tilde{\psi}_{2}(x)\rvert_{x=0} &= - \etwo \kappa_{2} C.
\end{align}
Using the relation in Eq. \eqref{eq:B=C+PsiD} and condition \ref{E_C4}, one obtains
\begin{align}
	B &= \frac{\varepsilon_{2}\kappa_{2}}{\varepsilon_{1}\kappa_{1} + \varepsilon_{2}\kappa_{2}}  \Psi_{D},
	\\
	C &= - \frac{\varepsilon_{1}\kappa_{1}}{\varepsilon_{1}\kappa_{1} + \varepsilon_{2}\kappa_{2}}  \Psi_{D}.
\end{align}
Given this, the final expression for the electrostatic potentials in the two media in the absence of any walls can now be stated as follows
\begin{shaded*}
\begin{align}
	\tpsi_1 (x) &=  \Psi_{D}  \frac{\varepsilon_{2}\kappa_{2}}{\varepsilon_{1} \kappa_{1} + \varepsilon_{2}\kappa_{2}} e^{\kappa_{1}x}, 
	\\
	\tpsi_2 (x) &=  \Psi_{D}  \qty[1- \frac{\varepsilon_{1}\kappa_{1}}{\varepsilon_{1}\kappa_{1} + \varepsilon_{2}\kappa_{2}} e^{-\kappa_{2} x}].
\end{align}
\end{shaded*}
\subsection{Two planar walls with fluid in between}

The electrostatic potential of two planar walls, filled with medium $i\in\qty{1,2}$,
as specified in sub-problem~\ref{enum:E_P2} and seen in Fig.~\ref{fig:e_problems}b is given by 
\begin{align}
	\phi(z) = 
	\begin{cases}
		\phi_1(z) \qquad x<0 \qquad \textrm{medium 1} 
		\\
		\phi_2(z) \qquad x>0 \qquad \textrm{medium 2}.
	\end{cases}
\end{align}
The corresponding Debye-Hückel equation for the electrostatic potential within this sub-problem reads
\begin{equation}
	\laplacian(\phi_{i}(z) - \Psi_{b,i}) = \kappa_{i}^{2} \qty(\phi_{i}(z) - \Psi_{b,i}),
\label{eq:DH_P2}
\end{equation}
with the Laplacian $\laplacian \equiv \dv*[2]{z}$.
As a next step one has to determine $\phi_i(z)$ for each medium $i$.

\subsubsection{Medium 1}
In medium 1 the bulk potential is given by $\Psi_{b,1} = 0$ and the Debye-Hückel equation 
in Eq.~\eqref{eq:DH_P2} therefore reads
\begin{equation}
	\laplacian \phi_{1}(z) = \kappa_{1}^{2} \phi_{1}(z),
\end{equation}
with the general solution 
\begin{equation}
	\phi_{1}(z) = A e^{- \kappa_{1} z} + B e^{\kappa_{1} z}.
\end{equation}
By using boundary condition~\ref{E_C5} one obtains
\begin{align}
	&\PsiP = A + B ,
	\\
	&\PsiP = A  e^{- \kappa_{1} L} + B  e^{\kappa_1 L}.
\end{align}
Subsequently, solving this set of linear equations leads to the solution for the integration constants 
\begin{align}
	A &= \Psi_{P} \qty[ 1 - \frac{1-e^{- \kappa_{1}L}}{2 \sinh(\kappa_{1}L)}],
	\\
	B &= \Psi_{P} \frac{1- e^{- \kappa_{1}L}}{2 \sinh(\kappa_{1}L)} .
\end{align}
Finally, the solution for the electrostatic potential of sub-problem~\ref{enum:E_P2} is given by
\begin{shaded*}
\begin{equation}
	\phi_{1}(z) =  \PsiP \qty[ \frac{\sinh(\kappa_{1}z)-\sinh(\kappa_{1}(z-L))}{\sinh(\kappa_{1}L)}].
\end{equation}
\end{shaded*}

\subsubsection{Medium 2}
The calculations for medium 2 are similar to those for medium 1. 
However, the bulk potential is now given by $\Psi_{b,2} = \PsiD$ instead so that one needs to solve 
\begin{equation}
	\laplacian(\phi_{2}(z) - \Psi_{D}) = \kappa_{2}^{2} \qty(\phi_{2}(z) - \Psi_{D}).
\end{equation}
Here, the general solution is given by
\begin{equation}
\phi_{2}(z) = \Psi_{D} + A e^{- \kappa_{2}z} + B e^{\kappa_{2}z}.
\end{equation}
Analogous to the solution for medium 1, condition~\ref{E_C5} leads to a system of linear equations for the integration constants
\begin{align}
	\PsiP &= A + B +\PsiD,
	\\
	\PsiP &=  A  e^{-\kappa_{2}L} + B  e^{\kappa_{2}L} + \PsiD.
\end{align}
Once again, the integration constants are obtained by solving the system of linear equations.
The integration constants read
\begin{align}
	B &= (\Psi_{P} - \Psi_{D}) \frac{1- e^{-\kappa_{2}L}}{2 \sinh(\kappa_{2}L)} ,
	\\
	A &= (\Psi_{P} - \Psi_{D}) \qty[ 1 - \frac{1-e^{-\kappa_{2}L}}{2 \sinh(\kappa_{2}L)}].
\end{align}
Therefore, the solution of sub-problem~\ref{enum:E_P2} for the electrostatic potential is obtained as
\begin{shaded*}
\begin{equation}
	\phi_{2}(z) = \Psi_{D} + (\Psi_{P} - \Psi_{D}) \qty[ \frac{\sinh(\kappa_{2}z)-\sinh(\kappa_{2}(z-L))}{\sinh(\kappa_{2}L)}].
\end{equation}
\end{shaded*}
\subsection{Correction function}
\label{subsec:E_correction}

As mentioned previously, the sum $\Psi^e_i(x,z) = \tpsi_i(x) + \phi_i(z)$ is a solution of the Debye-Hückel equation for medium $i\in \qty{1,2}$
due to its linear nature and fulfills almost all required boundary conditions, except the continuity of the 
electrostatic potential and electric displacement field at the fluid interface. 
By constructing a correction function $c_i(x,z)$ in similar fashion as described in Refs.~\cite{majee2014, majee2018},
it is possible to rectify this problem and restore the continuity at the fluid interface. 
Accordingly, the final expression for the electrostatic potential is then given by Eq.~\eqref{eq:E:TOTAL_POT}.
Therefore, the correction function also needs to be a solution of the Debye-Hückel equation 

\begin{equation}
	\laplacian c_{i}(x,z) = \kappa_{i}^{2} c_{i}(x,z),
\label{eq:E_DH_Cor}
\end{equation}
and has to keep the already fulfilled conditions unchanged.
To achieve all this, the correction function has to satisfy the following conditions.
\begin{enumerate}[wide, labelindent=0pt,label={{\color{mpidarkblue}$\vartriangleright$} \textit{Condition \arabic*}},ref=\arabic*]
	\item \label{S_C1}
		The electrostatic potential $\Psi_i^e(x,z)$ has to stay finite in the limit $x\to\pm\infty$ 
		and $\tpsi_1(x\to-\infty)= 0$, as well as $\tpsi_2(x\to\infty) = \PsiD$ has to stay unaltered. Therefore,
		\begin{align}
			c_1(x \rightarrow -\infty, z) = 0,
			\\
			c_2(x \rightarrow \infty, z) = 0.
		\end{align}	
	\item \label{S_C2}
		Since $\tpsi_1(x=0) = \tpsi_2(x=0)$ is already fulfilled and the electrostatic potential $\Psi^e_i(x,z)$ has to be continuous
		at the interface $x=0$,
		the correction function has to satisfy
		\begin{equation}		
			c_1(x=0,z) + \phi_1(z) = c_2(x=0,z) +\phi_2(z).
		\end{equation}	
	\item \label{S_C3}
		The surface potential at $z=0$ and $z=L$ has to stay unaltered $\phi_i(z=0) = \phi_i(z=L) = \PsiP$.
		Therefore, the correction function has to fulfill
		\begin{align}		
			c_i(x, z=0) + \tpsi_i(x) =0,
			\\
			c_i(x, z=L) + \tpsi_i(x) =0.
		\end{align}		
	\item \label{S_C4}
		The boundary condition for the electric displacement field at the fluid interface $x=0$ has to stay valid
		\begin{equation}		
		\eone \partial_x c_1(x,z)\rvert_{x=0} = \etwo \partial_x c_2(x,z)\rvert_{x=0}.
		\end{equation}
\end{enumerate}
Since the total electrostatic potential for each medium is obtained by superposing the solutions 
of the sub-problems, one has to be careful in considering the bulk potential $\PsiD$ only once for medium 2.
To end up with the correct bulk potential in the limit $x \rightarrow \infty$, the previous solution $\tpsi_i(x)$ 
of sub-problem~\ref{enum:E_P1} now reads
\begin{align}
	&\tilde{\psi}_1(x) = \frac{\varepsilon_{2}\kappa_{2}}{\varepsilon_{1} \kappa_{1} 
	+ \varepsilon_{2}\kappa_{2}} \PsiD \ e^{\kappa_{1}x}, 
	\\
	&\tilde{\psi}_2(x) = - \frac{\varepsilon_{1}\kappa_{1}}{\varepsilon_{1}\kappa_{1} 
	+ \varepsilon_{2}\kappa_{2}}  \PsiD \ e^{-\kappa_{2} x},
\end{align}
and the solution $\phi_i(z)$ of sub-problem~\ref{enum:E_P2} is given by
\begin{align}
	&\phi_{1}(z) =  \PsiP \qty[ \frac{\sinh(\kappa_{1}z)-\sinh(\kappa_{1}(z-L))}{\sinh(\kappa_{1}L)}]
	\label{eq:E_Cor_Phiz1},
	\\
	&\phi_{2}(z) = (\PsiP - \Psi_{D}) \qty[ \frac{\sinh(\kappa_{2}z)-\sinh(\kappa_{2}(z-L))}{\sinh(\kappa_{2}L)}
	\label{eq:E_Cor_Phiz2}].
\end{align}
The first step to determine the correction function is to expanded $c_i(x,z)$ into a Fourier series in $z \in \qty[0,2L]$
\begin{equation}
	c_{i}(x,z) = \frac{a_{0,i}(x)}{2} + \sum_{n=1}^{\infty} a_{n,i}(x) \sin\qty(\frac{n \pi z}{L}) 
	+ \sum_{n=1}^{\infty} b_{n,i}(x) \cos\qty(\frac{n \pi z}{L}).
\label{eq:E_cor_fourier_komplett}
\end{equation}
The symmetry of the system requires $c_i(x,z)$ to be symmetric about $z = L/2$. 
However there is no such constraint about $z=0$ or $z=L$.
Furthermore the system is expanded in such way that $c_i(x,z)$ is anti-symmetric about $z=L$,
which implies $b_{n,i} = 0$ in Eq.~\eqref{eq:E_cor_fourier_komplett}.
Therefore the Fourier series now reads
\begin{equation}
	c_{i}(x,y) = \frac{a_{0,i}(x)}{2} + \sum_{n=1}^{\infty} a_{n,i}(x) \sin\qty(\frac{n \pi z}{L}) . 
\label{eq:c_fourier}
\end{equation}
Inserting $c_i(x,z)$ into the Debye-Hückel equation in  Eq.~\eqref{eq:E_DH_Cor}, one obtains
\begin{multline}
	\laplacian c_{i}(x,y) = a_{0,i}^{''}(x) + \qty(\sum_{n=1}^{\infty} a_{n,i}^{''}(x) 
	\sin \qty(\frac{n \pi z}{L}))
	\\
	-\qty(\sum_{n=1}^{\infty} a_{n,i} \frac{n^{2} \pi^{2}}{L^{2}} \sin \qty(\frac{n \pi z}{L})) .  
\end{multline}
This implies
\begin{align}
	a_{0,i}^{''}(x) &= \kappa_{i}^{2} a_{0,i}(x), 
	\label{eq:a0}
	\\
	a_{n,i}^{''}(x) &= \qty[\frac{n^2 \pi^2}{L^2} + \kappa_{i}^{2}] a_{n,i}(x). 
	\label{eq:an}
\end{align}
For the sake of brevity, the following definition will be used in future calculations
\begin{equation}
	p_{n,i} := \sqrt{\frac{n^2\pi^2}{L^2} + \kappa^2_i}.
\end{equation}
The solution of Eq.~\eqref{eq:a0} is given by
\begin{equation}
	a_{0,i} = A_{0,i}  e^{- \kappa_{i}x} + B_{0,i}  e^{\kappa_i x}.
\end{equation}
As a result of condition \ref{S_C1} it is clear that $A_{0,1} = B_{0,2} = 0$, which provides 
the following expressions for the Fourier coefficients $a_{0,i}$ in medium $i\in\qty{1,2}$
\begin{align}
	a_{0,1} &= B_{0,1}  e^{\kone x}, 
	\label{eq:a01}
	\\
	a_{0,2} &= A_{0,2}  e^{- \ktwo x}. 
	\label{eq:a02}
\end{align}
Subsequently, the second differential equation in Eq.~\eqref{eq:an} is solved
via a similar exponential ansatz
\begin{align}
	a_{n,i}= A_{n,i}  e^{-p_{n,i}x} + B_{n,i}  e^{p_{n,i}x}. 
\end{align}
Once again due to conditions~\ref{S_C1} one has $A_{n,1} = B_{n,2}= 0$.
Therefore the Fourier coefficients now read 
\begin{align}
	a_{n,1} &= B_{n,1}  e^{p_{n,1}x}, 
	\label{eq:an1}
	\\
	a_{n,2} &= A_{n,2}  e^{-p_{n,2}x}. 
	\label{eq:an2}
\end{align}
Inserting all of the obtained expressions for the Fourier coefficients in Eqs.~\eqref{eq:a01},~\eqref{eq:a02},~\eqref{eq:an1} 
and~\eqref{eq:an2}, the Fourier series expansion of the correction function in Eq.~\eqref{eq:c_fourier} is now given by
\begin{align}
	c_1 (x,z) &= \frac{B_{0,1}}{2} e^{\kone x} 
	+ \sum_{n=1}^{\infty} B_{n,1}  e^{p_{n,1}x} \sin \qty(\frac{n \pi z}{L}), 
	\label{eq:c1_B}
	\\
	c_2 (x,z) &= \frac{A_{0,2}}{2} e^{-\ktwo x} 
	+ \sum_{n=1}^{\infty} A_{n,2}  e^{-p_{n,2}x} \sin \qty(\frac{n \pi z}{L}). 
	\label{eq:c1_A}
\end{align}
As a further step, the correction function as described in Eqs.~\eqref{eq:c1_B} 
and~\eqref{eq:c1_A} is evaluated at the interface $x=0$, which results in
\begin{align}
	c_1(x=0,z) &= \frac{B_{0,1}}{2} + \sum_{n=1}^{\infty} B_{n,1} \sin(\frac{n\pi z}{L})
	\label{eq:E_c1_x0},
	\\
	c_2(x=0,z) &= \frac{A_{0,2}}{2} + \sum_{n=1}^{\infty} A_{n,2} \sin(\frac{n\pi z}{L})
	\label{eq:E_c2_x0}.
\end{align}
Additionally, calculating the derivatives of Eqs.~\eqref{eq:c1_B} and~\eqref{eq:c1_A} with respect to $x$
one can easily write
\begin{align}
	\partial_x c_1(x,z)\rvert_{x=0} &= \kone\frac{B_{0,1} }{2} + \sum_{n=1}^{\infty}  p_{n,1}B_{n,1} \sin(\frac{n\pi z}{L}),
	\\
	\partial_x c_2(x,z)\rvert_{x=0} &= -\ktwo\frac{A_{0,2} }{2} + \sum_{n=1}^{\infty} p_{n,2} A_{n,2} \sin(\frac{n\pi z}{L}).
\end{align}
In order to determine the constants $B_{0,1}$ and $A_{0,2}$, the correction function $c_i(x,z)$ 
in Eqs.~\eqref{eq:c1_B} and~\eqref{eq:c1_A} is evaluated at the position of the walls $z=0$ and $z=L$, yielding 
\begin{align}
	c_1(x,z=0) &= c_1(x,z=L) = \frac{B_{0,1}}{2} e^{\kone x},
	\\
	c_2(x,z=0) &= c_2(x,z=L) = \frac{A_{0,2}}{2}  e^{-\ktwo x}.
\end{align}
Using these results in condition~\ref{S_C4}, one obtains
\begin{align}
	&A_{0,2} = \frac{2\eone \kone}{\eone \kone + \etwo \ktwo} \PsiD
	\label{eq:E_A_02},
	\\
	&B_{0,1} = - \frac{2\etwo \ktwo}{\eone \kone + \etwo \ktwo} \PsiD
	\label{eq:E_B_O1}.
\end{align}
By using the obtained expressions for $A_{0,2}$ and $B_{0,1}$ condition~\ref{S_C4} now yields
\begin{equation}
	\eone p_{n,1} B_{n,1} = - \etwo p_{n,2} A_{n,2}. 
	\label{eq:E_An1}
\end{equation}
As a further step, one can plug the expressions for $\phi_1(z)$ and $\phi_2(z)$ (Eqs.~\eqref{eq:E_Cor_Phiz1} and~\eqref{eq:E_Cor_Phiz2})
into condition~\ref{S_C2}.
Multiplying both sides with $\int_0^L \dd{z} \sin(\frac{m\pi z}{L})$ and using the relations taken from Ref.~\cite{gradshteyn}
\begin{align}
	&\int\limits_0^L \dd{z} \sin(\frac{m\pi z}{L}) = \frac{L - (-1)^m L}{m\pi},
	\\
	&\int\limits_0^L \dd{z} \sin(\frac{m\pi z}{L}) \sin(\frac{n\pi z}{L}) = \frac{L}{2} \delta_{n,m},
\end{align}
where $\delta_{n,m}$ denotes the Kronecker delta as defined in Eq.~\eqref{eq:kronecker_delta},
one obtains a second equation for the coefficients $B_{n,1}$ and $A_{n,2}$, which reads
\begin{align}
	&-\frac{\etwo \ktwo}{\eone\kone + \etwo\ktwo}  \PsiD  \frac{L - (-1)^n L}{n\pi} + B_{n,1}\frac{L}{2}
	\nonumber
	\\
	&+ \frac{\PsiP}{\sinh(\kone L)} \qty[\frac{-(-1)^n n\pi}{L p^2_{n,1}} \sinh(\kone L) 
	+ \frac{n\pi}{L p^2_{n,1}} \sinh(\kone L)]
	\nonumber
	\\
	&=
	\nonumber
	\\
	&\frac{\eone \kone}{\eone\kone + \etwo\ktwo}  \PsiD  \frac{L - (-1)^n L}{n\pi} 
	+ A_{n,2} \frac{L}{2}
	\nonumber
	\\
	&+ \frac{\PsiP-\PsiD}{\sinh(\ktwo L)} \qty[\frac{-(-1)^n n\pi}{L p^2_{n,2}} \sinh(\ktwo L) 
	+ \frac{n\pi}{L p^2_{n,2}}\sinh(\ktwo L)]. 
	\label{eq:E_An2}
\end{align}
Finally, the coefficients $A_{n,2}$ and $B_{n,1}$ are now obtained by 
solving the set of equations of Eqs.~\eqref{eq:E_An1} and~\eqref{eq:E_An2} as
\begin{align}
	&A_{n,2} = \qty[ \dfrac{\dfrac{2\PsiD}{n^2\pi^2} - \dfrac{2\PsiP}{n^2\pi^2+\kone^2 L^2} 
	+ \dfrac{2(\PsiP-\PsiD)}{n^2\pi^2+\ktwo^2 L^2}}{1 + \dfrac{\etwo p_{n,2}}{\eone p_{n,1}}} ]\qty{(-1)^n -1},
	\\
	\nonumber
	\\
	&B_{n,1} = \qty[ \dfrac{-\dfrac{2\PsiD}{n^2\pi^2} + \dfrac{2\PsiP}{n^2\pi^2+\kone^2 L^2} 
	- \dfrac{2(\PsiP-\PsiD)}{n^2\pi^2+\ktwo^2 L^2}}{1 + \dfrac{\eone p_{n,1}}{\etwo p_{n,2}}} ] \qty{(-1)^n -1}.
\end{align}
Having calculated all previously unknown coefficients of Eqs.~\eqref{eq:c1_B} and ~\eqref{eq:c1_A},
the correction function is now used to obtain the electrostatic potential.

\subsection{Result}
According to Eq.~\eqref{eq:E:TOTAL_POT} the final expression for the electrostatic potential 
in medium 1 within exact calculations is given by
\begin{shaded*}
\begin{align}
\label{eq:exact_potential_1}
	&\Psi_1^e(x,z) = \Psi_{P} \qty[ \frac{\sinh(\kappa_{1}z)-\sinh(\kappa_{1}(z-L))}{\sinh(\kappa_{1}L)}] 
	\\
	&+\sum_{n=1}^{\infty} \qty[ \frac{-\frac{2\PsiD}{n^2\pi^2} 
	+ \frac{2\PsiP}{n^2\pi^2+\kone^2 L^2} 
	- \frac{2(\PsiP-\PsiD)}{n^2\pi^2+\ktwo^2 L^2}}{1 + \frac{\eone p_{n,1}}{\etwo p_{n,2}}} ]
	n\pi \qty{(-1)^n -1}e^{p_{n,1}x} \sin(\frac{n\pi z }{L}),
	\nonumber
\end{align}
\end{shaded*}
\noindent
and the exact electrostatic potential in medium 2 reads
\begin{shaded*}
\begin{align}
	\label{eq:exact_potential_2}
	&\Psi_2^e(x,z) = \PsiD+ (\Psi_{P}-\PsiD) \qty[ \frac{\sinh(\kappa_{2}z)-\sinh(\kappa_{2}(z-L))}{\sinh(\kappa_2 L)}] 
	\\
	&+\sum_{n=1}^{\infty} \qty[ \frac{\frac{2\PsiD}{n^2\pi^2} 
	- \frac{2\PsiP}{n^2\pi^2+\kone^2 L^2} 
	+ \frac{2(\PsiP-\PsiD)}{n^2\pi^2+\ktwo^2 L^2}}{1 + \frac{\etwo p_{n,2}}{\eone p_{n,1}}} ]
	n\pi \qty{(-1)^n -1} e^{-p_{n,2}x} \sin(\frac{n\pi z }{L}).
	\nonumber
\end{align}
\end{shaded*}

\section{Superposition Approximation}
\label{sec:superposition}

The Debye-Hückel equation is a linear differential equation and thus, in principle,
a superposition approximation can be used to obtain the electrostatic potential for the system.
By superposing the electrostatic potential $\Psi_i^{\textrm{sin}}(x,z)$ in medium $i\in\qty{1,2}$ due to a single wall located
at $z=0$, with the electrostatic potential due to a single wall located at $z=L$ which is reflected about its new position, 
one obtains an approximation of the potential in the presence of both colloidal particles.
The electrostatic potential $\Psi_i^s(x,z)$ within this, widely known as, superposition approximation reads
\begin{equation}
	\Psi_i^s(x,z) = \Psi_i^{\textrm{sin}}(x,z) + \Psi_i^{\textrm{sin}}(x,-(z-L)).
	\label{eq:S_TOTAL_POTENTIAL}
\end{equation}
Therefore, one needs to first calculate the electrostatic potential due to a single wall in contact with the two fluid media.
To this end, the system is split into the following sub-problems similar to the exact calculation in Section~\ref{sec:exact_potential}.

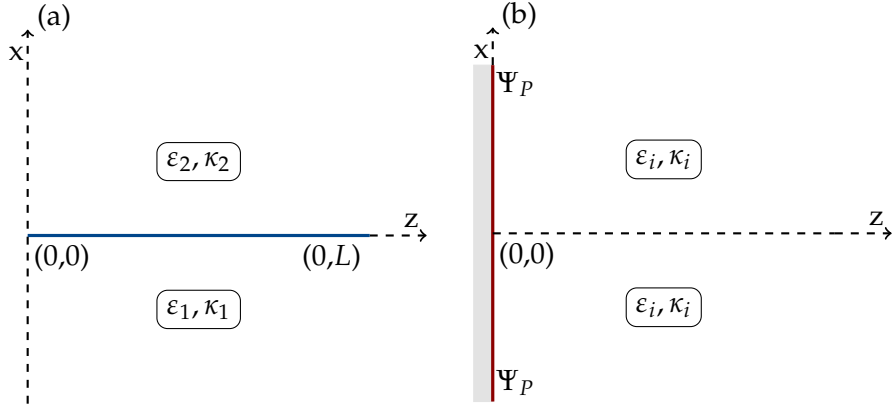
\begin{figure}[htbp]
	\centering	
	\hfil
	\input{figure/tikz/superposition_interface.tikz}
	\hfil
	\input{figure/tikz/superposition_wall.tikz}
	\hfil
	\caption
		{
		(a) Sketch of sub-problem~\ref{enum:E_P1}: Two fluids separated by a fluid-fluid interface 
		at $x=0$ in absence of any walls. Medium 1 $x<0$ and medium 2 $x>0$. 
		(b) Sketch of sub-problem~\ref{S_P2}: One flat wall located at $z=0$ filled by medium 
		$i \in \qty{1,2}$ in the half-space ($z>0$). 
		The wall carries the constant surface potential $\PsiP$ and is characterized by its 
		dielectric permittivity $\eone$, $\etwo$, and inverse Debye length $\kone$, $\ktwo$, respectively.
		}
	\label{fig:s_problems}
\end{figure}
\begin{enumerate}[wide, labelindent=0pt,label={{\color{mpidarkblue}$\vartriangleright$} \textit{Sub-problem \arabic*}},ref=\arabic*]
	\item \label{S_P1} 
		A fluid interface between the two media at $x=0$ without any walls (Fig.~\ref{fig:s_problems}a). 
		The obtained electrostatic potential $\tpsi_i(x)$ is a function of the $x$-coordinate, solves the Debye-Hückel
		equation for each corresponding medium $i\in\qty{1,2}$ and fulfills the boundary conditions at the interface 
		as well as in the limit of $x\to\pm\infty$.
	\item \label{S_P2} 
		A single flat wall at $z=0$ carrying the constant surface potential $\PsiP$, bounding the half-space ($z>0$) and filled by
		medium $i\in\qty{1,2}$. 
		The electrostatic potential $\phi_i(z)$ for each medium $i$ is a function of the $z$-coordinate only,
		solves the Debye-Hückel equation and fulfills the boundary conditions at the single wall ($z=0$) as well as 
		in the limit of $z\to\infty$.
\end{enumerate}
By adding each solution of subproblem~\ref{S_P1} and~\ref{S_P2}, one simply obtains $\Psi_i^s(x,z) = \tpsi_i(x) + \phi_i(z)$ 
for medium $i \in \qty{1,2}$. 
Analogous to the exact calculation in Section~\ref{sec:exact_potential}, this expression for the electrostatic potential 
fulfills the Debye-Hückel equation and the corresponding boundary conditions except the continuity at the fluid interface.
In similar fashion to Section~\ref{subsec:E_correction}, one can construct a correction function $c_i(x,z)$ which 
restores the continuity at the fluid interface and leaves already satisfied conditions unaltered.
The electrostatic potential due to a single wall within superposition approximation is therefore given by
\begin{equation}
	\Psi_i^{\textrm{sin}}(x,z) = \tpsi_i(x) + \phi_i(z) + c_i(x,z).
	\label{eq:S_POTENTIAL}
\end{equation}
\subsection{Fluid interface in the absence of the wall}

The solution of sub-problem~\ref{S_P1} within the superposition approximation is identical 
to sub-problem~\ref{enum:E_P1} in Section~\ref{subsec:E_interface}.
Therefore, the potential $\tpsi_i(x)$ for each medium $i\in \qty{1,2}$ is given by
\begin{shaded*}
\begin{align}
	\tpsi_1 (x) &= \Psi_{D}  \frac{\varepsilon_{2}\kappa_{2}}{\varepsilon_{1} \kappa_{1} 
	+ \varepsilon_{2}\kappa_{2}} \ e^{\kappa_{1}x}, \\
	\tpsi_2 (x) &= \Psi_{D}  \qty[1- \frac{\varepsilon_{1}\kappa_{1}}{\varepsilon_{1}\kappa_{1} 
	+ \varepsilon_{2}\kappa_{2}} \  e^{-\kappa_{2} x}].
\end{align}
\end{shaded*}
\subsection{A single wall in contact with a fluid} 

\subsubsection{Medium 1}

The electrostatic potential due to a single flat wall at $z=0$ carrying constant surface potential $\PsiP$ and 
in contact with medium 1 (see Fig.~\ref{fig:s_problems}b), is given by the solution of the Debye-Hückel equation
\begin{equation}
	\laplacian \phi_1(z) = \kone^2 \phi_1(z).
\end{equation}
Since $\phi_1(z)$ is a function of $z$ only, the Laplacian reads as $\laplacian \equiv \dv*[2]{z}$ in this case.
By using an exponential ansatz, the solution to this differential equation is obtained as
\begin{equation}
	\phi_1(z) = A  e^{-\kone z} + B  e^{\kone z}.
\end{equation}
The boundary condition $\phi_1(z \rightarrow \infty) = 0$ leads to $B=0$ and the boundary condition at the wall 
$\phi_1(z=0) = \PsiP$ provides $A=\PsiP$. Therefore, the final expression for medium 1 reads
\begin{shaded*}
\begin{equation}
	\phi_1 (z) = \PsiP e^{-\kone z}.
\end{equation}
\end{shaded*}

\subsubsection{Medium 2}

For a flat wall at $z=0$, which carries the constant surface potential $\PsiP$ and is in contact 
with medium 2 in the half-space $z>0$, the bulk potential is given by the Donnan-Potential 
$\Psi_{b,2} = \PsiD$ (see Fig.~\ref{fig:s_problems}b).
Therefore, the electrostatic potential $\phi_2(z)$ is obtained by solving the corresponding Debye-Hückel equation
\begin{equation}
	\laplacian \qty( \phi_2(z) - \PsiD) = \kone^2 \qty( \phi_2(z) - \PsiD).
\end{equation}
The general solution to this differential equation reads
\begin{equation}
	\phi_1(z) = A e^{-\ktwo z} + B e^{\ktwo z} + \PsiD.
\end{equation}
For the half-space $z>0$ filled by medium 2, the boundary conditions $\phi_2(z \rightarrow \infty) = \PsiD$ 
and $\phi_2(z=0)=\PsiP$ lead to $B=0$ and $A= (\PsiP-\PsiD)$, respectively. 
Therefore, the electrostatic potential in medium 2 reads
\begin{shaded*}
\begin{equation}
	\phi_2 (z) = (\PsiP -  \PsiD) e^{-\ktwo z} + \PsiD
\end{equation}
\end{shaded*}

\subsection{Correction function}

As rightly mentioned in the beginning of Section~\ref{sec:superposition}, the electrostatic potential obtained by simply adding the solutions 
of sub-problems~\ref{enum:E_P1} and~\ref{enum:E_P2} does not satisfy all conditions at the fluid interface.
This problem is again rectified by the construction of a correction function $c_i(x,z)$ similar to Section~\ref{subsec:E_correction}.
Accordingly, the electrostatic potential within superposition calculations is then given by Eq.~\eqref{eq:S_POTENTIAL}.
Therefore, the correction function solves the Debye-Hückel equation
\begin{equation}
	\laplacian c_i(x,z) = \kappa_i^2 c_i(x,z),
\label{eq:S_cor_DH}
\end{equation}
with the boundary conditions as follows.
\begin{enumerate}[wide, labelindent=0pt,label={{\color{mpidarkblue}$\vartriangleright$} \textit{Condition \arabic*}},ref=\arabic*]
	\item \label{S_COR1}
		The electrostatic potential of a single wall $\Psi_i^{\textrm{sin}}(x,z)$ has to stay finite in the limit of $x\to\pm\infty$ and 
		$z\to\infty$. Therefore,
		\begin{align}
			c_i(x \rightarrow \pm\infty,z) = 0,
			\\
			c_i(x, z\to\infty) = 0.
		\end{align}		
	\item \label{S_COR2}
		The correction function has to satisfy the continuity of the electric displacement field at the fluid interface $x=0$ such that
		\begin{equation}		
			\eone \partial_x c_1(x,z)\rvert_{x=0} = \etwo \partial_x c_2(x,z)\rvert_{x=0}.
		\end{equation}		
	\item \label{S_COR3}
		The surface potential at the wall located at $z=0$ has to stay unaltered $\phi_i(z=0) = \PsiP$.
		Therefore the correction function has to satisfy
		\begin{equation}
			c_i(x,z=0) +  \tpsi_i(x) = 0.
		\end{equation}
	\item \label{S_COR4}
		Since $\tpsi_1(x=0) = \tpsi_2(x=0)$ is already satisfied and the electrostatic potential has to be
		continuous at the fluid interface at $x=0$, the correction function has to satisfy
		\begin{equation}
			c_1(x=0,z) + \phi_1(z) = c_2(x=0,z) + \phi_2(z).
		\end{equation} 
\end{enumerate}
In order to treat condition \ref{S_COR3}, the correction function is constructed as
\begin{equation}
	c_i(x,z) = -\tpsi_i(x) + \frac{1}{2\pi} \int\limits_0^\infty \dd{q} \hat{c}_i(x,z) \sin(qz), 
	\label{eq:S_ci}
\end{equation}
which results in $c_i(x,z=0) = - \tpsi_i(x)$ at the the wall and therefore satisfies the condition.
In similar fashion to the exact calculations, one has to be careful in considering the 
bulk potential $\PsiD$ only once for medium 2. Thus, the previous results for sub-problem~\ref{S_P1} read
\begin{align}
	&\tpsi_1 (x) = \Psi_{D}  \frac{\varepsilon_{2}\kappa_{2}}{\varepsilon_{1} \kappa_{1} 
	+ \varepsilon_{2}\kappa_{2}} e^{\kappa_{1}x}, 
	\\
	&\tpsi_2 (x) = -\Psi_{D}\frac{\varepsilon_{1}\kappa_{1}}{\varepsilon_{1}\kappa_{1} 
	+ \varepsilon_{2}\kappa_{2}} e^{-\kappa_{2} x},
\end{align}
and those for sub-problem~\ref{S_P2} are given by
\begin{align}
	&\phi_1 (z) = \PsiP e^{-\kone z},
	\\
	&\phi_2 (z) = (\PsiP - \PsiD) e^{-\ktwo z}. 
\end{align}
By inserting Eq.~\eqref{eq:S_ci} into the Debye-Hückel equation in Eq.~\eqref{eq:S_cor_DH}, one obtains
\begin{multline}
	-\partial_x^2 \tpsi_i(x) + \frac{1}{2\pi} \int_0^\infty \dd{q} \partial_x^2 \hat{c}_i(x,z) \sin(qz) 
	- \frac{1}{2\pi} \int_0^\infty \dd{q} q^2 \hat{c}_i(x,z) \sin(qz) 
	\\
	= \kappa_i^2 \qty(-\tpsi_i(x) + \frac{1}{2\pi} \int_0^\infty \dd{q} \hat{c}_i(x,z) \sin(qz) ).
\label{eq:S_cor_laplace}
\end{multline}
Since $\tilde{\psi}_i(x)$ is already a solution of the Debye-Hückel equation, Eq.~\eqref{eq:S_cor_laplace} leads to 
\begin{equation}
	\partial_x^2 \hat{c}_i(x,z) = \qty(\kappa_i^2 + q^2) \hat{c}_i(x,z).
\end{equation}
The solution to this differential equation is given by
\begin{align}
	&\hat{c}_1(x,z) = A_1(q) e^{-p_1(q) x} + M_1(q) e^{p_1(q) x}, 
	\label{eq_chat1}
	\\
	&\hat{c}_2(x,z) = M_2(q) e^{-p_2(q) x} + A_2(q) e^{p_2(q) x}, 
	\label{eq_chat2}
\end{align}
where 
\begin{equation}
	p_i(q) := \sqrt{\kappa_i^2 + q^2},
\end{equation} 
and the coefficients $A_i(q)$ and $M_i(q)$ are functions of $q$. 
Now, by using condition \ref{S_COR1} and Eq.~\eqref{eq:S_ci} one obtains $A_1(q) = A_2(q) =0$. 
Therefore, Eqs.~\eqref{eq_chat1} and \eqref{eq_chat2} are reduced to 
\begin{align}
	&\hat{c}_1(x,z) =  M_1(q) e^{p_1(q) x}, 
	\label{eq:chat1}
	\\
	&\hat{c}_2(x,z) = M_2(q) e^{-p_2(q) x}. 
	\label{eq:chat2}
\end{align}
Applying condition~\ref{S_COR2} to Eqs.~\eqref{eq:chat1} and~\eqref{eq:chat2} yields
\begin{equation}
	\eone M_1(q)p_1(q) = - \etwo M_2(q) p_2(q). 
\label{eq:M1}
\end{equation}
In the next step the sine-transformation, as defined in Eq.~\eqref{eq:sine_trafo}, of $\phi_i(z)$ is calculated to 
proceed further with condition~\ref{S_COR4}. 
The results read
\begin{align}
	\hat{\phi}_1(q) &= \int\limits_0^\infty \dd{z} \phi_1(z) \sin(qz) 
	= \frac{q  \PsiP}{p_1^2(q)},
	\\
	\hat{\phi}_2(q) &= \int\limits_0^\infty \dd{z} \phi_2(z) \sin(qz) 
	= \frac{q  (\PsiP - \PsiD)}{p_2^2(q)}.
\end{align}
Furthermore, inserting the obtained sine-transformations into condition~\ref{S_COR4} and using the 
relation $\frac{2}{\pi} \int_0^\infty \dd{z} \sin(q)/q = 1$ results in
\begin{equation}
	\frac{q  \PsiP }{p_1^2(q)} + M_1(q) = \frac{\PsiD}{q} 
	+ \frac{q  (\PsiP - \PsiD)}{p_2^2(q)} + M_2(q).
\label{eq:M2}
\end{equation}
In order to determine the final expression for the correction function, the system of 
two linear equations in Eqs.~\eqref{eq:M1} and~\eqref{eq:M2} has to be solved.
Therefore, the expressions for $M_1(q)$ and $M_2(q)$ read 
\begin{align}
	M_1(q) &= -\frac{\etwo p_2(q)}{\eone p_1 + \etwo p_2}  \qty[\frac{q \PsiP}{p_1^2(q)} 
	-\frac{q  (\PsiP-\PsiD)}{p_2^2(q)} - \frac{\PsiD}{q} ],
	\\	
	M_2(q) &= \frac{\eone p_1(q)}{\eone p_1(q) + \etwo p_2} \qty[\frac{q \PsiP}{p_1^2(q)} 
	-\frac{q  (\PsiP -\PsiD)}{p_2^2(q)} - \frac{\PsiD}{q} ].
\end{align}
The correction function for medium 1, as constructed earlier in  Eq.~\eqref{eq:S_ci}, is then given by
\begin{multline}
	c_1(x,z) = -\frac{\etwo \ktwo}{\eone \kone + \etwo \ktwo}  \PsiD  e^{\kone x}
	\\
	-\frac{2}{\pi} \int\limits_0^\infty \dd{q} \frac{\etwo p_2(q)}{\eone p_1(q) + \etwo p_2(q)} 
	\qty[\frac{q  \PsiP}{p_1^2(q)} -\frac{q  (\PsiP-\PsiD)}{p_2^2(q)} - \frac{\PsiD}{q} ]  e^{p_1(q) x} \sin(qz),
\end{multline}
and for medium 2 as
\begin{multline}
	c_2(x,z) = \frac{\eone \kone}{\eone \kone + \etwo \ktwo}  \PsiD  e^{-\ktwo x}
	\\
	+\frac{2}{\pi} \int\limits_0^\infty \dd{q} \frac{\eone p_1(q)}{\eone p_1(q) + \etwo p_2(q)} 
	\qty[\frac{q \PsiP}{p_1^2(q)} -\frac{q (\PsiP-\PsiD)}{p_2^2(q)} - \frac{\PsiD}{q}]  
	e^{-p_2(q) x} \sin(qz).
\end{multline}

\subsection{Electrostatic potential for a single wall}

The electrostatic potential due to a single wall located at $z=0$ in contact with the two fluid 
media forming a fluid interface at $x=0$, is now calculated with Eq.~\eqref{eq:S_POTENTIAL}
and is therefore given in medium 1 by
\begin{multline}
	\Psi_1^{\textrm{sin}}(x,z) = \PsiP e^{-\kone z} 
	\\
	-\frac{2}{\pi} \int\limits_0^\infty \dd{q} \frac{\etwo p_2(q)}{\eone p_1(q) 
	+ \etwo p_2(q)} \qty[\frac{q \PsiP}{p_1^2(q)} -\frac{q (\PsiP-\PsiD)}{p_2^2(q)} 
	- \frac{\PsiD}{q} ]  e^{p_1(q) x} \sin(qz),
	\label{eq:s_single_1}
\end{multline}
and in medium 2 by
\begin{multline}
	\Psi_2^{\textrm{sin}}(x,z) = \PsiD + (\PsiP - \PsiD)  e^{- \ktwo z} 
	\\
	+\frac{2}{\pi} \int\limits_0^\infty \dd{q} \frac{\eone p_1(q)}{\eone p_1(q) + \etwo p_2(q)} 
	\qty[\frac{q \PsiP}{p_1^2(q)} -\frac{q (\PsiP-\PsiD)}{p_2^2(q)} - \frac{\PsiD}{q} ]  e^{-p_2(q) x} \sin(qz).
		\label{eq:s_single_2}
\end{multline}

\subsection{Result}

The electrostatic potential within the superposition approximation for a system featuring two flat walls (one at $z=0$ and $z=L$) is now finally 
obtained by superposing the solutions of the single wall problem as shown earlier in Eq.~\eqref{eq:S_TOTAL_POTENTIAL}.
Therefore, using Eq.~\eqref{eq:s_single_1}, the electrostatic potential distribution in medium 1 is given by 
\begin{shaded*}
\begin{align}
	&\Psi_1^s(x,z) = \PsiP  e^{-\kone z} + \PsiP  e^{\kone (z-L)}
	\nonumber	
	\\
	\nonumber
	&- \frac{2}{\pi} \int\limits_0^\infty \dd{q} \frac{\etwo p_2(q)}{\eone p_1(q) + \etwo p_2(q)} 
	\qty[\frac{q \PsiP}{p_1^2(q)} -\frac{q  (\PsiP-\PsiD)}{p_2^2(q)} - \frac{\PsiD}{q} ] e^{p_1(q) x} \sin(qz)
	\\
	\nonumber
	&+ \frac{2}{\pi} \int\limits_0^\infty \dd{q} \frac{\etwo p_2(q)}{\eone p_1(q) + \etwo p_2(q)} 
	\qty[\frac{q \PsiP}{p_1^2(q)} -\frac{q  (\PsiP-\PsiD)}{p_2^2(q)} - \frac{\PsiD}{q}] 
	\\
	&\times e^{p_1(q) x} \sin(q(z-L)).
	\label{eq:sup_potential_1}
\end{align}
\end{shaded*}
\newpage
\noindent
Analogous, the electrostatic potential in medium 2 is obtained by superposing Eq.~\eqref{eq:s_single_2} as mentioned above and reads
\begin{shaded*}
\begin{align}
	&\Psi_2^s (x,z) = 2 \PsiD + (\PsiP - \PsiD) (e^{-\ktwo z} -  e^{\ktwo(z-L)})
	\nonumber	
	\\
	\nonumber
	+& \frac{2}{\pi} \int\limits_0^\infty \dd{q} \frac{\eone p_1(q)}{\eone p_1(q) 
	+ \etwo p_2(q)} \qty[\frac{q  \PsiP}{p_1^2(q)} -\frac{q  (\PsiP-\PsiD)}{p_2^2(q)} - \frac{\PsiD}{q} ]  e^{-p_2(q) x} \sin(qz)
	\\
	\nonumber
	-& \frac{2}{\pi} \int\limits_0^\infty \dd{q} \frac{\eone p_1(q)}{\eone p_1(q) 
	+ \etwo p_2(q)} \qty[\frac{q  \PsiP}{p_1^2(q)} -\frac{q (\PsiP-\PsiD)}{p_2^2(q)} - \frac{\PsiD}{q} ] 
	\\
	\times&  e^{-p_2(q) x} \sin(q(z-L)).
	\label{eq:sup_potential_2}
\end{align}
\end{shaded*}

%% file: figure/tikz/exact_interface.tikz
\begin{tikzpicture}
		\draw[->,thick,dashed] (0.25,0) -- (0.25,4.96);
		\draw[->,thick,dashed] (4.75,2.23) -- (5.5,2.23);
		\draw[mpidarkblue,very thick] (0.25,2.23) -- (4.75,2.23);
		\draw (0.1,4.65) node {x};
		\draw (5.3,2.43) node {z};
		\draw (0.6,5.12) node {(a)};
		\draw (0.7,1.93) node{(0,0)};
		\draw (2.5,3.21) node[draw,rounded corners]{$\etwo, \ktwo$};
		\draw (2.5,1.25) node[draw,rounded corners]{$\eone, \kone$};
		\draw (4.25,1.93)	node{(0,$L$)};
\end{tikzpicture}

%% file: figure/tikz/exact_walls.tikz
\begin{tikzpicture}
		\filldraw[lightgrey] (0,0) -- (0.25,0) -- (0.25,4.46) -- (0,4.46) -- cycle;
		\filldraw[lightgrey] (4.75,0) -- (5,0) -- (5,4.46) -- (4.75,4.46) -- cycle;
		\draw[->,thick,dashed] (0.25,4.46) -- (0.25,4.96);
		\draw[->,thick,dashed] (4.75,2.23) -- (5.5,2.23);
		\draw[mpired,very thick] (0.25,0) -- (0.25,4.46);
		\draw[mpired,very thick] (4.75,0) -- (4.75,4.46);
		\draw[thick,dashed] (0.25,2.23) -- (4.75,2.23);
		\draw (0.1,4.65) node {x};
		\draw (5.3,2.43) node {z};
		\draw (0.6,5.12) node {(b)};
		\draw (0.7,1.93) node{(0,0)};
		\draw (2.5,3.21) node[draw,rounded corners]{$\epsilon_i, \kappa_i$};
		\draw (2.5,1.25) node[draw,rounded corners]{$\epsilon_i, \kappa_i$};
		\draw (4.25,1.93)	node{(0,$L$)};
		\draw (0.55,0.25) node{$\PsiP$};
		\draw (0.55,4.21) node{$\PsiP$};
		\draw (4.45,0.25) node{$\PsiP$};
		\draw (4.45,4.21) node{$\PsiP$};
\end{tikzpicture}

%% file: figure/tikz/superposition_interface.tikz
\begin{tikzpicture}
		\draw[->,thick,dashed] (0.25,0) -- (0.25,4.96);
		\draw[->,thick,dashed] (4.75,2.23) -- (5.5,2.23);
		\draw[mpidarkblue,very thick] (0.25,2.23) -- (4.75,2.23);
		\draw (0.1,4.65) node {x};
		\draw (5.3,2.43) node {z};
		\draw (0.6,5.12) node {(a)};
		\draw (0.7,1.93) node{(0,0)};
		\draw (2.5,3.21) node[draw,rounded corners]{$\etwo, \ktwo$};
		\draw (2.5,1.25) node[draw,rounded corners]{$\eone, \kone$};
		\draw (4.25,1.93)	node{(0,$L$)};
\end{tikzpicture}

%% file: figure/tikz/superposition_wall.tikz
\begin{tikzpicture}
		\filldraw[lightgrey] (0,0) -- (0.25,0) -- (0.25,4.46) -- (0,4.46) -- cycle;
		\draw[->,thick,dashed] (0.25,4.46) -- (0.25,4.96);
		\draw[->,thick,dashed] (4.75,2.23) -- (5.5,2.23);
		\draw[mpired,very thick] (0.25,0) -- (0.25,4.46);
		\draw[thick,dashed] (0.25,2.23) -- (4.75,2.23);
		\draw (0.1,4.65) node {x};
		\draw (5.3,2.43) node {z};
		\draw (0.6,5.12) node {(b)};
		\draw (0.7,1.93) node{(0,0)};
		\draw (2.5,3.21) node[draw,rounded corners]{$\epsilon_i, \kappa_i$};
		\draw (2.5,1.25) node[draw,rounded corners]{$\epsilon_i, \kappa_i$};
		\draw (0.55,0.25) node{$\PsiP$};
		\draw (0.55,4.21) node{$\PsiP$};
\end{tikzpicture}

%% file: chapters/04_Interaction_Energy.tex

\addtocontents{toc}{\protect\newpage}
\chapter{Interaction Energies}
\label{ch:4}

In this chapter, analytical expressions for the $L$-dependent interaction energies between the two 
plates as well as $L$-independent interactions (i.e. surface tensions or line tensions acting between 
the plates and the fluids or interfacial tension between the fluids) present in the system are derived.
Results are obtained for both, the exact calculations in Section~\ref{sec:energy_exact} and the 
superposition approximation in Section~\ref{sec:energy_sup}.

\section{Energy contributions}

At the end of Chapter~\ref{ch:2}, the grand potential corresponding to the system is expressed as a functional 
of the electrostatic potentials $\Psi_i(x,z)$ in medium $i \in \qty{1,2}$.
Having calculated $\Psi_i(x,z)$ in Chapter~\ref{ch:3}, one can now insert the results back into 
Eq.~\eqref{eq:INTERACTION_POTENTIAL} to obtain the grand potential as a function of the separation distance $L$, 
which then can be further decomposed into the following contributions
\begin{align}
	\tilde{\Omega}(L) 
	= 
	\sum\limits_{i\in \qty{1,2}} (\omega_{\gamma,i}(L) + \gamma_i) 2A_i 
	+ (\omega_{\tau}(L) + \tau) 2l 
	+ \gamma_{1,2} A_{1,2},
\label{eq:total_energies}
\end{align}
where

\begin{itemize}
	\item 
		$\omega_{\gamma,i}(L)$ is the surface interaction energy per total area of the two surfaces
		$2A_i$  in contact with medium $i \in \qty{1,2}$ at separation $L$.
	\item 
		$\gamma_i$ is the surface tension acting between medium $i \in \qty{1,2}$ and the 
		adjacent walls, with total area $2A_i$, at $z=0$ and $z=L$ respectively.
	\item 
		$\omega_\tau(L)$ is the line interaction energy density between the two three-phase contact lines
		separated by distance $L$ and expressed by their total length $2l$. 
	\item 
		$\tau$ is the line tension acting at the two three-phase contact lines (total length $2l$) formed at 
		$z=0$ and $z=L$ respectively. 
	\item 
		$\gamma_{1,2}$ is the interfacial tension acting between the two media $i \in \qty{1,2}$ 
		at the fluid interface at $x=0$. 
		The total area of the fluid interface is denoted by $A_{1,2}$.
\end{itemize}

\paragraph{Remark}

The expression in Eq.~\eqref{eq:total_energies} does not contain the bulk contribution to the free energy. 
The bulk energy density $\Omega_{b,i}$ in medium $i \in \qty{1,2}$ with volume $\V_i$ is given by 
the first term in Eq.~\eqref{eq:exp_DF}.
Therefore,

\begin{align}
	\beta \Omega_{b,1} 
	= \frac{1}{\V_1} \int\limits_{\V_1}\dd[3]{r} \sum\limits_{i=\pm} I_1 \qty[\ln(\frac{I_1}{\zeta_i})- 1
	+ \beta V_i(\vb{r})]
	= -2 I_1,
	\\
	\beta \Omega_{b,2} 
	= \frac{1}{\V_2} \int\limits_{\V_2}\dd[3]{r} \sum\limits_{i=\pm} I_2 \qty[\ln(\frac{I_2}{\zeta_i})- 1
	+ \beta V_i(\vb{r})]
	= -2 I_2.
\end{align}

\section{Method}
\label{sec:method}

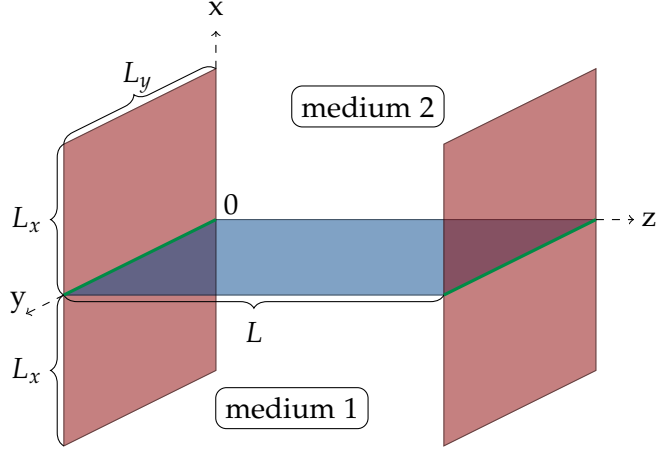
\begin{figure}[htbp]
	\centering
	\input{figure/tikz/model_system.tikz}
	\label{fig:energie_system}
	\caption
		{
		Sketch of the considered system volume $\V$ used to calculated the interaction parameters.
		The interfacial area is given in blue ({\color{mpidarkblue}$\bullet$}) and the three-phase contact 
		line is given in green ({\color{mpigreen}$\bullet$}).
		Each one of the two walls carrying constant surface potential $\PsiP$ is denoted 
		by the color red ({\color{mpired}$\bullet$}).
		The lower space ($x<0$) is filled by medium 1 and the upper space ($x>0$) by medium 2.
		}
\end{figure}
\noindent
The energy contributions are calculated by using a system as seen in Fig.~\ref{fig:energie_system}.
Here, the total system volume is given by $\V = \qty[-L_x,L_x] \cross \qty[0,L_y] \cross \qty[0,L]$.
Therefore the line and surface contribution given in Eq.~\eqref{eq:total_energies} now reads 
\begin{align}
	\tilde{\Omega}(L) 
	=
	\sum\limits_{i \in \qty{1,2}}(\omega_{\gamma,i}(L) + \gamma_i)2L_xL_y 
	+ (\omega_{\tau}(L) + \tau)2L_y 
	+ \gamma_{1,2}LL_y.
\label{eq:system_energies}
\end{align}
In order to extract all interaction parameters in Eq.~\eqref{eq:system_energies} the following procedure is used. 
First, the electrostatic potentials are inserted into the grand potential in 
Eq.~\eqref{eq:system_energies} and then terms proportional to $2L_xL_y$, $2L_y$ and $LL_y$ are separated.
Clearly, the coefficient of $LL_y$ is the interfacial tension $\gamma_{1,2}$.
The coefficients of $2L_xL_y$ and $2L_y$, however, include both the $L$- dependent and $L$-independent contributions.
In each case, the $L$-independent contribution is obtained by taking the limit of $L\to \infty$.
In other words, if $\overline{p}(L)$ is the quantity that includes both, the $L$-dependent and 
$L$-independent contributions, then $\overline{p}(L\to\infty)$ gives the $L$-independent 
contribution as the interaction between the plates vanishes in the limit of infinite separation.
Consequently, the $L$-dependent contribution of $\overline{p}(L)$ is then given by
\begin{equation}
	p(L) = \overline{p}(L) - \lim_{L\to\infty} \overline{p}(L).
\end{equation}

\section{Exact calculations}
\label{sec:energy_exact}

The contribution to the free energy as given in Eq.~\eqref{eq:total_energies}, combined with the exact expression of the electrostatic potential 
in medium $i\in\qty{1,2}$ provided by Eqs.~\eqref{eq:exact_potential_1} and~\eqref{eq:exact_potential_2}, reads
\begin{align}
	\tilde{\Omega}^e(L) =&  \ \eone L_y \PsiP \int\limits_{-L_x}^0 \dd{x} \partial_z \Psi_1^e(x,z)\rvert_{z=0} 
	+ (\PsiP-\PsiD) \etwo L_y \int\limits_{0}^{L_x} \dd{x} \partial_z \Psi_2^e(x,z)\rvert_{z=0}
	\nonumber
	\\
	&- \frac{\eone \PsiD L_y}{2} \int\limits_0^L \dd{z} \partial_x \Psi_1^e(x,z)\rvert_{x=0}.
\label{eq:E_Omega_Int}
\end{align}
To end up with the free energy as a function of separation $L$, one now has to first calculate the derivatives 
of the electrostatic potentials as needed for Eq.~\eqref{eq:E_Omega_Int}. 
Evaluating those derivatives at $x=0$ and $z=0$ respectively results in
\begin{align}
	\partial_x \Psi_1^e(x,z)\rvert_{x=0} =& \sum\limits_{n=1}^\infty \chi_{1,n}(L) n\pi  \qty{(-1)^n-1 }p_{n,1} \sin(\frac{n\pi z}{L}), 
	\label{eq:tOmega^e1}
	\\
	\nonumber
	\\
	\partial_z \Psi_1^e(x,z)\rvert_{z=0} =& \ \kone \PsiP \qty[\frac{1-\cosh(\kone L)}{\sinh(\kone L)}]
	\nonumber
	\\
	&+\sum\limits_{n=1}^\infty \chi_{1,n}(L) \frac{n^2\pi^2}{L} \qty{(-1)^n-1}e^{p_{n,1}x},
	\label{eq:tOmega^e2}
	\\
	\nonumber
	\\
	\partial_z \Psi_2^e(x,z)\rvert_{z=0} =& \ \ktwo (\PsiP-\PsiD) \qty[\frac{1-\cosh(\ktwo L)}{\sinh(\ktwo L)}]
	\nonumber
	\\
	&+ \sum\limits_{n=1} \chi_{2,n} (L) \frac{n^2\pi^2}{L}  \qty{(-1)^n-1}e^{-p_{n,2}x},
	\label{eq:tOmega^e3}
\end{align}
where for the sake of brevity the following notations are used
\begin{align}
	&p_{n,i} \coloneqq \sqrt{\frac{n^2\pi^2}{L^2} + \kappa_i^2}  \qquad i \in \qty{1,2},
	\label{eq:p_ni} 
	\\
	&\chi_{1,n}(L) \coloneqq \qty[ \dfrac{-\dfrac{2\PsiD}{n^2\pi^2} 
	+ \dfrac{2\PsiP}{n^2\pi^2+\kone^2 L^2} - \dfrac{2(\PsiP-\PsiD)}{n^2\pi^2+\ktwo^2 L^2}}{1 + \dfrac{\eone p_{n,1}}{\etwo p_{n,2}}} ],
	\label{eq:Chi1n}
	\\
	&\chi_{2,n}(L) \coloneqq  \qty[ \dfrac{\dfrac{2\PsiD}{n^2\pi^2} - \dfrac{2\PsiP}{n^2\pi^2+\kone^2 L^2} 
	+ \dfrac{2(\PsiP-\PsiD)}{n^2\pi^2+\ktwo^2 L^2}}{1 + \dfrac{\etwo p_{n,2}}{\eone p_{n,1}}} ].
	\label{eq:Chi2n}
\end{align}
With the assumption that $L_x \gg 1$, one furthermore obtains the following expression for the free energy contribution
\begin{align}
	\widetilde{\Omega}^e(L) =&\ \qty[ \obar_{\gamma,1}(L)  + \obar_{\gamma,2}(L)] 2L_xL_y
	\nonumber	
	\\
	&+ \qty[\obar_{\tau}^1(L)  + \obar_{\tau}^2(L)  + \obar_{\tau,\gamma_{1,2}}(L)] 2L_y,
\end{align}
in which different terms are separated by their proportionality as outlined in Section~\ref{sec:method}.
\begin{align}
	&\obar_{\gamma,1}(L) := \frac{\eone \kone}{2} \PsiP^2 \qty[\frac{1-\cosh(\kone L)}{\sinh(\kone L)}],
	\label{eq:E_COMBINED_SURFACE1}
	\\
	\nonumber	
	\\
	&\obar_{\gamma,2}(L) := \frac{\etwo \ktwo}{2} (\PsiP-\PsiD)^2 \qty[\frac{1-\cosh(\ktwo L)}{\sinh(\ktwo L)}],
	\label{eq:E_COMBINED_SURFACE2}
	\\
	\nonumber	
	\\
	&\obar_{\tau}^1(L) := \frac{\eone \PsiP}{2} \sum\limits_{n=1}^\infty \chi_{1,n}(L) \frac{n^2\pi^2}{L} \frac{\qty{(-1)^n -1}}{p_{n,1}},
	\label{eq:obar_tau1}
	\\
	\nonumber
	\\
	&\obar_{\tau}^2(L) := \frac{\etwo (\PsiP - \PsiD)}{2}\sum\limits_{n=1}^\infty \chi_{2,n}(L) 
	\frac{n^2\pi^2}{L}\frac{\qty{(-1)^n -1}}{p_{n,2}},
	\label{eq:E_obar_tau2}
	\\
	\nonumber	
	\\
	&\obar_{\tau,\gamma_{1,2}}(L) := \frac{\eone\PsiD}{4} \sum\limits_{n=1}^\infty \chi_{1,n}(L) \qty{(-1)^n-1}^2 Lp_{n,1}. 
	\label{eq:E_obar_tau_gamma}
\end{align}
\paragraph{Remark}

So far, none of the occurring terms can be assigned to the interfacial tension. 
Further calculation of Eq.~\eqref{eq:E_obar_tau_gamma} will however result in a 
term proportional to the interfacial area $LL_y$ later on.

\subsection{Surface interaction energy densities and surface tensions}

Surface contributions to the free energy are given by coefficients proportional to $2L_xL_y$. 
Consequently, to separate the $L$-independent contributions from the $L$-dependent contributions, the procedure as 
specified in Section~\ref{sec:method} is applied.

\subsubsection{Surface tensions}

$L$-dependent contributions to the surface part vanish in the limit of infinite separation $L\to \infty$. 
Therefore, $L$-independent contributions to the surface interaction 
are identified by taking the limit $L\to\infty$ of $\obar_{\gamma,i}(L)$ in medium  $i \in \qty{1,2}$. 
The behavior in medium 1 for this infinite separation limit is given by
\begin{align}
	\lim_{L\to\infty} \obar_{\gamma,1} (L)&= \lim_{L\to\infty} \frac{\eone \kone}{2}\PsiP^2 
	\qty[\frac{1- \cosh(\kone L)}{\sinh(\kone L)}]
	\nonumber
	\\
	&= - \frac{\eone \kone}{2}\PsiP^2,
\label{eq:E_lim_gamma1}
\end{align}
and for medium 2 analogous as,
\begin{align}
	\lim_{L\to\infty} \obar_{\gamma,2}(L)&= \lim_{L\to\infty} \frac{\etwo\ktwo}{2} (\PsiP-\PsiD)^2 
	\qty[\frac{1- \cosh(\ktwo L)}{\sinh(\ktwo L)}]
	\nonumber
	\\
	&= - \frac{\etwo \ktwo}{2}(\PsiP-\PsiD)^2.
\label{eq:E_lim_gamma2}
\end{align}
Therefore, the energy contribution due to the surface tension (proportional to $2L_xL_y$) 
within exact calculation is given by
\begin{shaded*}
\begin{align}
	&\gamma_{1}^e = - \frac{\eone \kone}{2} \PsiP^2,
	\\
	&\gamma_2^e = - \frac{\etwo \ktwo}{2} (\PsiP-\PsiD)^2.
\end{align}
\end{shaded*}

\subsubsection{Surface interaction energy densities}

Subsequently, the $L$-dependent contributions to the surface interactions are obtained by subtracting 
Eqs.~\eqref{eq:E_lim_gamma1} and~\eqref{eq:E_lim_gamma2} from the terms in Eqs.~\eqref{eq:E_COMBINED_SURFACE1} 
and \eqref{eq:E_COMBINED_SURFACE2}. 
As a result, one obtains for medium 1
\begin{align}
	\omega_{\gamma,1}(L) &= \obar_{\gamma,1}(L) - \lim_{L\to\infty} \obar_{\gamma,1} (L)
	\nonumber
	\\
	&= \frac{\eone \kone}{2}\PsiP^2 \qty[1 + \qty(\frac{1- \cosh(\kone L)}{\sinh(\kone L)})].
\end{align}
and for medium 2
\begin{align}
	\omega_{\gamma,2}(L) &= \obar_{\gamma,2}(L) - \lim_{L\to\infty} \obar_{\gamma,2}(L)
	\nonumber
	\\
	&= \frac{\etwo \ktwo}{2} (\PsiP-\PsiD)^2 
	\qty[1+ \qty(\frac{1- \cosh(\ktwo L)}{\sinh(\ktwo L)})].
\end{align}
The surface interaction energy per total surface area $2L_xL_y$ within the exact calculation is therefore given by
\begin{shaded*}
\begin{align}
	\omega^e_{\gamma,1}(L) &= \frac{\eone \kone}{2}\PsiP^2 
	\qty[1 + \qty(\frac{1- \cosh(\kone L)}{\sinh(\kone L)})]
	\label{eq:E_SURFACE1_ENERGY},
	\\
	\omega^e_{\gamma,2}(L) &= \frac{\etwo \ktwo}{2}(\PsiP-\PsiD)^2 
	\qty[1 + \qty(\frac{1- \cosh(\ktwo L)}{\sinh(\ktwo L)})]
	\label{eq:E_SURFACE2_ENERGY}.
\end{align}
\end{shaded*}

\subsection{Line interaction energy density, line tension and interfacial tension}

For a system as shown in Fig.~\ref{fig:energie_system}, the line contributions to the free energy are proportional 
to the total length of the three-point lines $2L_y$.
To separate the $L$-dependent from the $L$-independent contributions, the procedure of 
Section~\ref{sec:method} is applied once again.

\subsubsection{Line tension and interfacial tension}

The line tension $\tau$ is characterized by $L$-independent terms, which are proportional to $2L_y$.
As outlined earlier, those contributions are identified in the limit of infinite separation $L\to\infty$.
First, the sum occurring in Eqs.~\eqref{eq:obar_tau1},~\eqref{eq:E_obar_tau2} and~\eqref{eq:E_obar_tau_gamma} is 
simplified since all even numbers yield zero viz.
\begin{align}
	\sum\limits_{n=1}^\infty \frac{\pi}{L} \qty{(-1)^n-1} 
	= 
	\sum\limits_{n=1,3,5,..}^\infty -\frac{2\pi}{L}.	
\end{align}
Furthermore, by defining the new variable $x \coloneqq \frac{n\pi}{L}$ one obtains
\begin{align}
	\lim\limits_{L\to\infty}  \sum\limits_{n=1,3,5,..}^\infty \frac{2\pi}{L} \to \int\limits_0^\infty \dd{x},
\label{eq:E_sum_to_integral}
\end{align}
and $p_{n,i}$ as defined in Eq.~\eqref{eq:p_ni} now reads $p_{x,i} \coloneqq \sqrt{x^2 + \kappa_i^2}$.
Subsequently, $\chi_{1,n}(L)$ and $\chi_{2,n}(L)$ (Eqs.~\eqref{eq:Chi1n} and~\eqref{eq:Chi2n}) transform in the limit
$L\to\infty$ as follows
\begin{align}
\lim\limits_{L\to\infty} \chi_{i,n}(L) = \frac{1}{L^2} \chi_{i,x} ,
\end{align}
where
\begin{align}
	&\chi_{1,x} \coloneqq \qty[ \dfrac{-\dfrac{2\PsiD}{x^2} 
	+ \dfrac{2\PsiP}{x^2+\kone^2} - \dfrac{2(\PsiP-\PsiD)}{x^2+\ktwo^2}}{1 + \dfrac{\eone p_{x,1}}{\etwo p_{x,2}}} ],
	\\
	&\chi_{2,x} \coloneqq \qty[ \dfrac{\dfrac{2\PsiD}{x^2} 
	- \dfrac{2\PsiP}{x^2+\kone^2} + \dfrac{2(\PsiP-\PsiD)}{x^2+\ktwo^2}}{1 + \dfrac{\etwo p_{x,2}}{\eone p_{x,1}}} ].
\end{align}
As a result of these properties, Eq.~\eqref{eq:obar_tau1} yields
\begin{align}
	\lim_{L\to\infty} \obar_{\tau}^{1}(L) 
	&= \lim_{L\to\infty} \frac{\eone \PsiP}{2} \sum\limits_{n=1}^\infty \chi_{1,n} 
	\frac{n^2\pi^2}{L} \frac{\qty{(-1)^n -1}}{p_{n,1}}
	\nonumber
	\\
	&= -\frac{\eone\PsiP}{2}\int\limits_0^\infty \dd{x} \chi_{1,x} \frac{x^2}{\pi p_{x,1}},
\label{eq:E_Lim_OmBar_Tau_1}
\end{align}
whereas Eq.~\eqref{eq:E_obar_tau2} is reduced to
\begin{align}
	\lim_{L\to\infty} \obar_{\tau}^{2}(L) 
	&= \lim_{L\to\infty}  \frac{\etwo (\PsiP-\PsiD)}{2} \sum\limits_{n=1}^\infty \chi_{2,n} 
	\frac{n^2\pi^2}{L} \frac{\qty{(-1)^n-1}}{p_{n,2}}
	\nonumber
	\\
	&= -\frac{\etwo (\PsiP-\PsiD)}{2}\int\limits_0^\infty \dd{x} \chi_{2,x} \frac{x^2}{\pi p_{x,2}}.
\label{eq:E_Lim_OmBar_Tau_2}
\end{align}
As mentioned earlier, Eq.~\eqref{eq:E_obar_tau_gamma} contains terms 
proportional to $2L_y$ (line contribution) as well as $LL_y$ (interfacial contribution). 
By further simplification, it is possible to recover the following expression
\begin{align}
	\obar_{\tau,\gamma_{1,2}}(L) 
	= \obar_{\tau}^{3}(L)  
	+  \frac{\gamma_{1,2}}{2}  L.
\label{eq:E_gamma_tau_aufteilung}
\end{align}
Thereby, the contribution to the line interaction, which is proportional to $2L_y$ reads
\begin{align}
	\obar_{\tau}^{3}(L)=&\ \frac{\PsiD \eone \etwo}{L} \sum\limits_{n=1,3,5,..}^{\infty} \frac{1}{\eone p_{n,1} + \etwo p_{n,2}}
	\label{eq:E_gamma_bar_tau_3}	
	\\
	& \cross  \Biggl[ \frac{-2\PsiD \ktwo^2}{\eone\kone+\etwo\ktwo} 
	\Biggl(\frac{(\ktwo^2-\kone^2)(\eone\kone+\etwo\ktwo)}{\ktwo^2p_{n,2}^2\qty[(p_{n,1}/p_{n,2})+(\kone/\ktwo)]} 
	\nonumber
	\\
	\nonumber	
	&- \frac{\kone}{\ktwo} \Biggl(\frac{\eone}{p_{n,1} + \kone} + \frac{\etwo}{p_{n,2} 
	+ \ktwo}\Biggr)\Biggr)+ \frac{2\PsiP(\ktwo^2-\kone^2)}{p_{n,1}p_{n,2}} \Biggr].
\end{align}
By taking the limit $L\to\infty$ of Eq.~\eqref{eq:E_gamma_bar_tau_3} and simultaneously using the relation 
provided in Eq.~\eqref{eq:E_sum_to_integral}, one obtains
\begin{align}
	\lim_{L\to\infty} \obar_{\tau}^{3}(L) =& \lim_{L\to\infty} \frac{\PsiD \eone \etwo}{L} 
	\sum\limits_{n=1,3,5,..}^{\infty} \frac{1}{\eone p_{n,1} + \etwo p_{n,2}}
	\nonumber	
	\\
	& \times \Biggl[ \frac{-2\PsiD \ktwo^2}{\eone\kone+\etwo\ktwo} 
	\Biggl(\frac{(\ktwo^2-\kone^2)(\eone\kone+\etwo\ktwo)}{\ktwo^2p_{n,2}^2\qty[(p_{n,1}/p_{n,2})+(\kone/\ktwo)]} 
	\nonumber
	\\
	&- \frac{\kone}{\ktwo} \Biggl(\frac{\eone}{p_{n,1} + \kone} + \frac{\etwo}{p_{n,2} + \ktwo}\Biggr)\Biggr)
	+ \frac{2\PsiP(\ktwo^2-\kone^2)}{p_{n,1}p_{n,2}} \Biggr]
	\nonumber	
	\\
	=& \ \frac{\PsiD\eone\etwo}{2\pi} \int\limits_0^\infty \dd{x} \frac{1}{\eone p_{x,1} + \etwo p_{x,2}}
	\nonumber
	\\
	& \times  \Biggl[ \frac{-2\PsiD \ktwo^2}{\eone\kone+\etwo\ktwo} 
	\Biggl(\frac{(\ktwo^2-\kone^2)(\eone\kone+\etwo\ktwo)}{\ktwo^2p_{x,2}^2\qty[(p_{x,1}/p_{x,2})+(\kone/\ktwo)]} 
	\nonumber
	\\
	&- \frac{\kone}{\ktwo} \Biggl(\frac{\eone}{p_{x,1}+ \kone} + \frac{\etwo}{p_{x,2} 
	+ \ktwo}\Biggr)\Biggr)+ \frac{2\PsiP(\ktwo^2-\kone^2)}{p_{x,1}p_{x,2}} \Biggr].
	\label{eq:E_Limit_OmBar_Tau_3}
\end{align}
Finally, the sum of Eqs.~\eqref{eq:E_Lim_OmBar_Tau_1},~\eqref{eq:E_Lim_OmBar_Tau_2} and~\eqref{eq:E_Limit_OmBar_Tau_3}
results in the line tension
\begin{align}
	\tau 
	= \lim_{L\to\infty} \obar_{\tau}^{1}(L) 
	+ \lim_{L\to\infty} \obar_{\tau}^{2}(L)  
	+\lim_{L\to\infty} \obar_{\tau}^{3}(L),
\end{align}
which is therefore given within exact calculations as
\begin{shaded*}
\begin{align}
	\tau^e =& -\frac{\eone \PsiP}{2}\int\limits_0^\infty \dd{x} \chi_{1,x} \frac{x^2}{\pi p_{x,1}} 
	\nonumber
	\\	
	&-\frac{\etwo (\PsiP-\PsiD)}{2}\int\limits_0^\infty \dd{x} \chi_{2,x} \frac{x^2}{\pi p_{x,2}}
	\nonumber
	\\
	&+\frac{\PsiD\eone\etwo}{2\pi} \int\limits_0^\infty \dd{x} \frac{1}{\eone p_{x,1} + \etwo p_{x,2}}
	\nonumber
	\\
	&\times  \Biggl[ \frac{-2\PsiD \ktwo^2}{\eone\kone+\etwo\ktwo} 
	\Biggl(\frac{(\ktwo^2-\kone^2)(\eone\kone+\etwo\ktwo)}{\ktwo^2p_{x,2}^2\qty[(p_{x,1}/p_{x,2})+(\kone/\ktwo)]} 
	\nonumber
	\\
	&- \frac{\kone}{\ktwo} 
	\Biggl(\frac{\eone}{p_{x,1}+ \kone} + \frac{\etwo}{p_{x,2} + \ktwo}\Biggr)\Biggr)
	+ \frac{2\PsiP(\ktwo^2-\kone^2)}{p_{x,1}p_{x,2}} \Biggr].
\end{align}
\end{shaded*}
\noindent
Now, the only term left is proportional to the interfacial area $LL_y$ and is therefore identified as the interfacial tension.
The expression is given in Eq.~\eqref{eq:E_gamma_tau_aufteilung} and reads
\begin{align}
	\gamma_{1,2} &= -\frac{4}{\pi^2}\frac{\eone \etwo \kone\ktwo}{(\eone \kone + \etwo \ktwo)}  \PsiD^2
	\sum\limits_{n=1,3,5,..}^\infty \frac{1}{n^2}.
\label{eq:interfacial_tension_exact}
\end{align}
By using the Riemann zeta function 
\begin{equation}
	\zeta(s) = \sum\limits_{n=1}^{\infty} \frac{1}{n^s},
\end{equation}
the sum in Eq.~\eqref{eq:interfacial_tension_exact} is calculated as
\begin{align}
	\sum\limits_{n=1,3,5,..}^\infty \frac{1}{n^2} 
	&= \sum\limits_{n=1}^\infty \frac{1}{n^2} - \sum\limits_{n=1}^\infty \frac{1}{(2n)^2}
	\nonumber
	\\
	&= \qty(1 - \frac{1}{4}) \sum\limits_{n=1}^\infty \frac{1}{n^2}
	\nonumber
	\\
	&= \frac{3}{4} \zeta(2)
	\nonumber
	\\
	&= \frac{\pi^2}{8}.
\end{align}
Therefore, the final result of the interfacial tension within exact calculations, which acts between the two media $i \in \qty{1,2}$, 
is given by
\begin{shaded*}
\begin{align}
	\gamma^e_{1,2} = -\frac{\eone \etwo \kone \ktwo}{2(\eone \kone + \etwo \ktwo)}\PsiD^2.
\label{eq:E_INTERFACE_TENSION}
\end{align}
\end{shaded*}
\paragraph{Remark}

In principle, the interfacial tension should not depend on the nature of the particle surface since it is a property of the 
fluid-fluid interface only.
This behavior is consistent with the results of Ref.~\cite{majee2018}, where an identical expression for the interfacial tension is obtained 
for a different type of colloids and surface boundaries.

\subsubsection{Line interaction energy density}

As outlined in Section \ref{sec:method}, the $L$-dependent contributions to the line interaction energy 
density are now obtained by subtracting the constant $L$-independet terms, which are given in Eqs.~\eqref{eq:E_Lim_OmBar_Tau_1}, 
\eqref{eq:E_Lim_OmBar_Tau_2},~\eqref{eq:E_Limit_OmBar_Tau_3} and~\eqref{eq:E_INTERFACE_TENSION}, from the terms in Eqs.~\eqref{eq:obar_tau1}, 
\eqref{eq:E_obar_tau2} and~\eqref{eq:E_obar_tau_gamma}, effectively resulting in 
\begin{align}
	\omega_{\tau}(L) = \qty( \obar_{\tau}^1(L) + \obar_{\tau}^2(L) + \obar_{\tau, \gamma_{1,2}}(L) ) - \tau - \frac{\gamma_{1,2}}{2}L.
\end{align}
Therefore, the line interaction energy density within exact calculation reads
\begin{shaded*}
\begin{align}
	\omega_{\tau}^e(L) =~&\frac{\eone \PsiP}{2} \sum\limits_{n=1}^\infty \chi_{1,n}(L) 
	\frac{n^2\pi^2}{L} \frac{\qty{(-1)^n-1}}{p_{n,1}}
	\nonumber
	\\
	&+ \frac{\etwo (\PsiP-\PsiD) }{2}\sum\limits_{n=1}^\infty \chi_{2,n}(L) 
	\frac{n^2\pi^2}{L} \frac{\qty{(-1)^n-1}}{p_{n,2}}
	\nonumber
	\\
	&+ \frac{\eone\PsiD}{4} \sum\limits_{n=1}^\infty \chi_{1,n}(L)\qty{(-1)^n-1}^2 L p_{n,1}	
	\nonumber
	\\
	&+ \frac{\eone\PsiP}{2}\int\limits_0^\infty \dd{x} \chi_{1,x} \frac{x^2}{\pi p_{x,1}}
	\nonumber
	\\
	&+ \frac{\etwo (\PsiP-\PsiD)}{2}\int\limits_0^\infty \dd{x} \chi_{2,x} \frac{x^2}{\pi p_{x,2}}
	\nonumber
	\\
	&+\frac{\eone\etwo\kone\ktwo}{4(\eone\kone+\etwo\ktwo)} \PsiD^2L
	\nonumber
	\\
	&-\frac{\PsiD\eone\etwo}{2\pi} \int\limits_0^\infty \dd{x} 
	\frac{1}{\eone p_{x,1} + \etwo p_{x,2}}
	\nonumber
	\\
	& \times  \Biggl[ \frac{-2\PsiD \ktwo^2}{\eone\kone+\etwo\ktwo} 
	\Biggl(\frac{(\ktwo^2-\kone^2)(\eone\kone+\etwo\ktwo)}{\ktwo^2p_{x,2}^2\qty[(p_{x,1}/p_{x,2})+(\kone/\ktwo)]} 
	\nonumber
	\\
	&- \frac{\kone}{\ktwo} \Biggl(\frac{\eone}{p_{x,1}+ \kone} + 
	\frac{\etwo}{p_{x,2} + \ktwo}\Biggr)\Biggr)+ \frac{2\PsiP(\ktwo^2-\kone^2)}{p_{x,1}p_{x,2}} \Biggr].
\label{eq:E_LINE_ENERGY}
\end{align}
\end{shaded*}
\section{Superposition approximation} 
\label{sec:energy_sup}

As a next step, expressions for the interaction parameters, such as surface and line interaction energy densities,
are derived by using the electrostatic potential obtained within the superposition approximation.
Analogous to the exact calculations, expressions for the surface, line and interfacial tension are calculated as well.

It is important to note, that by calculating the superposition approximation expressions for the line and surface interactions, no further 
approximation method is applied.
Here, superposition approximation simply states that the quantities are calculated by using the electrostatic potential obtained by superposing the potential 
of one single wall, as carried out in Section~\ref{sec:superposition}.

\noindent
The contribution to the free energy given in Eq.~\eqref{eq:system_energies}, inserted with the approximated
electrostatic potential in Eqs.~\eqref{eq:sup_potential_1} and~\eqref{eq:sup_potential_2}, reads
\begin{align}
	\widetilde{\Omega}^s(L) =~&\eone \PsiP L_y \int\limits_{-L_x}^0 \dd{x} \partial_z \Psi_1^s(x,z)\rvert_{z=0}
	+ \etwo (\PsiP - \PsiD) L_y \int\limits_{0}^{L_x} \dd{x} \partial_z \Psi_2^s(x,z)\rvert_{z=0}
	\nonumber
	\\
	&- \frac{\eone \PsiD L_y}{2} \int\limits_0^{L} \dd{z} \partial_x \Psi_1^s(x,z)\rvert_{x=0}.
\label{eq:S_omega_int}
\end{align}
In similar fashion to Section \ref{sec:energy_exact}, the introduction of the following definitions 
allow for a clearer presentation
\begin{align}
	&p_i(q) \coloneqq \sqrt{\kappa_i^2 + q^2} \qquad \textrm{with} \qquad i \in \qty{1,2},
	\\
	&\Xi(q) \coloneqq \qty[\frac{q^2\PsiP}{p_1^2(q)} -\frac{q^2 (\PsiP -\PsiD)}{p_2^2(q)} -\PsiD ].
\end{align}
The derivatives of the electrostatic potential within superposition approximation
in Eq.~\eqref{eq:S_omega_int}, evaluated at $x=0$ and $z=0$ respectively, are given by
\begin{align}
	\partial_z \Psi_1^s(x,z)\rvert_{z=0} =& - \kone \PsiP +\kone \PsiP e^{-\kone L}
	\nonumber
	\\
	&-\frac{2}{\pi} \int\limits_0^\infty \dd{q} \frac{\etwo p_2(q)}{\eone p_1(q) + \etwo p_2(q)}  \Xi(q) e^{p_{1}(q)x}
	\nonumber
	\\
	&+ \frac{2}{\pi} \int\limits_0^\infty \dd{q} \frac{\etwo p_2(q)}{\eone p_1(q) + \etwo p_2(q)} \Xi(q) \ e^{p_1(q)x} \cos(qL),
	\label{eq:S_Derivative}
	\nonumber
	\\
	\partial_z \Psi_2^s(x,z)\rvert_{z=0} =& - \ktwo (\PsiP-\PsiD) + \ktwo (\PsiP-\PsiD) e^{-\ktwo L}
	\nonumber
	\\
	&+\frac{2}{\pi} \int\limits_0^\infty \dd{q} \frac{\eone p_1(q)}{\eone p_1(q) + \etwo p_2(q)} \Xi(q)\ e^{-p_2(q)x}
	\nonumber
	\\
	&-\frac{2}{\pi} \int\limits_0^\infty \dd{q} \frac{\eone p_1(q)}{\eone p_1(q) + \etwo p_2(q)} \Xi(q)\ e^{-p_2(q)x} \cos(qL),
	\nonumber
	\\
	\partial_x \Psi_1^s(x,z)\rvert_{x=0} =& 
	-\frac{2}{\pi} \int\limits_0^\infty \dd{q} \frac{\etwo p_2(q) p_1(q)}{\eone p_1(q) + \etwo p_2(q)}  \frac{\Xi(q)}{q} \sin(qz)
	\nonumber
	\\
	&+\frac{2}{\pi} 
	\int\limits_0^\infty \dd{q}  \frac{\etwo p_2(q) p_1(q}{\eone p_1(q) + \etwo p_2(q)}\frac{\Xi(q)}{q} \sin(qz-qL).
\end{align}
To apply the procedure outlined in Section~\ref{sec:method}, the free energy contribution in Eq.~\eqref{eq:E_Omega_Int} 
is separated into  the following terms 
\begin{align}
	\widetilde{\Omega}^s(L) =& \qty[\obar_{\gamma,1}(L) + \obar_{\gamma,2}(L)]2L_xL_y
	\nonumber
	\\
	&+\qty[\obar_{\tau}^{1}(L) + \obar_{\tau}^{2}(L)+ \obar_{\tau,\gamma_{1,2}}(L)]2L_y.
\label{eq:S_ENERGY_CONTR}
\end{align}
As a further step, the terms in Eq.~\eqref{eq:S_ENERGY_CONTR} are separated by their proportionality as explained earlier and read
\begin{align}
	&\obar_{\gamma,1}(L) := -\frac{\eone \kone}{2} \PsiP^2 + \frac{\eone \kone}{2} \PsiP e^{-\kone L},
	\label{eq:S_Omega_Gamma_1_Combined}
	\\
	\nonumber
	\\
	&\obar_{\gamma,2}(L) :=  -\frac{\etwo \ktwo}{2} (\PsiP-\PsiD)^2 + \frac{\etwo \ktwo}{2} (\PsiP-\PsiD)^2  e^{-\ktwo L},
	\label{eq:S_Omega_Gamma_2_Combined}
	\\
	\nonumber
	\\
	&\obar_{\tau}^{1}(L) := - \frac{\eone \PsiP}{\pi} \int\limits_0^\infty \dd{q} \frac{\etwo p_2(q)}{\eone p_1 + \etwo p_2}  
	\frac{\Xi(q)}{p_1(q)}
	\nonumber
	\\
	&\qquad\qquad + \frac{\eone \PsiP}{\pi} \int\limits_0^\infty \dd{q} \frac{\etwo p_2(q)}{\eone p_1(q) 
	+ \etwo p_2(q)} \frac{\Xi(q)\cos(qL)}{p_1(q)},
	\label{eq:s_barline_1}
	\\
	\nonumber
	\\
	&\obar_{\tau}^{2}(L): = \frac{\etwo (\PsiP-\PsiD)}{\pi} \int\limits_0^\infty \dd{q} 
	\frac{\eone p_1(q)}{\eone p_1(q) + \etwo p_2(q)} \frac{\Xi(q)}{p_2(q)}
	\nonumber
	\\
	&\qquad\qquad -\frac{\etwo (\PsiP-\PsiD)}{\pi} \int\limits_0^\infty \dd{q} 
	\frac{\eone p_1(q)}{\eone p_1(q) + \etwo p_2(q)}  \frac{\Xi(q)\cos(qL)}{p_2(q)},
	\label{eq:s_barline_2}
	\\
	\nonumber
	\\
	&\obar_{\tau,\gamma_{1,2}}(L) := \frac{2\eone \etwo \PsiD}{\pi} \int\limits_0^\infty \dd{q} 
	\frac{p_1(q) p_2(q)}{\eone p_1(q) + \etwo p_2(q)} \frac{\Xi(q) \sin^2(\frac{qL}{2})}{q^2}.
	\label{eq:s_obar_tau_gamma}
\end{align}

\subsection{Surface interaction energy densities and surface tensions}

As before, the surface contributions to the free energy scale with $2L_xL_y$. 
Subsequently, to separate the $L$-independent from the $L$-dependent contributions the procedure 
outlined in Section~\ref{sec:method} is applied.

\subsubsection{Surface tension}

Starting with the surface tension, one has to consider the limit $L\to\infty$, which corresponds to infinite inter-particle separation
and thus, all L-dependent contributions vanish.
Therefore, the $L$-independent quantity for to medium 1 in Eq.~\eqref{eq:S_Omega_Gamma_1_Combined} reads
\begin{align}
	\lim_{L\to\infty} \obar_{\gamma,1}^s(L) 
	&= \lim_{L\to\infty} -\frac{\eone\kone}{2}\PsiP^2\qty(1 + e^{-\kone L})
	\nonumber
	\\
	&= -\frac{\eone \kone}{2}\PsiP^2,
\end{align} 
and Eq.~\eqref{eq:S_Omega_Gamma_2_Combined} provides the contribution for medium 2 as
\begin{align}
	\lim_{L\to\infty} \obar_{\gamma,2}(L)
	&= \lim_{L\to\infty} - \frac{\etwo\ktwo}{2}(\PsiP - \PsiD)^2\qty(1 + e^{-\ktwo L})
	\nonumber
	\\
	&= -\frac{\etwo \ktwo}{2}(\PsiP - \PsiD)^2.
\end{align}
Consequently, the surface tensions within the superposition approximation are given by
\begin{shaded*}
\begin{align}
	\gamma_1^s &= -\frac{\eone \kone}{2} \PsiP^2,
	\label{eq:S_surface_tension1}
	\\
	\gamma_2^s &= -\frac{\etwo \ktwo}{2} (\PsiP - \PsiD)^2.
	\label{eq:S_surface_tension}
\end{align}
\end{shaded*}

\subsubsection{Surface interaction energy density}

Since all L-independent parts of the surface interaction are identified, the L-dependent contributions
are now obtained as outlined in Section~\ref{sec:method}.
Thereby, the surface interaction energy densities, which act between the walls and medium $i \in \qty{1,2}$, 
are calculated by subtracting the surface tensions in Eqs.~\eqref{eq:S_surface_tension1} 
and~\eqref{eq:S_surface_tension} from the terms in Eqs.~\eqref{eq:S_Omega_Gamma_1_Combined} 
and~\eqref{eq:S_Omega_Gamma_2_Combined}
\begin{align}
	&\obar_{\gamma,1}(L) - \lim_{L\to\infty} \obar_{\gamma,1}(L) = \frac{\eone \kone}{2} \PsiP^2 e^{-\kone L},
	\\
	&\obar_{\gamma,2}(L)- \lim_{L\to\infty} \obar_{\gamma,2}(L) = \frac{\etwo \ktwo}{2} (\PsiP-\PsiD)^2 e^{-\ktwo L}.
\end{align}
As a result, the surface interaction energy densities within the superposition approximation are given by
\begin{shaded*}
\begin{align}
	&\omega_{\gamma,1}^s(L) = \frac{\eone \kone}{2}\PsiP^2 e^{-\kone L}\label{eq:S_SURFACE1_ENERGY},
	\\
	&\omega_{\gamma,2}^s(L) = \frac{\etwo \ktwo}{2} (\PsiP-\PsiD)^2 e^{-\ktwo L}\label{eq:S_SURFACE2_ENERGY}.
\end{align}
\end{shaded*}

\subsection{Line interaction energy density, line tension and interfacial tension}

The line interaction energy density, as well as the line tension, are characterized by their 
proportionality to the total length of the three-contact lines given by $2L_y$ (see Fig.~\ref{fig:energie_system}).
By applying the same procedure used earlier, it is possible to separate $L$-independent 
from $L$-dependent terms and by doing so, one obtains the line interaction energy density and the line tension. 
Additionally, the $L$-independent interfacial tension is identified as term proportional to the interfacial area $LL_y$ later on.

\subsubsection{Line tension and interfacial tension}

The first contribution to the line tension is identified by taking the limit $L\to\infty$
of Eq.~\eqref{eq:s_barline_1}, which results in
\begin{align}
	\lim_{L\to\infty} \obar_{\tau}^{1}(L) 
	=& \lim_{L\to\infty} \Biggl[-\frac{\eone \PsiP}{\pi} 
	\int\limits_0^\infty \dd{q} \frac{\etwo p_2(q)}{\eone p_1 + \etwo p_2} \frac{\Xi(q)}{p_1(q)}
	\nonumber
	\\
	&+\frac{\eone \PsiP}{\pi} \int\limits_0^\infty \dd{q} \frac{\etwo p_2(q)}{\eone p_1(q) 
	+ \etwo p_2(q)} \frac{\Xi(q)\cos(qL)}{p_1(q)} \Biggr]
	\nonumber
	\\
	=& - \frac{\eone \PsiP}{\pi} \int\limits_0^\infty \dd{q} 
	\frac{\etwo p_2(q)}{\eone p_1(q) + \etwo p_2(q)} \frac{\Xi(q)}{p_1(q)}.
\label{eq:s_lim_obar_tau_1}
\end{align}
Analogous to the line tension, the second contribution is obtained in the limit $L\to\infty$ of Eq.~\eqref{eq:s_barline_2}
\begin{align}
	\lim_{L\to\infty} \obar_{\tau}^{2}(L)  =&
	\lim_{L\to\infty} \Biggl[ \frac{\etwo (\PsiP-\PsiD)}{\pi} \int\limits_0^\infty \dd{q} 
	\frac{\eone p_1(q)}{\eone p_1(q) + \etwo p_2(q)} \frac{\Xi(q)}{p_2(q)}
	\nonumber
	\\
	&-\frac{\etwo (\PsiP-\PsiD)}{\pi} \int\limits_0^\infty \dd{q} 
	\frac{\eone p_1(q)}{\eone p_1(q) + \etwo p_2(q)}  \frac{\Xi(q)\cos(qL)}{p_2(q)} \Biggr]
	\nonumber
	\\
	=& \frac{\etwo (\PsiP-\PsiD)}{\pi} \int\limits_0^\infty \dd{q} 
	\frac{\eone p_1(q)}{\eone p_1(q) + \etwo p_2(q)} \frac{\Xi(q)}{p_2(q)}.
\label{eq:s_lim_obar_tau_2}
\end{align}
By further simplification of Eq.~\eqref{eq:s_obar_tau_gamma}, it is possible to obtain the following terms
\begin{align}
	\obar_{\tau,\gamma_{1,2}}(L) = \obar_\tau^{3}(L) + \frac{\gamma_{1,2}}{2} L.
\label{eq:S_LINETENSION_INTERFACETENSION}
\end{align}
Here, the first term contributes to the line interaction and reads
\begin{align}
	\obar_{\tau}^{3}(L)=~&\frac{\PsiD \eone \etwo}{\pi} 
	\int\limits_0^\infty \dd{q} \frac{1}{\eone p_1(q) + \etwo p_2(q)}
	\nonumber	
	\\
	& \times  \Biggl[ \frac{-\PsiD \ktwo^2}{\eone\kone+\etwo\ktwo} 
	\Biggl(\frac{(\ktwo^2-\kone^2)(\eone\kone+\etwo\ktwo)}{\ktwo^2p_2^2(q)\qty[(p_1(q)/p_2(q))+(\kone/\ktwo)]} 
	\nonumber
	\\
	&- \frac{\kone}{\ktwo} \Biggl(\frac{\eone}{p_1(q) + \kone} + 
	\frac{\etwo}{p_2(q) + \ktwo}\Biggr)+ \frac{\PsiP(\ktwo^2-\kone^2)}{p_1(q)p_2(q)} \Biggr]
	\nonumber
	\\
	&\times \qty{1-\cos(qL)}.
\label{eq:s_barlinne_3}
\end{align}
Since this term is proportional to the total line length $2L_y$, taking the limit $L\to\infty$ yields the contribution to the line tension 
\begin{align}
	\lim_{L\to\infty} \obar_{\tau}^{3}(L) =&  \lim_{L\to\infty} 
	\Biggl[ \frac{\PsiD \eone \etwo}{\pi} 
	\int\limits_0^\infty \dd{q} \frac{1}{\eone p_1(q) + \etwo p_2(q)}
	\nonumber	
	\\
	& \times  \Biggl[ \frac{-\PsiD \ktwo^2}{\eone\kone+\etwo\ktwo} 
	\Biggl(\frac{(\ktwo^2-\kone^2)(\eone\kone+\etwo\ktwo)}{\ktwo^2p_2^2(q)\qty[(p_1(q)/p_2(q))+(\kone/\ktwo)]} 
	\nonumber
	\\
	&- \frac{\kone}{\ktwo} \Biggl(\frac{\eone}{p_1(q) + \kone} + 
	\frac{\etwo}{p_2(q) + \ktwo}\Biggr)+ \frac{\PsiP(\ktwo^2-\kone^2)}{p_1(q)p_2(q)} \Biggr]
	\nonumber
	\\
	&\times \qty{1-\cos(qL)} \Biggr]
	\nonumber
	\\
	=~&\frac{\PsiD \eone \etwo}{\pi} \int\limits_0^\infty \dd{q} 
	\frac{1}{\eone p_1(q) + \etwo p_2(q)}
	\nonumber
	\\
	&\cross  \Biggl[ \frac{-\PsiD \ktwo^2}{\eone\kone+\etwo\ktwo} 
	\Biggl(\frac{(\ktwo^2-\kone^2)(\eone\kone+\etwo\ktwo)}{\ktwo^2p_2^2(q)\qty[(p_1(q)/p_2(q))+(\kone/\ktwo)]} 
	\nonumber
	\\
	&-\frac{\kone}{\ktwo} \Biggl( \frac{\eone}{p_1(q) + \kone} + 
	\frac{\etwo}{p_2(q) + \ktwo}\Biggr)\Biggr)+ \frac{\PsiP(\ktwo^2-\kone^2)}{p_1(q)p_2(q)} \Biggr].
\label{eq:s_lim_obar_tau3}
\end{align}
Finally, the sum of all three L-independent terms, given in Eqs.~\eqref{eq:s_lim_obar_tau_1},~\eqref{eq:s_lim_obar_tau_2}
and~\eqref{eq:s_lim_obar_tau3}, results in the line tension
\begin{align}
	\tau = 
	\lim\limits_{L\to\infty} \obar^{1}_\tau(L) 
	+ \lim\limits_{L\to\infty} \obar^{2}_\tau(L) 
	+ \lim\limits_{L\to\infty} \obar^{3}_\tau(L).
\end{align}
Therefore, the final expression for the line tension within the superposition approximation reads
\begin{shaded*}
\begin{align}
	\tau^s =&- \frac{\eone \PsiP}{\pi} \int\limits_0^\infty \dd{q} 
	\frac{\etwo p_2(q)}{\eone p_1(q) + \etwo p_2(q)}  \frac{\Xi(q)}{p_1(q)}
	\nonumber
	\\
	&+  \frac{\etwo (\PsiP-\PsiD)}{\pi} \int\limits_0^\infty \dd{q} 
	\frac{\eone p_1(q)}{\eone p_1(q) + \etwo p_2(q)} \frac{\Xi(q)}{p_2(q)}
	\nonumber
	\\
	&+ \frac{\PsiD \eone \etwo}{\pi} \int\limits_0^\infty \dd{q} 
	\frac{1}{\eone p_1(q) + \etwo p_2(q)}
	\nonumber
	\\
	&\times  \Biggl[ \frac{-\PsiD \ktwo^2}{\eone\kone+\etwo\ktwo} 
	\Biggl(\frac{(\ktwo^2-\kone^2)(\eone\kone+\etwo\ktwo)}{\ktwo^2p_2^2(q)\qty[(p_1(q)/p_2(q))+(\kone/\ktwo)]} 
	\nonumber
	\\
	&-\frac{\kone}{\ktwo} \Biggl( \frac{\eone}{p_1(q) + \kone} 
	+ \frac{\etwo}{p_2(q) + \ktwo}\Biggr)\Biggr)+ \frac{\PsiP(\ktwo^2-\kone^2)}{p_1(q)p_2(q)} \Biggr].
\end{align}
\end{shaded*}
\noindent
Clearly, the remaining term in Eq.~\eqref{eq:S_LINETENSION_INTERFACETENSION} is proportional to the 
interfacial area $LL_y$ (see Fig.~\ref{fig:energie_system}) and thus provides the expression for the interfacial tension
\begin{align}
	\gamma_{1,2} &= - \frac{2 \eone \kone \etwo \ktwo}{\pi(\eone\kone + \etwo \ktwo)}\PsiD^2 
	\int\limits_{0}^{\infty} \dd{q} \frac{\sin^2(qL/2)}{(qL/2)^2} \frac{L}{2}.
\label{eq:s_gamma_12_2}
\end{align}
The integral in Eq. \eqref{eq:s_gamma_12_2} is solved by using the substitution 
\begin{equation}
	u = \frac{qL}{2} \qquad \textrm{and} \qquad \dd{q} = \frac{2}{L}\dd{u}.
\end{equation}
Therefore, its solution reads 
\begin{equation}
	\int\limits_{0}^{\infty} \dd{q} \frac{\sin^2(qL/2)}{(qL/2)^2} \frac{L}{2} = \int\limits_{0}^\infty \dd{u} \frac{\sin^2(u)}{u^2}= \frac{\pi}{2}.
\end{equation}
Finally, the interfacial tension within the superposition approximation is given by
\begin{shaded*}
\begin{equation}
	\gamma_{1,2}^s = -  \frac{\eone\kone \etwo\ktwo}{\eone\kone+\etwo\ktwo} \PsiD^2.
	\label{eq:s_interface_tension}
\end{equation}
\end{shaded*}
\paragraph{Remark}

A comparison with the result within the exact calculation in Eq.~\eqref{eq:E_INTERFACE_TENSION} shows, that the superposition approximation
fails to predict the interfacial tension correctly and overestimates it by a factor of 2. 
Since the interfacial tension is a property solely determined by the fluid-interface, the behavior is consistent with the 
results obtained in Ref.~\cite{majee2018}.

\subsubsection{Line interaction energy density}

Analogous to the exact calculations, the superposition approximation for the line interaction energy density is obtained
by applying the method outlined in Section~\ref{sec:method}. 
Thus, subtracting the L-independent contributions (line and interfacial tension) in 
Eqs.~\eqref{eq:s_lim_obar_tau_1},~\eqref{eq:s_lim_obar_tau_2},~\eqref{eq:s_lim_obar_tau3} and~\eqref{eq:s_interface_tension} 
from Eqs.~\eqref{eq:s_barline_1},~\eqref{eq:s_barline_2} and~\eqref{eq:s_barlinne_3} results in 
\begin{align}
	\omega_{\tau}(L) = \qty( \obar_{\tau}^1(L) + \obar_{\tau}^2(L) + \obar_{\tau, \gamma_{1,2}}(L) ) - \tau - \frac{\gamma_{1,2}}{2}L.
\end{align}
Therefore, the line interaction energy density within superposition approximation is given by

\begin{shaded*}
\begin{align}
	\omega_\tau^s(L) =~  
	&\frac{\eone \PsiP}{\pi} \int\limits_0^\infty \dd{q}
	\frac{\etwo p_2(q)}{\eone p_1(q)+ \etwo p_2(q)} \frac{\Xi(q)\cos(qL)}{p_1(q)}
	\nonumber
	\\
	&- \frac{\etwo (\PsiP-\PsiD)}{\pi} \int\limits_0^\infty \dd{q}
	\frac{\eone p_1(q)}{\eone p_1(q)+ \etwo p_2(q)}  \frac{\Xi(q)\cos(qL)}{p_2(q)}
	\nonumber
	\\
	&-\frac{\PsiD\eone\etwo}{\pi} \int\limits_0^\infty \dd{q} 
	\frac{1}{\eone p_1(q) + \etwo p_2(q)}
	\nonumber
	\\
	& \times  \Biggl[ \frac{-\PsiD \ktwo^2}{\eone\kone+\etwo\ktwo} 
	\Biggl(\frac{(\ktwo^2-\kone^2)(\eone\kone+\etwo\ktwo)}{\ktwo^2p_2^2(q)\qty[(p_1(q)/p_2(q))+(\kone/\ktwo)]} 
	\nonumber
	\\
	&- \frac{\kone}{\ktwo} \Biggl(\frac{\eone}{p_1(q)+ \kone} 
	+ \frac{\etwo}{p_2(q) + \ktwo}\Biggr)\Biggr)+ \frac{\PsiP(\ktwo^2-\kone^2)}{p_1(q)p_2(q)} \Biggr]
	\nonumber
	\\
	&\times \cos(qL).
\label{eq:S_LINE_ENERGY}
\end{align}
\end{shaded*}
%

%% file: figure/tikz/model_system.tikz
\begin{tikzpicture}
		\filldraw[mpired,opacity=0.5] (0,0) -- (2,1) -- (2,5) -- (0,4) --cycle;
		\draw[->,dashed] (2,5) -- (2,5.5);
		\filldraw[mpidarkblue,opacity=0.5] (0,2) -- (5,2) -- (7,3) -- (2,3) -- cycle;
		\draw[->,dashed] (7,3) -- (7.5,3);
		\draw[black,opacity=0.5] (0,0) -- (2,1) -- (2,5) -- (0,4) --cycle;
		\draw[black,opacity=0.5] (0,2) -- (5,2) -- (7,3) -- (2,3) -- cycle;
		\filldraw[mpired,opacity=0.5] (5,0) -- (7,1) -- (7,5) -- (5,4) -- cycle;
		\draw[black,opacity=0.5] (5,0) -- (7,1) -- (7,5) -- (5,4) --cycle;
		\draw[->,dashed] (0,2) -- (-0.5,1.75);
		\draw (2,5.8) node{x};
		\draw (7.7,3) node{z};
		\draw (-0.6,1.9) node{y};
		\draw [black,decorate,decoration={brace,amplitude=5pt}] (0,2)  -- (0,4) node[xshift=-14pt,midway]{$L_x$};
		\draw [black,decorate,decoration={brace,amplitude=5pt}] (0,0)  -- (0,2) node[xshift=-14pt,midway]{$L_x$};
		\draw [black,decorate,decoration={brace,amplitude=5pt}] (0,4)  -- (2,5) node[yshift=11pt,xshift=-1,midway]{$L_y$};
		\draw [black,decorate,decoration={brace,amplitude=5pt}] (5,2)  -- (0,2) node[yshift=-14pt,midway]{$L$};
		\draw (2.2,3.2) node{0};
		\draw (3,0.5) node[draw, rounded corners]{medium 1};
		\draw (4,4.5) node[draw, rounded corners]{medium 2};
		\draw[mpigreen,very thick] (5,2) -- (7,3);	
		\draw[mpigreen,very thick] (0,2) -- (2,3);
\end{tikzpicture}

%% file: chapters/05_Discussion.tex

\chapter{Discussion}

In this chapter, the analytical expressions for the electrostatic potential $\Psi(x,z)$, the surface interaction energy densities
$\omega_{\gamma,i}(L)$ in medium $i \in \qty{1,2}$ and the line interaction energy density $\omega_\tau(L)$
are discussed for different system parameters. 
The standard configuration and variation of said parameters is given in Tab.~\ref{tab:parameter}. 
Section~\ref{sec: EvS} compares the analytical expressions obtained by exact calculation with the expressions of the 
superposition approximation within the standard configuration.
In Section~\ref{sec:VAR} the effects of variations of parameters on the interaction energies and electrostatic potential are compared 
by using the exact expressions. 
Finally, Section~\ref{sec:VAR_EvS} compares the expression within exact calculation with the superposition approximation for 
variations of the parameters.

Unless stated otherwise, the separation length is scaled as $\kone L$, the electrostatic potential is expressed 
in the units of $1/\beta e $, the surface interaction energy density in the units of $\kone^2/\beta$ 
and the line interaction energy density in the units of $\kone/\beta$.

\begin{table}[htbp]
	\centering
	\begin{tabular}{@{}lllllll@{}}
		\toprule
 			& $\beta e\PsiP$ & $\beta e\PsiD$ & $\epsilon_{r,1}$ & $\epsilon_{r,2}$ & $\kone$ & $\ktwo$ \\ \midrule
			standard & 5 & 1 & 80 & 2 & 0.1 $\textrm{nm}^{-1}$ & 0.01 $\textrm{nm}^{-1}$ \\
 			&  &  &  &  &  &  \\
			variation & 2 & -1 & - & 6 & 0.5 $\textrm{nm}^{-1}$& 0.005 $\textrm{nm}^{-1}$ \\
 			& 10 & 3 & - & - & - & - \\
 			&-&5&-&-&-&-
 			\\ \bottomrule
	\end{tabular}
	\label{tab:parameter}
	\caption{
	Standard configuration and variation of system parameters used to discuss the electrostatic potential, 
	the surface interaction energy densities and the line interaction energy density.
	}
\end{table}

\newpage

\section{Standard parameter set}
\label{sec: EvS}

This section compares the analytical expressions for the electrostatic potential, the surface interaction energy densities 
and the line interaction energy density obtained within the exact calculation and those using the superposition approximation 
for the standard configuration specified in Tab.~\ref{tab:parameter}.

\subsection{Electrostatic potential}

\begin{figure}[htbp]
	\centering
	\includegraphics[width=\linewidth]{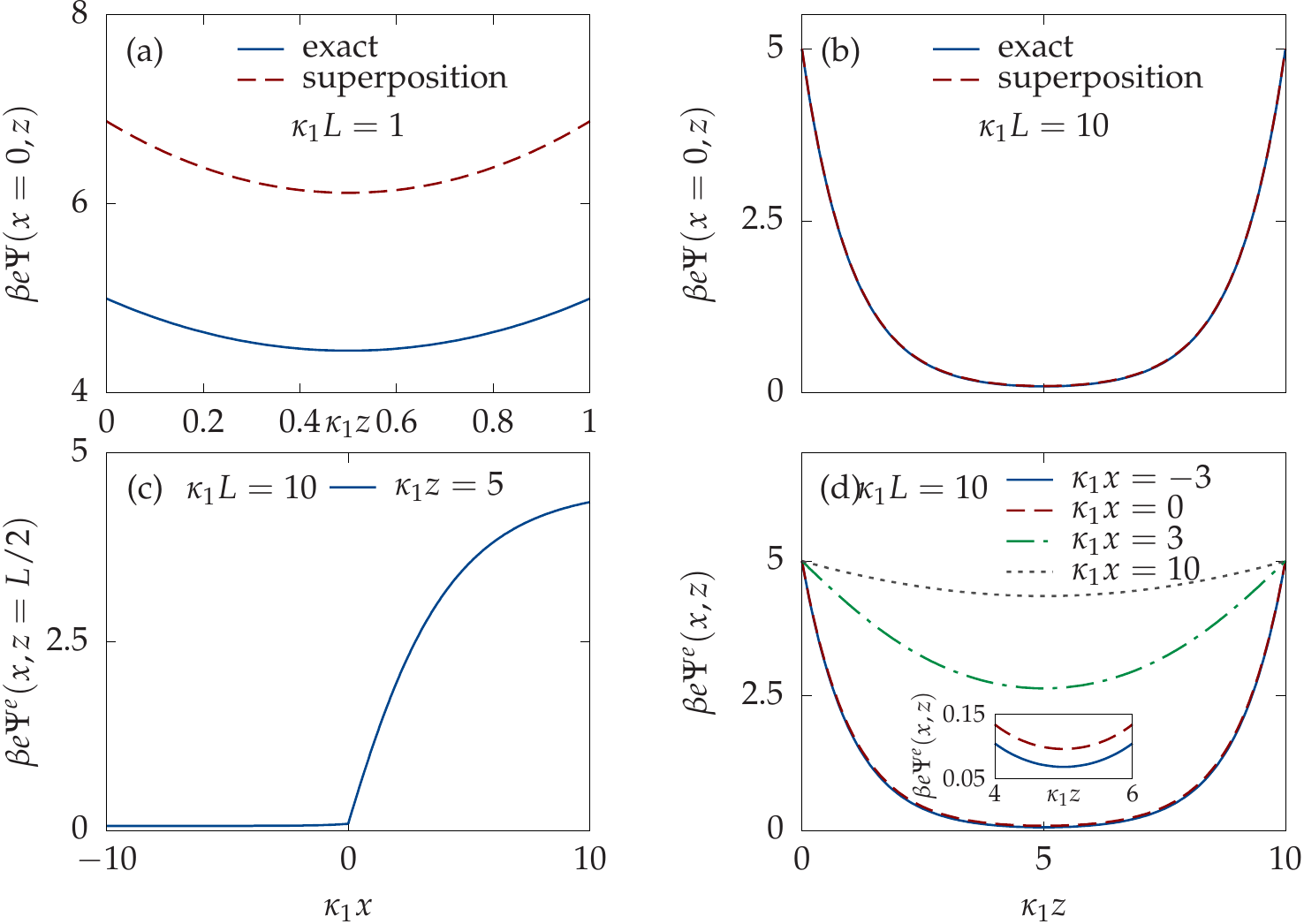}
	\caption
			[
			Electrostatic potential within exact and superposition calculation for standard system parameters. 
			Additional cross-sections for $x\neq0$ and $z=L/2$ are shown
			]
			{
			(a)(b) Electrostatic potential $\Psi(x,z)$ within exact and superposition calculations expressed in the units 
			of $1/\beta e$, as functions of $\kone z$ for standard configuration specified in Tab.~\ref{tab:parameter}. 
			The separation length between the two walls is given by $\kone L = 1$ in (a) and $\kone L = 10$ in (b), respectively. 
			Clearly, (a) shows that the superposition fails to predict the electrostatic potential for small slit widths properly. 
			However, upon increasing $\kone L$, the differences between exact and superposition calculations decreases as expected. 
			(c) The exact electrostatic potential $\Psi^e(x,z=L/2)$ expressed in the units of $1/\beta e$, as a function 
			of $\kone x$ with slit width $\kone L =10$. 
			(d) Exact electrostatic potential $\Psi^e(x\neq0,z)$ in the units of $1/\beta e $, 
			as function of $\kone z$ and slit width $\kone L =10$.
			}
	\label{fig:EVS_S_POT}
\end{figure}

A comparison between the exact expression and superposition approximation of the electrostatic potential $\Psi(x,z)$ 
(obtained in Chapter~\ref{ch:3}) at the fluid-fluid interface at $x=0$ is shown in Fig.~\ref{fig:EVS_S_POT}a and Fig.~\ref{fig:EVS_S_POT}b. 
The potentials are illustrated as functions of $\kone z$ for two different slit widths $\kone L$. 
As seen in Fig.~\ref{fig:EVS_S_POT}a, the potential within the superposition approximation differs significantly 
from the exact expression for small separation lengths $\kone L$. 
Since the superposition approximation is obtained by superposing the electrostatic potential of two single walls 
(which decay $\sim e^{-\kappa_i z}$), it overestimates the electrostatic potential at narrow widths, i.e. the boundary condition 
of the constant surface potential $\PsiP$ on the walls is violated.
Upon increasing the separation length as seen in Fig.~\ref{fig:EVS_S_POT}b, the difference between exact calculation and 
superposition approximation decreases as expected and the boundary condition for the surface potential at the walls is restored.
Additionally, Fig.~\ref{fig:EVS_S_POT}c shows the exact electrostatic potential for $z=L/2$ as a function 
of $\kone x$ with slith width $\kappa_1 L = 10$.
Additionally, Fig.~\ref{fig:EVS_S_POT}d shows the behavior of the exact electrostatic potential for $x \neq 0$ (away from the interface).
The electrostatic potential increases for $x>0$ (moving in direction of medium 2) and decreases for $x<0$ (moving in direction of medium 1).

\newpage
\null
\vfill

\vfill
\newpage

\subsection{Surface interaction energy densities}

The surface interaction energy densities $\omega_{\gamma,i}(L)$ (calculation in Chapter~\ref{ch:4}), acting in medium $i \in \qty{1,2}$ 
is given within the exact calculation in Eqs.~\eqref{eq:E_SURFACE1_ENERGY} and~\eqref{eq:E_SURFACE2_ENERGY} and superposition approximation 
in Eqs.~\eqref{eq:S_SURFACE1_ENERGY} and~\eqref{eq:S_SURFACE2_ENERGY}.
Fig.~\ref{fig:EVS_S_SI} compares the surface interaction energies for both cases.
As one can see, the interaction is repulsive and decays monotonically. 
Nevertheless, the superposition approximation underestimates the surface energy density for all separation lengths. 
Additionally, the effective force per unit surface area $F_i$ in medium $i\in\qty{1,2}$ is shown in the insets of Fig.~\ref{fig:EVS_S_SI}a 
and Fig.~\ref{fig:EVS_S_SI}c in the units of $\kone^3/\beta$ and as function of $\kone L$, where it is 
defined as $ -\partial\omega_{\gamma,i}(L)/ \partial L$.
As the plots show, expressions within the exact and superposition calculations are both positive, therefore repulsive and 
decay monotonically for large separation.
However, the superposition approximation fails to predict the overall behavior correctly. 
In the limit of vanishing separation $L\to0$ it significantly overestimates the results obtained by exact calculation,
but then decays faster if the separations grow larger.

\subsubsection{Behavior for separations $L\to 0$}

In the limit of vanishing separations $L \to 0$, both $\omega_{\gamma,i}^e(L)$ and $\omega_{\gamma,i}^s(L)$ result in
a non-divergent, finite and repulsive surface interaction.
Moreover, the exact calculation and superposition approximation result in the same expression, which reads
\begin{align}
	&\lim\limits_{L\to 0} \omega_{\gamma,1}^e(L) = \lim\limits_{L\to 0} \omega_{\gamma,1}^s(L) 
	= \frac{\eone\kone}{2}\PsiP^2,
	\label{eq:SI_1}
	\\
	&\lim\limits_{L\to 0} \omega_{\gamma,2}^e(L) = \lim\limits_{L\to 0} \omega_{\gamma,2}^s(L) 
	= \frac{\etwo\ktwo}{2}(\PsiP-\PsiD)^2.
	\label{eq_SI_2}
\end{align}

\subsubsection{Asymptotic behavior}

\begin{figure}[htbp]
	\centering
	\includegraphics[width=\linewidth]{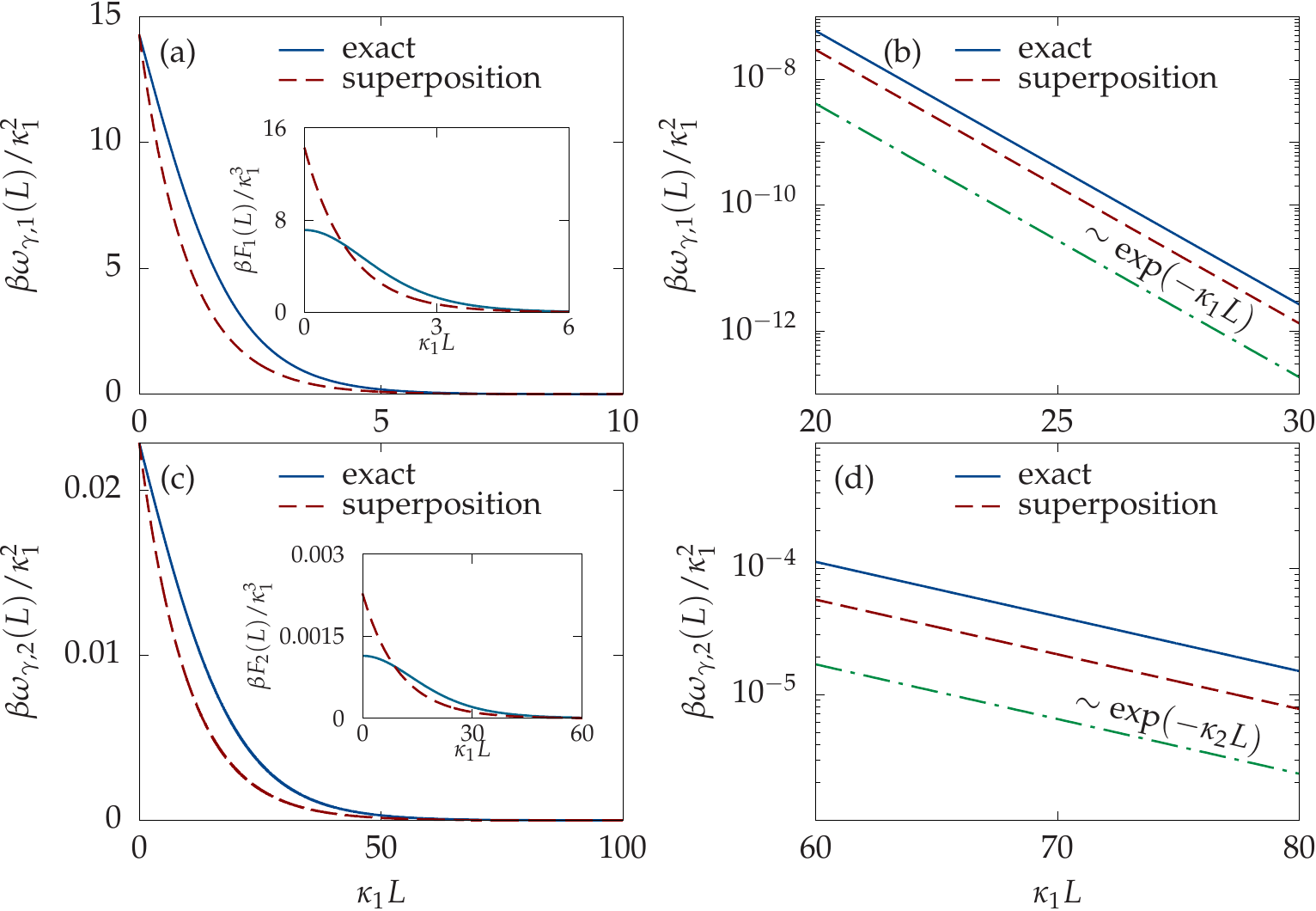}
	\caption
			[
			Surface interaction energy densities within exact and superposition calculation for system parameters in standard configuration
			]
			{
			Surface interaction energy densities $\omega_{\gamma,i}(L)$ in medium $i \in \qty{1,2}$ within exact and 
			superposition calculations expressed in of units $\kone^2/\beta$, as functions of the separation $\kone L$ for the 
			standard configuration specified in Tab.~\ref{tab:parameter}. 
			As the plots show, both surface interaction energy densities decay monotonically,
			but the superposition approximation fails to capture the overall behavior properly. 
			However, in the limit of vanishing separations, both the exact and superposition expressions reach the same 
			finite value as seen in (a) and (c). The semi-log plots in (b) and (d) show that although the superposition approximation 
			predicts the exponential decay $\sim e^{-\kappa_i L}$ correctly, it always underestimates the surface 
			interactions by a factor of 2, seen in the offset. 
			In the inset of (a) and (c) the effective forces $F_i(L)$ per unit area in medium $i\in\qty{1,2}$ are shown. 
			For the exact and superposition calculations, the force is positive and therefore repulsive. 
			However, the superposition approximation fails to capture the overall behavior properly. 
			It overestimates the force for small separation lengths and underestimates it for larger separation.
			}
	\label{fig:EVS_S_SI}
\end{figure}

The surface interaction energy densities within the exact calculation and the superposition approximation 
both decay  $\sim e^{-\kappa_i L}$ in the large asymptotic limit $L\gg 1$ in each corresponding medium $i \in \qty{1,2}$. 
Although the superposition approximation predicts the decay for larger separations $L$ correctly, 
it is too small by a factor of 2 compared to the exact expression as seen in the offset between the two 
curves in Fig.~\ref{fig:EVS_S_SI}b and Fig.~\ref{fig:EVS_S_SI}d. 
In the asymptotic limit the exact expression is given with $\omega^e_{\gamma,i}(L) \simeq \epsilon_i \kappa_i e^{-\kappa_i L}$ 
and the superposition approximation reads $\omega^s_{\gamma,i}(L) \simeq \frac{\epsilon_i \kappa_i}{2} e^{-\kappa_i L}$. 
Therefore, one obtains
\begin{equation}
	\lim\limits_{L\to \infty} \frac{\omega_{\gamma,i}^e(L)}{\omega_{\gamma,i}^s(L)} = 2.
\end{equation}

\newpage
\null
\vfill

\vfill
\newpage

\subsection{Line interaction energy density}

The line interaction energy density $\omega_\tau(L)$ is calculated in Chapter~\ref{ch:4} within exact calculation 
Eq.~\eqref{eq:E_LINE_ENERGY} and superposition approximation Eq.~\eqref{eq:S_LINE_ENERGY}.
Fig.~\ref{fig:EVS_S_LI} shows the line interaction energy density for the standard parameter set 
specified in Tab.~\ref{tab:parameter}.
Both $\omega^e_\tau(L)$ and $\omega^s_\tau(L)$ stay finite and behave repulsive for small separation. 
Upon increasing the separation length $L$ both expressions decay, show a minimum and vanish in the limit of $L\to\infty$.
Although the superposition approximation predicts the overall behavior correctly, it reaches the minimum sooner 
compared to the exact expression.

\subsubsection{Behavior for separation $L\to 0$}

In the limit of vanishing separations, the exact and superposition calculations yield 
the same finite line interaction energy density (seen in Fig.~\ref{fig:EVS_S_LI}a). 

\subsubsection{Asymptotic behavior}

\begin{figure}[htbp]
	\centering
	\includegraphics[width=\linewidth]{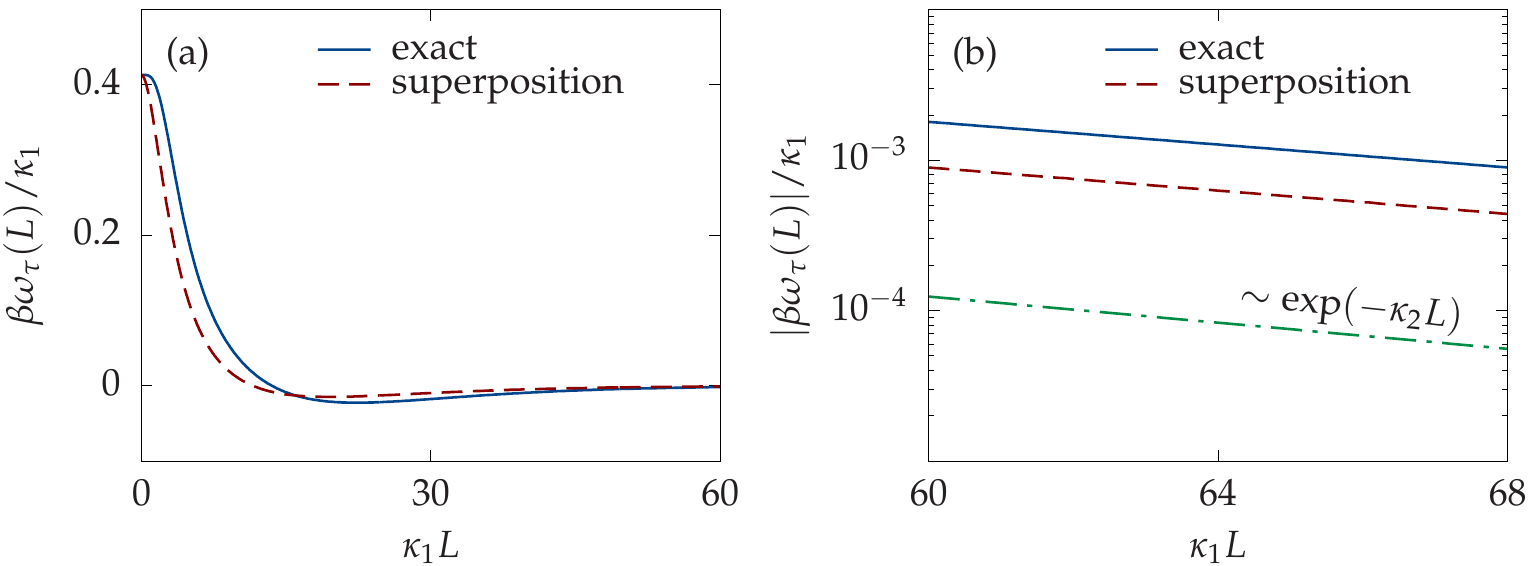}
	\caption
			[
			Line interaction energy density within exact and superposition calculation for system parameters in standard configuration.
			]
			{
			Line interaction energy density $\omega_\tau(L)$ within exact and superposition calculations expressed 
			in units of $\kone/\beta$, as functions of the separation $\kone L$ for standard configuration specified 
			in Tab.~\ref{tab:parameter}. Clearly, (a) shows that the superposition approximation fails to capture the correct behavior. 
			Although in the limit of vanishing separations, $\omega_\tau^e(L)$ and $\omega_\tau^s(L)$ reach the same finite value. 
			Additionally, the superposition approximation predicts the exponential decay $\sim e^{-\ktwo L}$ correctly, however, it
			underestimates the exact line interaction by a factor of 2.
			}
	\label{fig:EVS_S_LI}
\end{figure}

The decay at large separations is proportional to $\sim e^{-\ktwo L}$ since $\ktwo < \kone$ (see Tab.~\ref{tab:parameter}) 
and shown by the two parallel lines in the semi-logarithmic plot in Fig.~\ref{fig:EVS_S_LI}b.
Although the superposition approximation predicts the overall exponential decay at large separations correctly, it is too small compared to 
the exact expression for the line interaction energy destiny by a factor of 2 at large separations, as 
seen in  Fig.~\ref{fig:EVS_S_LI}.

\newpage

\section{Variation of parameters}
\label{sec:VAR}

This section compares the exact expressions for the electrostatic potential, the surface interaction energy densities and the line 
interaction energy density for different variations of system parameters specified in Tab.~\ref{tab:parameter}.

\subsection{Inverse Debye length in medium 1}

Fig.~\ref{fig:var_k1} shows the variation of the inverse Debye length $\kone$ in medium 1. 
To compare the variations of $\kone$, the separation length is scaled with $\ktwo L$ in this specific case.
Additionally, the line interaction energy density $\omega_\tau^e(L)$ is expressed in units of $\ktwo/\beta$ and the surface line 
interaction energy densities $\omega_{\gamma,i}^e(L)$ in medium $i \in \qty{1,2}$ in units of $\ktwo^2/\beta$.
Other parameters stay in the standard configuration specified in Tab.~\ref{tab:parameter}.

\subsubsection{Electrostatic potential}

Fig.~\ref{fig:var_k1}a shows the electrostatic potential $\Psi^e(x=0,z)$ for varying $\kone$ and the separation length $\ktwo L= 10$.
Clearly, increasing $\kone$ leads to stronger screening and faster decay of the electrostatic potential.

\subsubsection{Line interaction energy density $\omega_\tau(L)$}

The line interaction energy density is shown in Fig.~\ref{fig:var_k1}b for two different $\kone$ values. 
In the limit of vanishing separation, increasing $\kone$ results in an increase of $\omega_\tau(L)$ itself. 
As seen in the inset of Fig.~\ref{fig:var_k1}b, a larger inverse Debye length $\kone$ results in a faster decrease and deeper minimum, 
which is additionally shifted to smaller separation $L$.
However, the asymptotic decay is given by $\sim e^{-\ktwo L}$ and therefore not affected by variation in $\kone$.

\subsubsection{Surface interaction energy densities $\omega_{\gamma,i}(L)$}

\begin{figure}[htbp]
	\centering
	\includegraphics[width=\linewidth]{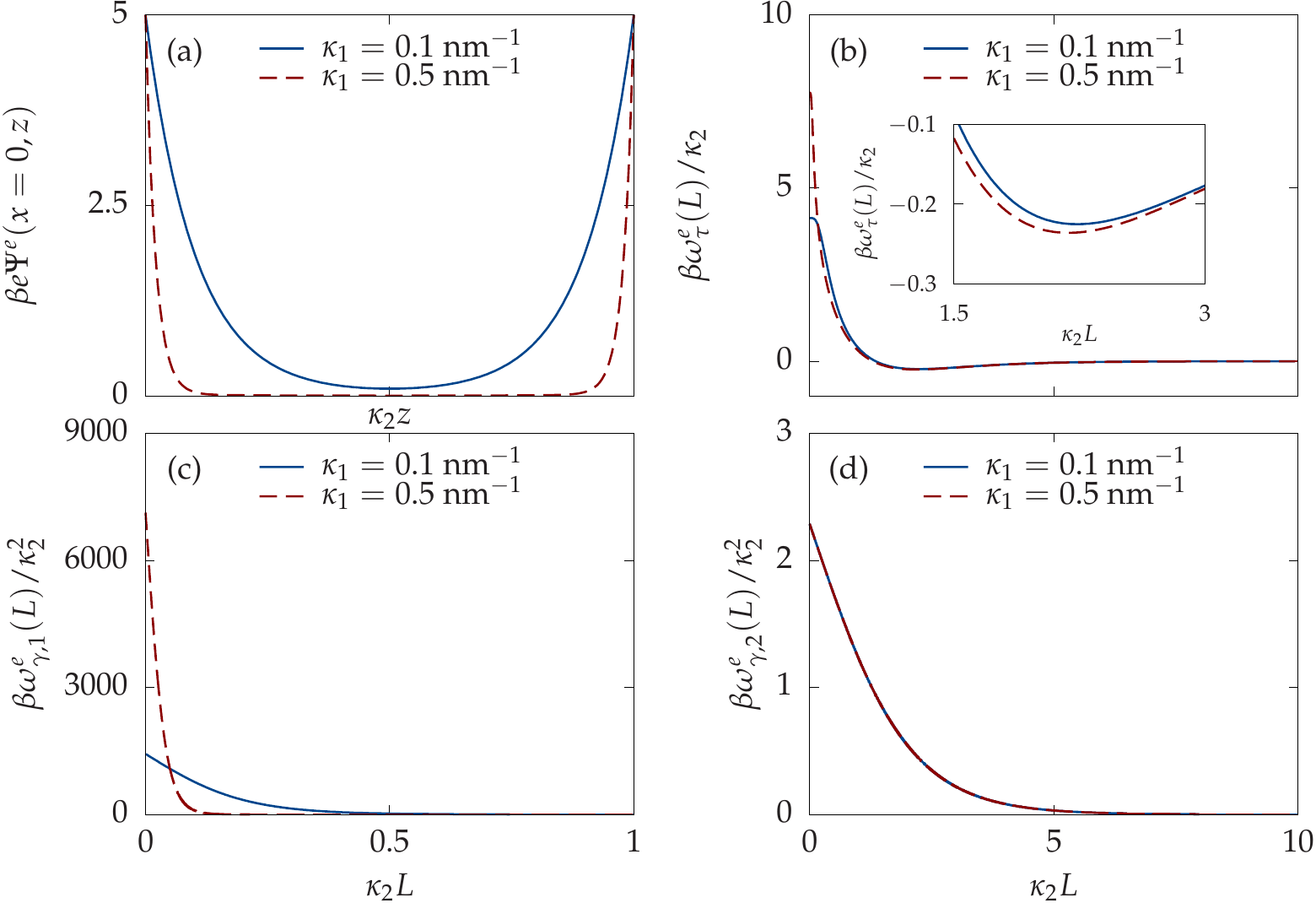}
	\caption
			[
			Exact electrostatic potential $\Psi^e(x,z)$, surface interaction energy densities $\omega^e_{\gamma,i}(L)$ 
			in medium $i\in\qty{1,2}$ and line interaction energy density $\omega^e_\tau(L)$ 
			for variation in $\kone = 0.5 \ \textrm{nm}^{-1}$
			]
			{
			Variation of inverse Debye length $\kone$ in medium 1, other parameters are in standard configuration specified
			in Tab.~\ref{tab:parameter}. 
			(a) Exact electrostatic potential $\Psi^e(x,z)$ expressed in units of $1/\beta e $, 
			as function of $\ktwo z$. The separation length between the walls is given by $\ktwo L =10$. 
			As seen in the plot, increasing $\kone$ results in a stronger screening of the electrostatic potential. 
			(b) Exact line interaction energy density $\omega_\tau^e(L)$ expressed in units of $\ktwo/\beta$, as function of separation 
			length $\ktwo L$. Increase in $\kone$ leads to a larger line interaction in the limit of vanishing separations, 
			increases the magnitude of the minimum and shifts it to smaller separations. 
			(c) Exact surface interaction energy density $\omega_{\gamma,1}^e(L)$ in medium 1 expressed in units of $\ktwo^2/\beta$, 
			as function of separation length $\ktwo L$. 
			Increasing $\kone$ results in larger surface interaction for vanishing separations, but also a faster decay. 
			(d) Exact surface interaction energy density $\omega_{\gamma,2}^e(L)$ in medium 2 expressed in units of $\ktwo^2/\beta$, 
			as function of separation length $\ktwo L$. Since $\kone$ is a property of medium 1 $\omega_{\gamma,2}^e(L)$ stays unchanged.
			}
	\label{fig:var_k1}
\end{figure}

Fig.~\ref{fig:var_k1}c shows the surface interaction energy density $\omega_{\gamma,1}^e(L)$ in medium 1 and 
Fig.~\ref{fig:var_k1}d shows $\omega_{\gamma,2}^e(L)$ in medium 2.
Here, $\omega_{\gamma,2}^e(L)$ is independent of the inverse Debye length $\kone$ in medium 1 and therefore remains unaffected.
Increasing $\kone$ results in larger surface interaction energy density $\omega_{\gamma,1}^e(L)$ in the limit
of small inter-particle separation $L\to\infty$, as seen in Fig.~\ref{fig:var_k1}c and predicted by Eq.~\eqref{eq_SI_2}.

\newpage
\null
\vfill

\vfill
\newpage

\subsection{Inverse Debye length in medium 2}

Fig.~\ref{fig:var_k2} shows the variation of inverse Debye length $\ktwo$ in medium 2. 
Other parameters stay in standard configuration specified in Tab.~\ref{tab:parameter}.

\subsubsection{Electrostatic potential $\Psi(x,z)$}

The electrostatic potential is shown for variable $\ktwo$ at separation length $\kone L =10$ in Fig.~\ref{fig:var_k2}a.
Clearly, variation in $\ktwo$ has no significant effect on the electrostatic potential as seen in the inset.

\subsubsection{Line interaction energy density $\omega_\tau(L)$}

The line interaction energy density for variable $\ktwo$ is shown in Fig.~\ref{fig:var_k1}b. 
In the limit of vanishing separation, an increase in $\ktwo$ results in a decrease of $\omega_\tau^e(L)$.
As seen in the inset of Fig.~\ref{fig:var_k2}(b) the minimum of the line interaction energy $\omega_\tau^e(L)$ 
shifts to larger separation for smaller inverse Debye length $\ktwo$.
Additionally, the asymptotic decay is given by $\sim e^{-\ktwo L}$ and
therefore, the line interaction energy density decays slower for smaller $\ktwo$.

\subsubsection{Surface interaction energy densities $\omega_{\gamma,i}(L)$}

\begin{figure}[htbp]
	\centering
	\includegraphics[width=\linewidth]{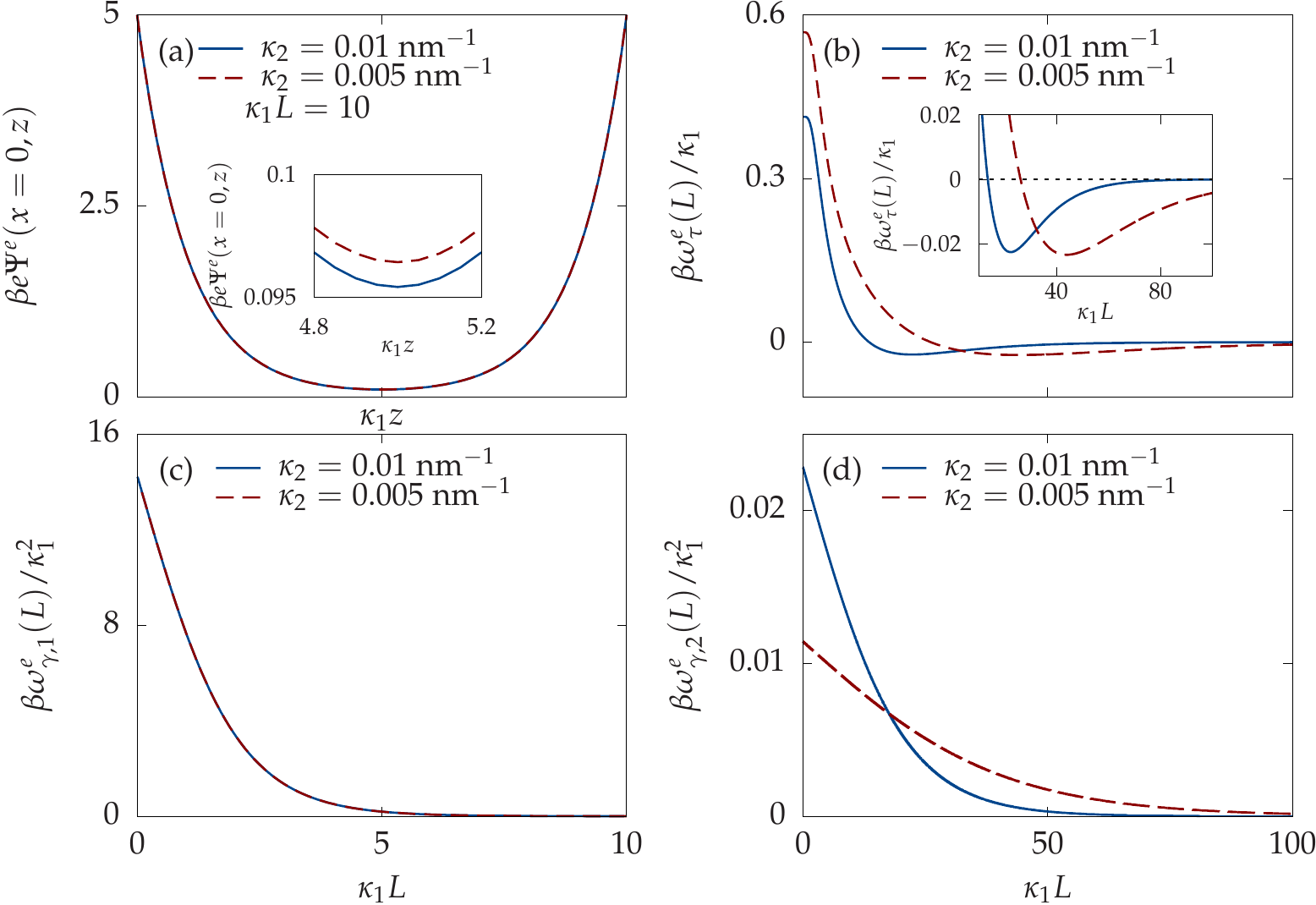}
	\caption
			[
			Exact electrostatic potential $\Psi^e(x,z)$, surface interaction energy densities $\omega^e_{\gamma,i}(L)$ 
			in medium $i\in\qty{1,2}$ and line interaction energy density $\omega^e_\tau(L)$ for 
			variation in $\ktwo= 0.005 \ \textrm{nm}^{-1}$
			]
			{
			Variation of inverse Debye length $\ktwo$ in medium 2, other parameters are in standard configuration specified in 
			Tab.~\ref{tab:parameter}. 
			(a) Exact electrostatic potential $\Psi^e(x,z)$ expressed in units of $1/\beta e $, as function of $\kone z$. 
			The separation length between the walls is given by $\kone L =10$. 
			As seen in the plot, variation in $\ktwo$ has no significant impact. 
			(b) Exact line interaction energy density $\omega_\tau^e(L)$ expressed in units of $\kone/\beta$, as function of 
			separation length $\kone L$. 
			Decreasing $\kone$ leads to a larger line interaction in the limit of vanishing separations, increases the magnitude 
			of the minimum and shifts it to larger separations, as seen in the inset. 
			(c) Exact surface interaction energy density $\omega_{\gamma,1}^e(L)$ in medium 1 expressed in units of $\kone^2/\beta$, 
			as function of separation length $\kone L$. 
			Since $\ktwo$ is a property of medium 2, $\omega_{\gamma,1}^e(L)$ stays unchanged. 
			(d) Exact surface interaction energy density $\omega_{\gamma,2}^e(L)$ in medium 2 expressed in 
			units of $\kone^2/\beta$, as function of separation length $\kone L$. 
			A decrease of $\ktwo$ results in smaller surface interaction for vanishing separations and slower decay also.
			}
	\label{fig:var_k2}
\end{figure}

Fig.~\ref{fig:var_k2}c shows the surface interaction energy density $\omega_{\gamma,1}^e(L)$ in medium 1, 
and Fig.~\ref{fig:var_k2}d shows $\omega_{\gamma,2}^e(L)$ in medium 2.
$\omega_{\gamma,1}^e(L)$ is independent of the inverse Debye length $\ktwo$  
and therefore remains unaffected by variation of $\ktwo$, as seen in Fig.~\ref{fig:var_k2}c.
Eq.~\eqref{eq_SI_2} predicts the increase of the surface interaction energy density  $\omega_{\gamma,2}^e(L)$ for 
larger $\ktwo$ in the limit of small separations, as seen in Fig.~\ref{fig:var_k1}c. 
Additionally, variation in $\ktwo$ affects the asymptotic decay since $\omega_{\gamma,2}^e(L)\sim e^{-\ktwo L}$.

\newpage
\null
\vfill

\vfill
\newpage

\subsection{Relative permittivity}

Fig.~\ref{fig:var_e2} shows the variation of relative permittivity $\epsilon_{r,2}$ for medium 2. 
Other parameters stay in standard configuration specified in Tab.~\ref{tab:parameter}.

\subsubsection{Electrostatic potential $\Psi(x,z)$}

Increasing the relative permittivity $\epsilon_{r,2}$ results in an increased electrostatic potential as seen in Fig.~\ref{fig:var_e2}a
for separation length $\kone L =10$.

\subsubsection{Line interaction energy density $\omega_\tau(L)$}

The line interaction energy density for different values of $\epsilon_{r,2}$ is shown in Fig.~\ref{fig:var_e2}b.
Increasing $\epsilon_{r,2}$ results in an increase of the interaction 
in the limit of vanishing separations $L\to0$.
Additionally, the minimum of the line interaction energy density shifts to larger separations 
and increases in magnitude as seen in the inset in Fig.~\ref{fig:var_e2}b.
The overall asymptotic behavior with decay $\sim e^{-\kappa_i L}$ stays unchanged.

\subsubsection{Surface interaction energy densities $\omega_{\gamma,i}(L)$}

\begin{figure}[htbp]
	\centering
	\includegraphics[width=\linewidth]{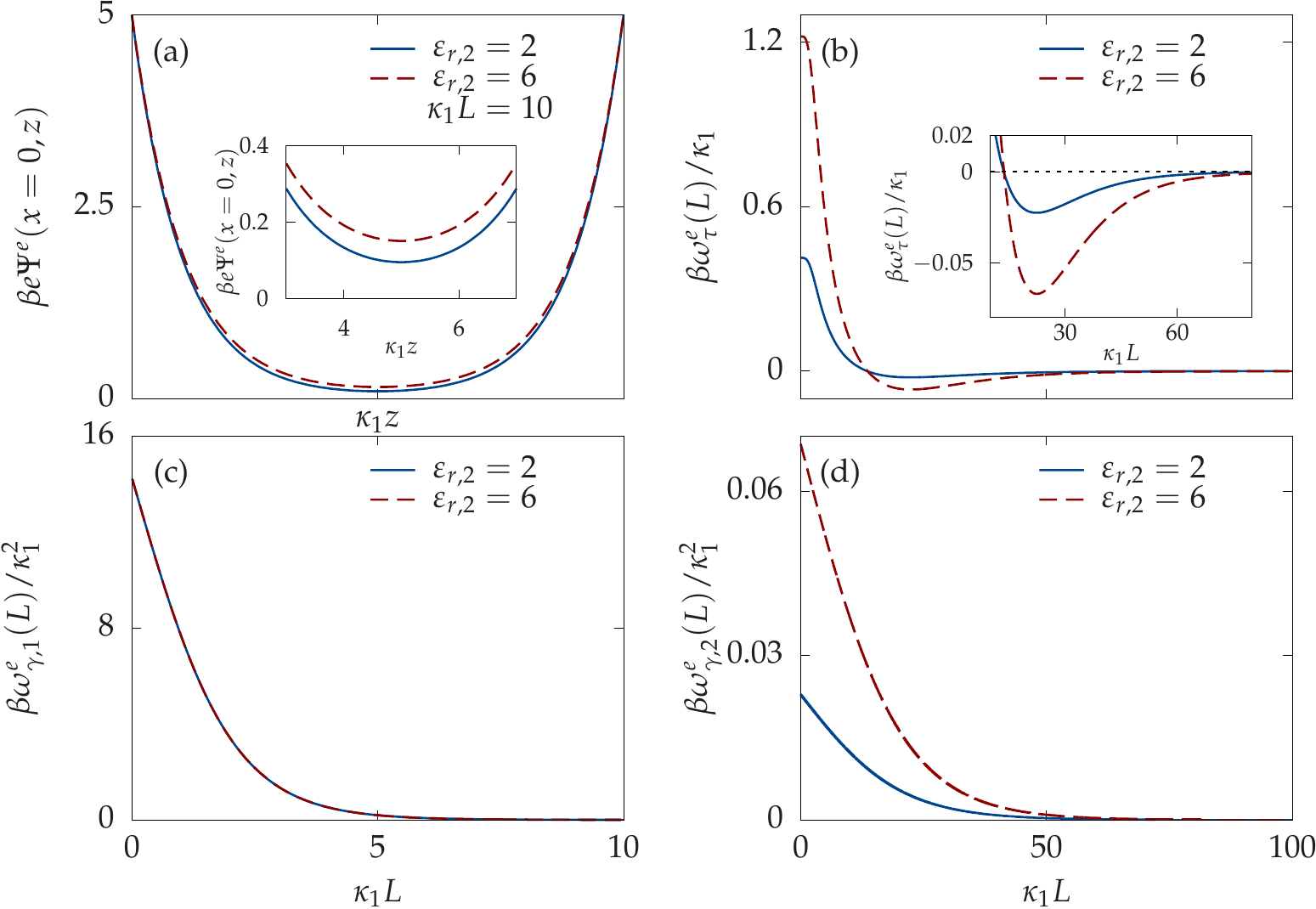}
	\caption
			[
			Exact electrostatic potential $\Psi^e(x,z)$, surface interaction energy densities $\omega^e_{\gamma,i}(L)$ 
			in medium $i\in\qty{1,2}$ and line interaction energy density $\omega^e_\tau(L)$ for variation 
			in $\epsilon_{r,2}=6$
			]
			{
			Variation of the relative permittivity $\epsilon_{r,2}$ in medium 2, other parameters are in standard 
			configuration specified in Tab.~\ref{tab:parameter}. 
			(a) Exact electrostatic potential $\Psi^e(x,z)$ expressed in units of $1/\beta e $, as function of $\kone z$.
			The separation length between the walls is given by $\kone L =10$. 
			As seen in the plot, increasing $\etwo$ also slightly increases the electrostatic potential. 
			(b) Exact line interaction energy density $\omega_\tau^e(L)$ expressed in units of $\kone/\beta$, as 
			function of separation length $\kone L$. 
			Increase in $\etwo$ leads to a larger line interaction in the limit 
			of vanishing separations, increases the magnitude of the minimum and shifts it to larger separations, 
			as seen in the inset. 
			(c) Exact surface interaction energy density $\omega_{\gamma,1}^e(L)$ in medium 1 expressed in 
			units of $\kone^2/\beta$, as function of separation length $\kone L$. 
			Since $\etwo$ is a property of medium 2 , $\omega_{\gamma,1}^e(L)$ stays unchanged. 
			(d) Exact surface interaction energy density $\omega_{\gamma,2}^e(L)$ in medium 2 expressed in 
			units of $\kone^2/\beta$, as function of separation length $\kone L$. 
			Increasing $\etwo$ results in larger surface interaction for vanishing separations.
			}
	\label{fig:var_e2}
\end{figure}

The variation for the surface interaction energy density in medium 1 is illustrated in 
Fig.~\ref{fig:var_e2}c and for medium 2 in Fig.~\ref{fig:var_e2}d. 
Since $\epsilon_{r,2}$ is a property of medium 2, $\omega_{\gamma,1}^e(L)$ stays unchanged.
However, $\omega_{\gamma,2}^e(L)$ is proportional to $\epsilon_{r,2}$ in the limit of 
small separations as given in Eq.~\eqref{eq_SI_2} and, therefore, increases for larger $\epsilon_{r,2}$.
The overall monotonic decay, however, stays unchanged.

\newpage
\null
\vfill

\vfill
\newpage

\subsection{Donnan-Potential}

The variation of the Donnan potential $\PsiD$ is shown in Fig.~\ref{fig:var_psiD}. 
Other parameters stay in the standard configuration specified in Tab.~\ref{tab:parameter}.

\subsubsection{Electrostatic potential $\Psi(x,z)$}
		
The electrostatic potential, as shown in Fig.~\ref{fig:var_psiD}a, is not significantly 
affected by the variation of $\PsiD$ for a separation length of $\kone L =10$.

\subsubsection{Line interaction energy density $\omega_\tau(L)$}

Fig.~\ref{fig:var_psiD}b shows the line interaction energy density for variable $\PsiD$.
In the cases of $\beta e \PsiD = 5$ and $\beta e \PsiD = 3 $, the line interaction energy 
density decays monotonically. 
However, for $\beta e \PsiD = 1$ and $\beta e \PsiD = -1$ the decay becomes non-monotonic.
Here, the line interaction forms a minimum which increases and shifts to smaller separation for smaller $\PsiD$.

\subsubsection{Surface interaction energy densities $\omega_{\gamma,i}(L)$}

\begin{figure}[htbp]
	\centering
	\includegraphics[width=\linewidth]{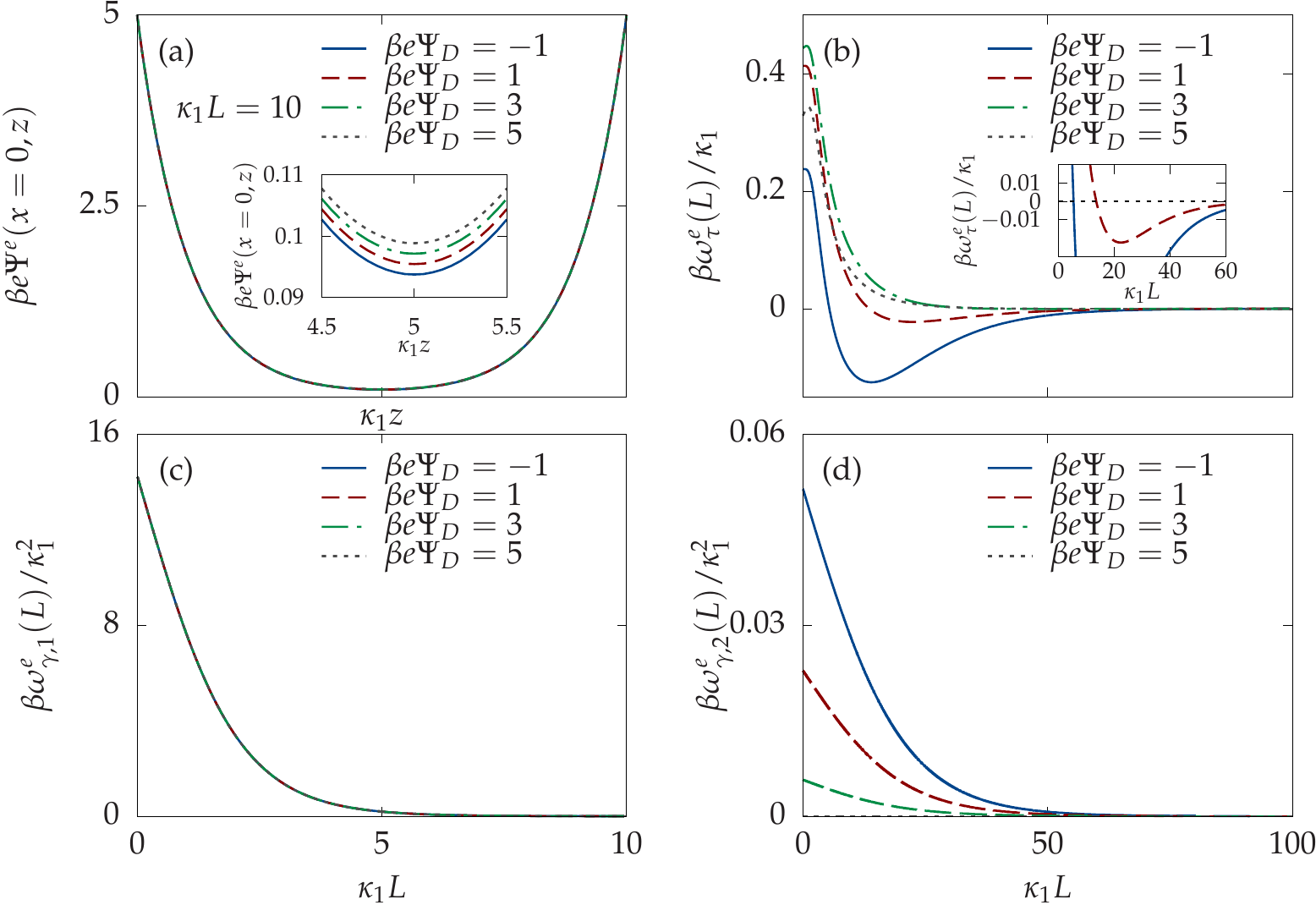}
	\caption
			[
			Exact electrostatic potential $\Psi^e(x,z)$, surface interaction energy densities 
			$\omega^e_{\gamma,i}(L)$ in medium $i\in\qty{1,2}$ and line interaction energy density 
			$\omega^e_\tau(L)$ for variation in $\beta e \PsiD$
			]{
			Variation of the Donnan potential $\PsiD$ in medium 2, other parameters are in standard 
			configuration specified in Tab.~\ref{tab:parameter}. 
			(a) Exact electrostatic potential $\Psi^e(x,z)$ expressed in units of $1/\beta e $, as 
			function of $\kone z$. The separation length between the walls is given by $\kone L =10$. 
			As seen in the inset, variation of $\PsiD$ has no significant impact. 
			Although, in general the electrostatic potential slightly increases for larger values of $\PsiD$. 
			(b) Exact line interaction energy density $\omega_\tau^e(L)$ expressed in units of $\kone/\beta$, 
			as function of separation length $\kone L$. 
			In the cases $\beta e \PsiD = -1$ and 
			$\beta e \PsiD = 1$ the line interaction energy density decays non-monotonically 
			(deeper minimum for $\beta e \PsiD=-1$).  
			However, for values of $\beta e \PsiD=3$ and $\beta e \PsiD = 5$ the decay is monotonic. 
			(c) Exact surface interaction energy density $\omega_{\gamma,1}^e(L)$ in medium 1 expressed 
			in units of $\kone^2/\beta$, as function of separation length $\kone L$. 
			Since $\PsiD$ is a property of medium 2, $\omega_{\gamma,1}^e(L)$ stays unchanged. 
			(d) Exact surface interaction energy density $\omega_{\gamma,2}^e(L)$ in medium 2 expressed 
			in units of $\kone^2/\beta$, as function of separation length $\kone L$. 
			Decreasing $\PsiD$ results in a larger surface interaction energy density, 
			especially in the limit of small separation.
			In the special case of $\beta e \PsiD = 5$ the surface interaction energy density vanishes.
			}
	\label{fig:var_psiD}
\end{figure}

Figs.~\ref{fig:var_psiD}c and~\ref{fig:var_psiD}d show the surface interaction energy 
density acting in medium $i \in \qty{1,2}$. 
Since $\omega_{\gamma,1}^e(L)$, as calculated in Eq.~\eqref{eq:SI_1}, is independent in $\PsiD$, 
a variation in such has no effect.
However, $\omega_{\gamma,2}^e(L)$ is proportional to the factor $(\PsiP - \PsiD)^2$ and thus changes 
its behavior.
As seen in Fig.~\ref{fig:var_psiD}d and as Eq.~\eqref{eq_SI_2} predicts, an increase in the 
absolute value $\abs{\PsiP-\PsiD}$ leads to an increase in the surface interaction energy density 
and a decrease in $\abs{\PsiP-\PsiD}$ leads to a decrease in the surface interaction energy density respectively.
In the special case of $\PsiD = 5 $ the above-mentioned absolute value vanishes and $\omega_{\gamma,2}^e(L)$ therefore too.

\newpage
\null
\vfill

\vfill
\newpage

\subsection{Surface potential}

Fig.~\ref{fig:var_psiP} shows the variation of the constant surface potential $\PsiP$, the other parameters are 
in standard configuration specified in Tab.~\ref{tab:parameter}.

\subsubsection{Electrostatic potential $\Psi(x,z)$}

Fig.~\ref{fig:var_psiP}a shows the electrostatic potential with separation length $\kone L = 10$ 
for variable surface potentials $\PsiP$.
As expected, the corresponding values of the electrostatic potentials at the walls $\kone z=0$ and $\kone z = 10$ 
are given by each value of $\PsiP$ respectively.

\subsubsection{Line interaction energy density $\omega_\tau(L)$}

\begin{figure}[htbp]
	\centering
	\includegraphics[width=\linewidth]{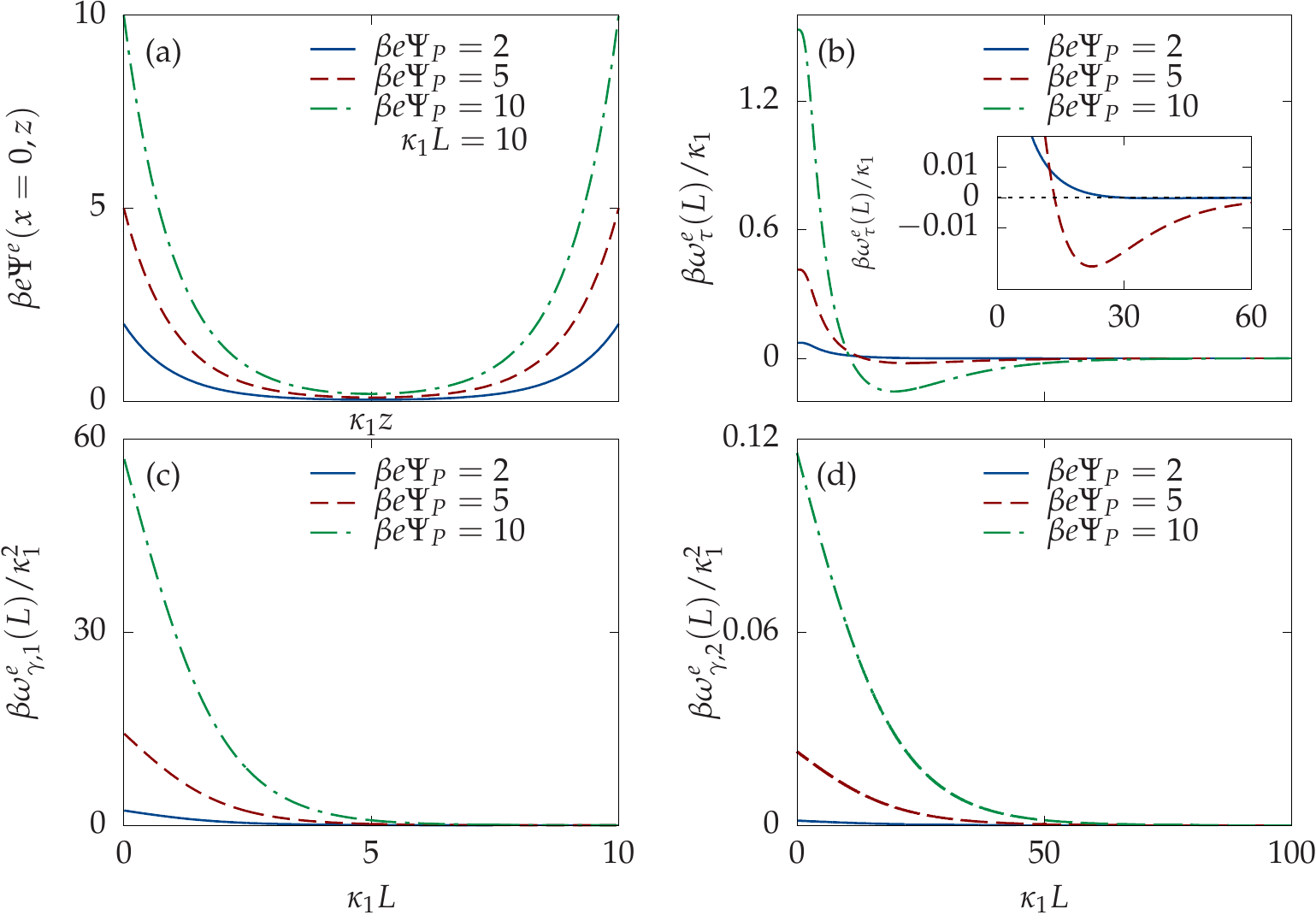}
	\caption
			[
			Exact electrostatic potential $\Psi^e(x,z)$, surface interaction energy densities $\omega^e_{\gamma,i}(L)$ 
			in medium $i\in\qty{1,2}$ and line interaction energy density $\omega^e_\tau(L)$ for variation 
			in $\beta e \PsiP$
			]
			{
			Variation of the surface potential $\PsiP$, other parameters are in standard configuration 
			specified in Tab.~\ref{tab:parameter}. 
			(a) Exact electrostatic potential $\Psi^e(x,z)$ expressed in units of $1/\beta e $, as function 
			of $\kone z$. 
			The separation length between the walls is given by $\kone L =10$. 
			As seen in the plot, variation in $\PsiP$ leads to the corresponding surface potentials at the boundaries. 
			(b) Exact line interaction energy density $\omega_\tau^e(L)$ expressed in units of $\kone/\beta$, 
			as function of separation length $\kone L$. 
			Increasing $\PsiP$ leads to a larger line interaction in the limit of vanishing separations, increases the 
			magnitude of the minimum and shifts it to larger separations. 
			In the case of $\beta e \PsiP = 2$ no minimum is present, as seen in the inset. 
			(c) Exact surface interaction energy density $\omega_{\gamma,1}^e(L)$ in medium 1 expressed in units 
			of $\kone^2/\beta$, as function of separation length $\kone L$. 
			Increasing $\PsiP$ results in larger surface interaction for vanishing separations. 
			(d) Exact surface interaction energy density $\omega_{\gamma,2}^e(L)$ in medium 2 expressed in units 
			of $\kone^2/\beta$, as function of separation length $\kone L$. 
			Increasing $\PsiP$ results in larger surface interaction for vanishing separations.
			}
	\label{fig:var_psiP}
\end{figure}

Fig.~\ref{fig:var_psiP}b shows the line interaction energy density for variable $\PsiP$.
Increasing $\PsiP$ results in larger line interaction energy densities for vanishing separations. 
For a smaller surface potential value, in this case $\beta e \PsiP = 2$, the line interaction energy density 
$\omega_\tau^e(L)$ decays monotonically, implying a repulsive behavior of the line interaction. 
However, as the surface potential increases, $\omega_\tau^e(L)$ 
becomes non-monotonic with a minimum taking place at roughly $\kone L \approx 20$.
By increasing $\beta e \PsiP$ this minimum becomes deeper.

\subsubsection{Surface interaction energy densities $\omega_{\gamma,i}(L)$}

Both surface interaction energy densities in media $i \in \qty{1,2}$ behave
similarly in variation of the surface potential $\PsiP$.
In the limit of vanishing separations, as given by Eqs.~\eqref{eq:SI_1} and~\eqref{eq_SI_2}, 
$\omega_{\gamma,i}^e(L)$ increases upon increasing $\PsiP$.
However, the asymptotic decays says unchanged.

\newpage
\null
\vfill

\vfill
\newpage

\section{Comparision of the variation for system parameters}
\label{sec:VAR_EvS}

\begin{figure}[htbp]
	\centering
	\includegraphics[width=\linewidth]{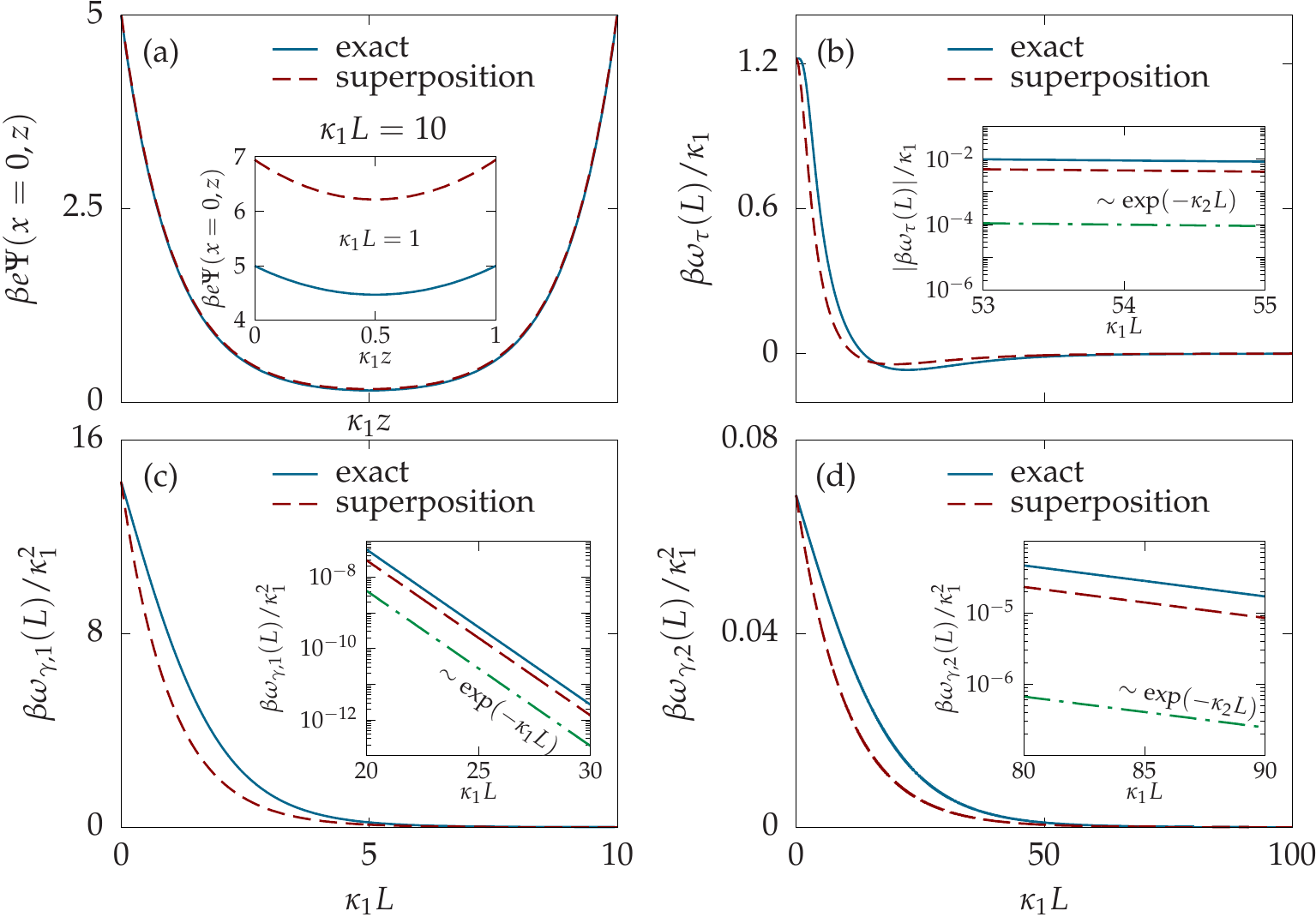}
	\caption
			[
			Comparision between the electrostatic potential, the surface interaction energy densities and 
			line interaction energy density between exact and superposition approximation for $\epsilon_{r,2}=6$
			]
			{
			Variation of relative permittivity $\epsilon_{r,2} = 6$, other parameters are given by their standard configuration 
			specified in Tab.~\ref{tab:parameter}.  
			(a) Electrostatic potential $\Psi(x,z)$ within both exact calculation and superposition approximation 
			expressed in the units of $1/\beta e$, as functions of $\kone z$.
			The separation length between the two walls is given by $\kone L = 10$ and $\kone L = 1$ in the inset. 
			The superposition approximation fails to predict the electrostatic potential correctly for small separations. 
			Upon increasing the separation, as expected, the differences between exact and superposition calculations decrease. 
			(b) Line interaction energy density $\omega_\tau(L)$ within the exact and superposition calculations expressed 
			in the units of $\kone/\beta$, as functions of the separation $\kone L$.
			As the plot shows, the superposition approximation fails to capture the correct behavior, 
			although both calculations show non-monotonic decay. 
			However, in the limit of vanishing separations $\omega_\tau^e(L)$ and $\omega_\tau^s(L)$ reach the same finite value. 
			Additionally, the superposition expression predicts the exponential decay for larger separation correctly, 
			but underestimates the line interaction always by a factor of 2, as seen by the offset in the inset. 
			(c)(d) Surface interaction energy density $\omega_{\gamma,i}(L)$ in medium $i \in \qty{1,2}$ within 
			both exact and superposition calculations expressed in units of $\kone^2/\beta$, as functions of the 
			separation $\kone L$.
			As the plots show, the superposition approximation fails to predict the correct behavior properly, although 
			both expressions show monotonic decay. In the limit of vanishing separations the exact calculation and 
			superposition approximation of the surface interaction reach the same finite value. 
			Additionally, the superposition approximation predicts the exponential decay for larger separation correctly, 
			but underestimates the surface interaction by a factor of 2 as seen in the offset 
			between exact and superposition calculations in the inset.
			}
	\label{fig:ES_V_E2_6}
\end{figure}

\begin{figure}[htbp]
	\centering
	\includegraphics[width=\linewidth]{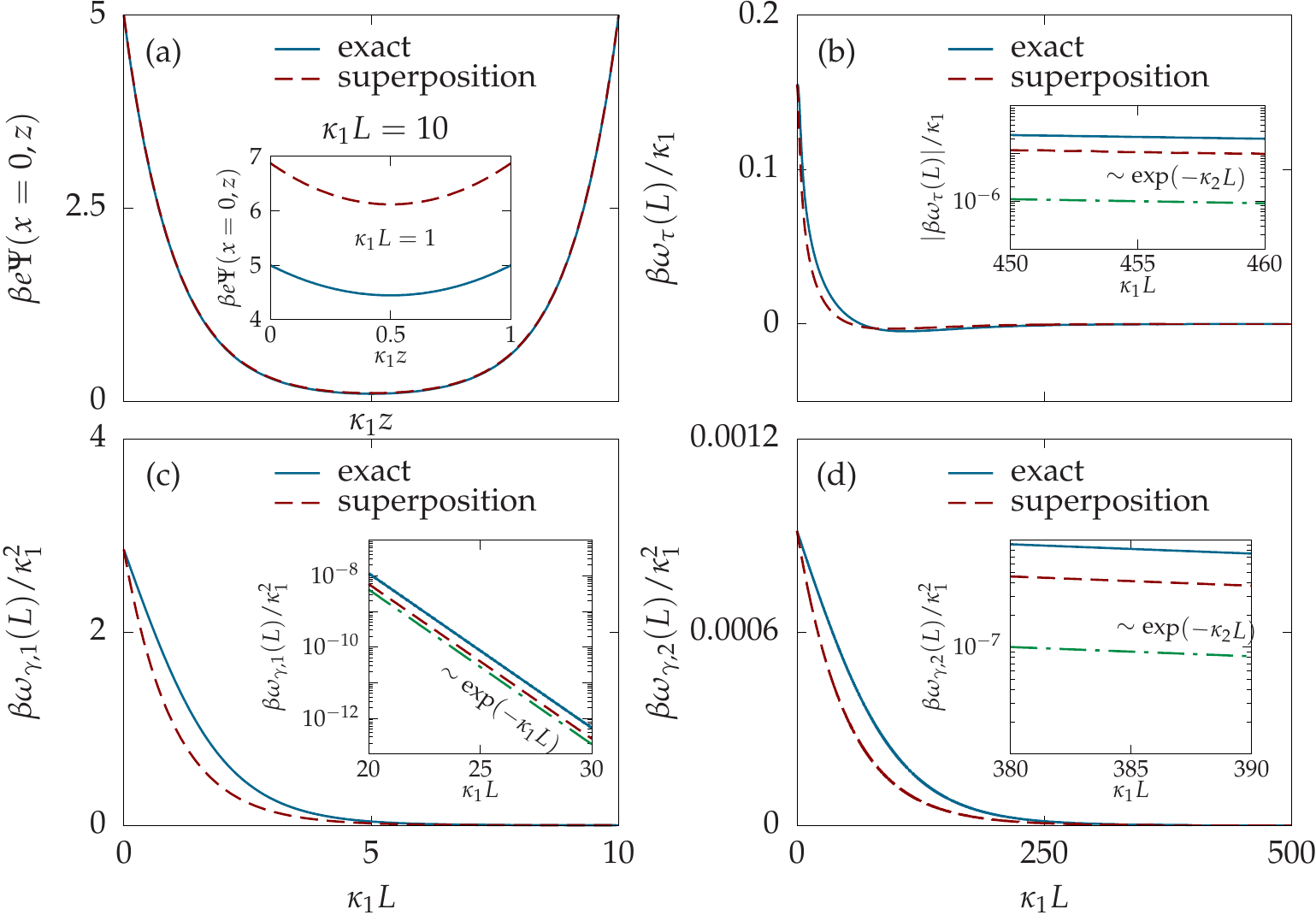}
	\caption
			[
			Comparison between the electrostatic potential, the surface interaction energy densities 
			and line interaction energy density between exact and superposition 
			approximation for $\kone = 0.5 \ \textrm{nm}^{-1}$
			]
			{
			Variation of inverse Debye length $\kone = 0.5 \ \textrm{nm}^{-1}$ in medium 1, other parameters are given by 
			their standard configuration specified in Tab.~\ref{tab:parameter}.  
			(a) Electrostatic potential $\Psi(x,z)$ within the exact and superposition calculations expressed 
			in the units of $1/\beta e$, as functions of $\kone z$.
			The separation length between the two walls is given by $\kone L = 10$ and $\kone L = 1$ 
			in the inset.
			The superposition approximation fails to predict the electrostatic potential correctly for small separations. 
			Upon increasing the separation, as expected, the differences between exact and superposition calculations decrease. 
			(b) Line interaction energy density $\omega_\tau(L)$ within the exact and superposition calculations 
			expressed in the units of $\kone/\beta$, as functions of the separation $\kone L$.
			As the plot shows, the superposition approximation fails to capture the correct behavior, but both 
			expressions decay non-monotonic. 
			In the limit of vanishing separations $\omega_\tau^e(L)$ and $\omega_\tau^s(L)$ reach the same finite value. 
			Additionally, the superposition expression predicts the exponential decay for 
			larger separation correctly, but underestimates the line interaction always by a factor 
			of 2 as seen by the offset in the inset. 
			(c)(d) Surface interaction energy density $\omega_{\gamma,i}(L)$ in medium $i \in \qty{1,2}$ within both 
			the exact and superposition calculations expressed in the units of $\kone^2/\beta$, as functions of the 
			separation $\kone L$.
			As the plots show, the superposition approximation fails to predict the correct behavior properly,
			but both expressions decay monotonic. 
			In the limit of vanishing separations, the exact calculation and superposition approximation predict the same 
			surface interaction.
			Although the superposition approximation predicts the exponential decay for larger correctly, 
			it underestimates the surface interaction by a factor of 2 as seen in the offset between
			exact and superposition calculations in the inset.
			}
	\label{fig:ES_V_K1_05}
\end{figure}

\begin{figure}[htbp]
	\centering
	\includegraphics[width=\linewidth]{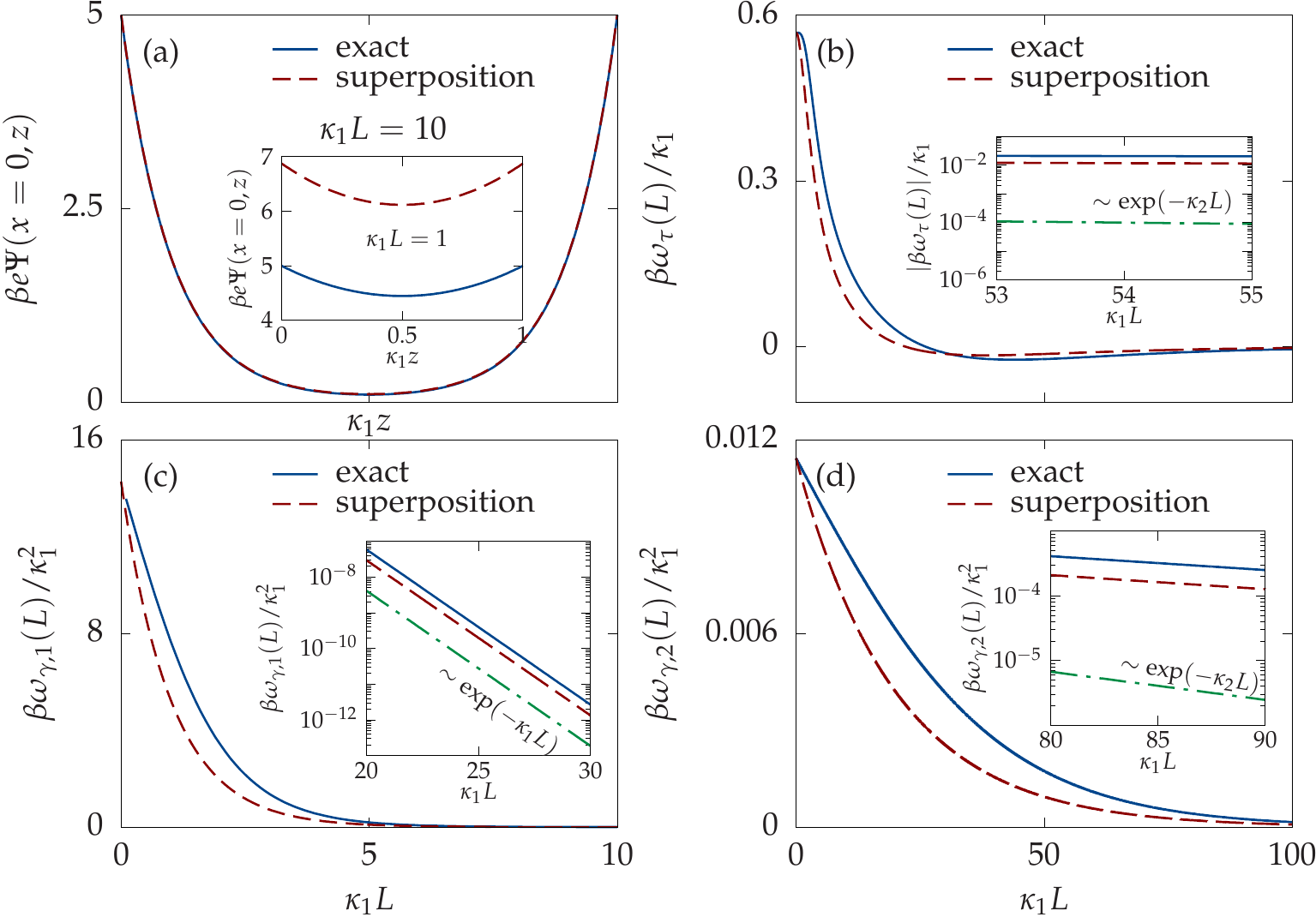}
	\caption
			[
			Comparision between the electrostatic potential, the surface interaction energy densities and 
			line interaction energy density between exact and superposition approximation 
			for $\ktwo = 0.005 \ \textrm{nm}^{-1}$
			]
			{
			Variation of inverse Debye length $\ktwo = 0.005\ \textrm{nm}^{-1}$ in medium 2, 
			other parameters are given by their standard configuration specified in Tab.~\ref{tab:parameter}.   
			(a) Electrostatic potential $\Psi(x,z)$ within both exact and superposition calculations 
			expressed in the units of $1/\beta e$, as functions of $\kone z$.
			The separation length between the two walls is given by $\kone L = 10$ and $\kone L = 1$ in the inset. 
			Clearly, the superposition approximation fails to predict the potential correctly for small separations. 
			Upon increasing the separation, as expected, the differences between exact and superposition calculations decrease. 
			(b) Line interaction energy density $\omega_\tau(L)$ within the exact and superposition calculations 
			expressed in the units of $\kone/\beta$, as functions of the separation $\kone L$.
			As the plot shows, the superposition approximation fails to capture the correct behavior but predicts the 
			non-monotonic behavior correctly. 
			In the limit of vanishing separations $\omega_\tau^e(L)$ and $\omega_\tau^s(L)$ reach the same finite value. 
			Additionally, the superposition expression predicts the exponential decay for larger separation correctly, 
			but underestimates the line interaction always by a factor of 2 as seen by the offset in the inset. 
			(c)(d) Surface interaction energy density $\omega_{\gamma,i}(L)$ in medium $i \in \qty{1,2}$ within the 
			exact and superposition calculations expressed in the units of $\kone^2/\beta$, as functions of the separation 
			$\kone L$. 
			As the plots show, the superposition approximation fails to predict the correct behavior properly
			but captures the monotonic decay. 
			In the limit of vanishing separations the exact calculation and superposition approximation 
			predict the same surface interaction. 
			Although the superposition approximation predicts the exponential decay for large separation correctly,
			it underestimates the surface interaction by a factor of 2 as seen in the offset 
			between exact and superposition calculations in the inset.
			}
	\label{fig:ES_V_K2_0005}
\end{figure}

\begin{figure}[htbp]
	\centering
	\includegraphics[width=\linewidth]{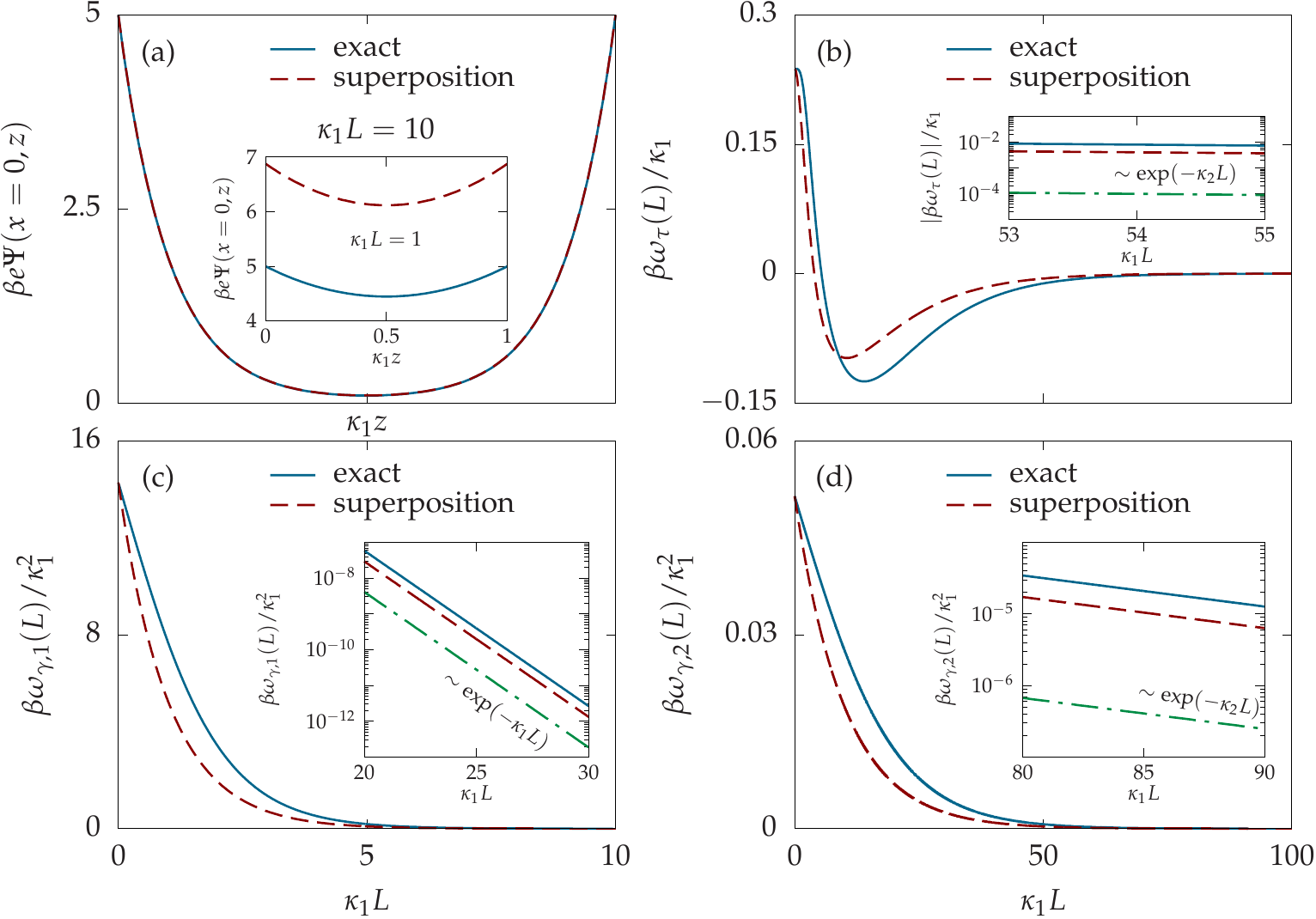}
	\caption
			[
			Comparison between the electrostatic potential, the surface interaction energy densities and 
			line interaction energy density between exact and superposition approximation for $\beta  e \PsiD = -1$
			]
			{
			Variation of the Donnan potential $\beta e \PsiD = -1$ in medium 2,
			other parameters are given by their standard configuration specified in Tab.~\ref{tab:parameter}.   
			(a) Electrostatic potential $\Psi(x,z)$ within both exact and superposition calculations expressed 
			in the units of $1/\beta e$, as functions of $\kone z$.
			The separation length between the two walls is given by $\kone L = 10$ and $\kone L = 1$ in the inset. 
			Clearly, the superposition fails to predict the electrostatic potential correctly for small separations. 
			Upon increasing the separation, as expected, the differences between exact and superposition calculations decrease. 
			(b) Line interaction energy density $\omega_\tau(L)$ within the exact and superposition calculations 
			expressed in the units of $\kone/\beta$, as functions of the separation $\kone L$.
			As the plot shows, the superposition approximation fails to capture the correct behavior but 
			captures the non-monotonic behavior. 
			In the limit of vanishing separations $\omega_\tau^e(L)$ and $\omega_\tau^s(L)$ reach the same finite value. 
			Additionally, the superposition expression predicts the exponential decay for large separation correctly, 
			but underestimates the line interaction always by a factor of 2 as seen by the offset in the inset. 
			(c)(d) Surface interaction energy density $\omega_{\gamma,i}(L)$ in medium $i \in \qty{1,2}$ within 
			the exact and superposition calculations expressed in the units of $\kone^2/\beta$, as functions of the 
			separation $\kone L$. 
			As the plots show, the superposition approximation fails to predict the correct behavior properly, 
			but captures the monotonic decay correctly. 
			In the limit of vanishing separations both the exact calculation and superposition approximation 
			predict the same surface interaction.
			Although the superposition approximation predicts the exponential decay for 
			large separation correctly, it underestimates the surface interaction by a factor of 2 as seen in 
			the offset between exact and superposition calculations in the inset.
			}
	\label{fig:ES_V_PSID-1}
\end{figure}

\begin{figure}[htbp]
	\centering
	\includegraphics[width=\linewidth]{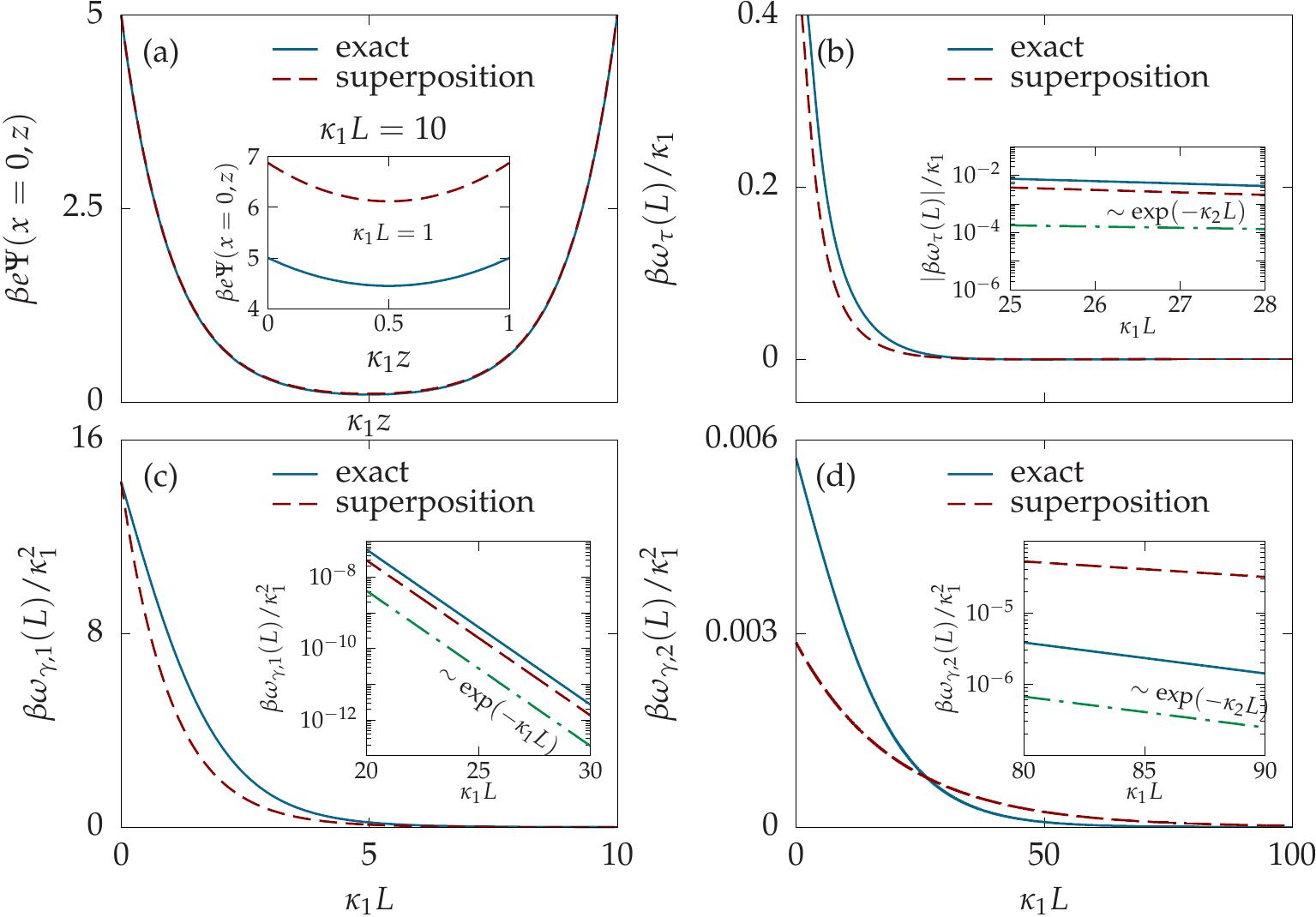}
	\caption
			[
			Comparison between the electrostatic potential, the surface interaction energy 
			densities and line interaction energy density between exact and superposition approximation 
			for $\beta e \PsiD = 3$
			]
			{
			Variation of the Donnan potential $\beta e \PsiD = 3$ in medium 2,
			other parameters are given by their standard configuration specified in Tab.~\ref{tab:parameter}.
			(a) Electrostatic potential $\Psi(x,z)$ within the exact and superposition calculations expressed 
			in the units of $1/\beta e$, as functions of $\kone z$.
			The separation length between the two walls is given by $\kone L = 10$ and $\kone L = 1$ in the inset. 
			Clearly, the superposition approximation fails to predict the electrostatic potential correctly for small separations. 
			Upon increasing the separation, as expected, the differences between exact and superposition calculations decrease. 
			(b) Line interaction energy density $\omega_\tau(L)$ within the exact and superposition calculations 
			expressed in the units of $\kone/\beta$, as functions of the separation $\kone L$.
			As the plot shows, the superposition approximation fails to capture the correct behavior. 
			In the limit of vanishing separations $\omega_\tau^e(L)$ and $\omega_\tau^s(L)$ reach the same finite value 
			and both decay monotonically. 
			Additionally, the superposition expression predicts the exponential decay for large separation correctly, 
			but underestimates the line interaction always by a factor of 2 as seen by the offset in the inset.
			(c)(d) Surface interaction energy density $\omega_{\gamma,i}(L)$ in medium $i \in \qty{1,2}$ within 
			the exact and superposition calculations expressed in the units of $\kone^2/\beta$, as functions of the separation 
			$\kone L$. 
			As the plots show, the superposition approximation fails to predict the correct behavior properly. 
			However, in the limit of vanishing separations the exact calculation and superposition approximation 
			predict the same surface interaction.
			Although the superposition approximation predicts the monotonic exponential 
			decay correctly, it underestimates the surface interaction by a factor of 2 as seen in the offset between exact 
			and superposition calculations in the inset.
			}
	\label{fig:ES_V_PSID3}
\end{figure}

\begin{figure}[htbp]
	\centering
	\includegraphics[width=\linewidth]{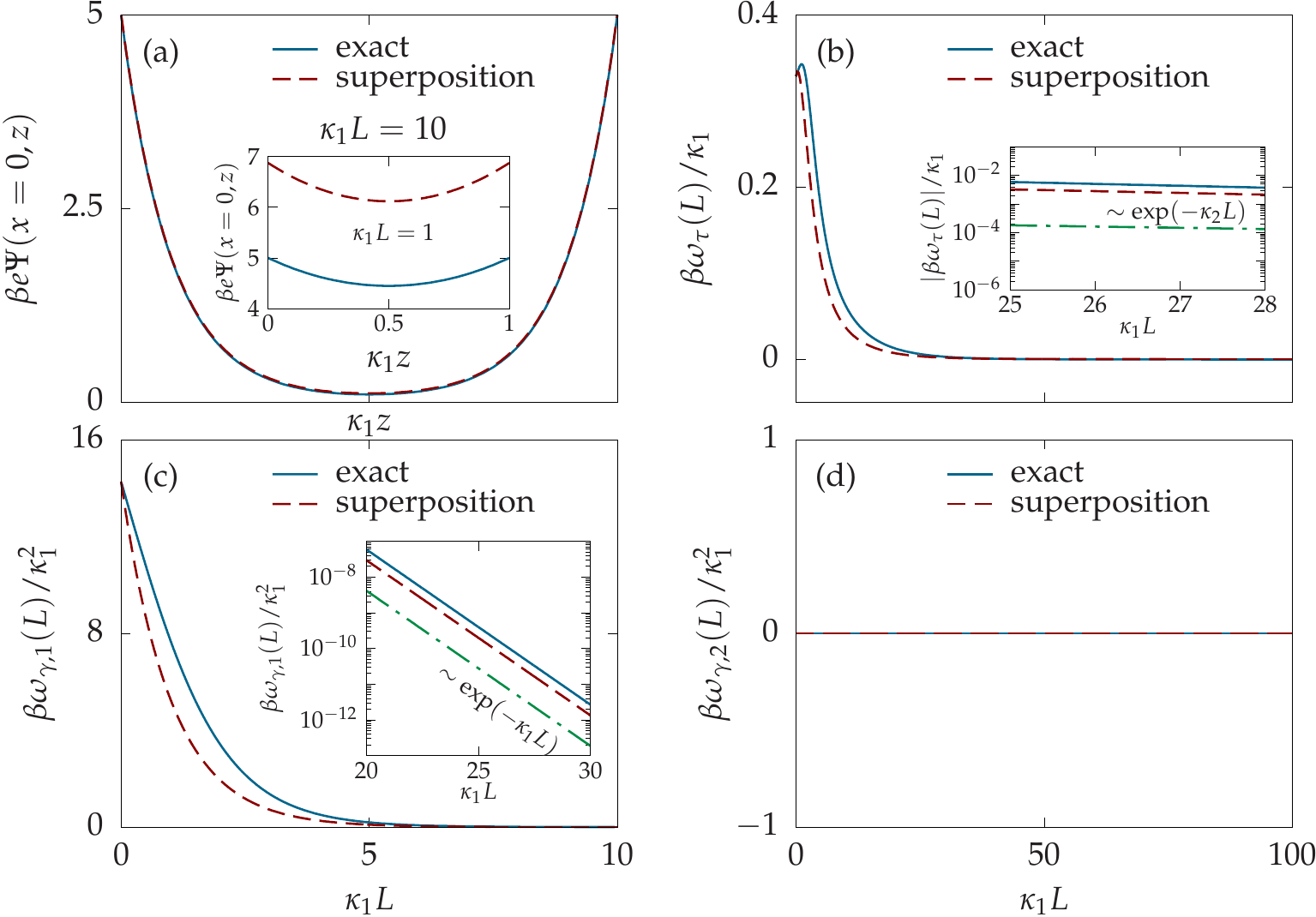}
	\caption
			[
			Comparision between the electrostatic potential, the surface interaction energy densities and 
			line interaction energy density between exact and superposition approximation for $\beta e \PsiD = 5$
			]
			{
			Variation of the Donnan potential $\beta e \PsiD = 5$ in medium 2,
			other parameters are given by their standard configuration specified in Tab.~\ref{tab:parameter}.
			(a) Electrostatic potential $\Psi(x,z)$ within the exact and superposition calculations expressed 
			in the units of $1/\beta e$, as functions of $\kone z$.
			The separation length between the two walls is given by $\kone L = 10$ and $\kone L = 1$ in the inset. 
			Clearly, the superposition approximation fails to predict the electrostatic potential correctly for small separations. 
			Upon increasing the separation, as expected, the differences between exact and superposition calculations decrease. 
			(b) Line interaction energy density $\omega_\tau(L)$ within exact and superposition calculations expressed 
			in units of $\kone/\beta$ , as functions of the separation $\kone L$ .
			As the plot shows, the superposition approximation fails to predict the correct behavior, 
			but captures the monotonic decay properly. 
			In the limit of vanishing separations $\omega_\tau^e(L)$ and $\omega_\tau^s(L)$ reach the same finite value. 
			Additionally, the superposition expression predicts the exponential decay for large separation correctly, 
			but underestimates the line interaction always by a factor of 2 as seen by the offset in the inset. 
			(c) Surface interaction energy density $\omega_{\gamma,1}(L)$ in medium 1 within the exact and 
			superposition calculations expressed in the units of $\kone^2/\beta$, as functions of the separation $\kone L$.  
			As the plot shows, the superposition approximation fails to predict the correct behavior properly but captures the 
			monotonic behavior correctly. 
			In the limit of vanishing separations the exact calculation and superposition 
			predict the same surface interaction.
			Although the superposition approximation predicts the exponential decay correctly, it underestimates 
			the surface interaction by a factor of 2 as seen in the offset between exact and superposition calculations in the inset.
			(d) In the case of $\beta e \PsiD = -1$ the surface interaction energy density $\omega_{\gamma,2}(L)$ in
			medium 2 is zero for the exact and superposition calculations.
			}
	\label{fig:ES_V_PSID5}
\end{figure}

\begin{figure}[htbp]
	\centering
	\includegraphics[width=\linewidth]{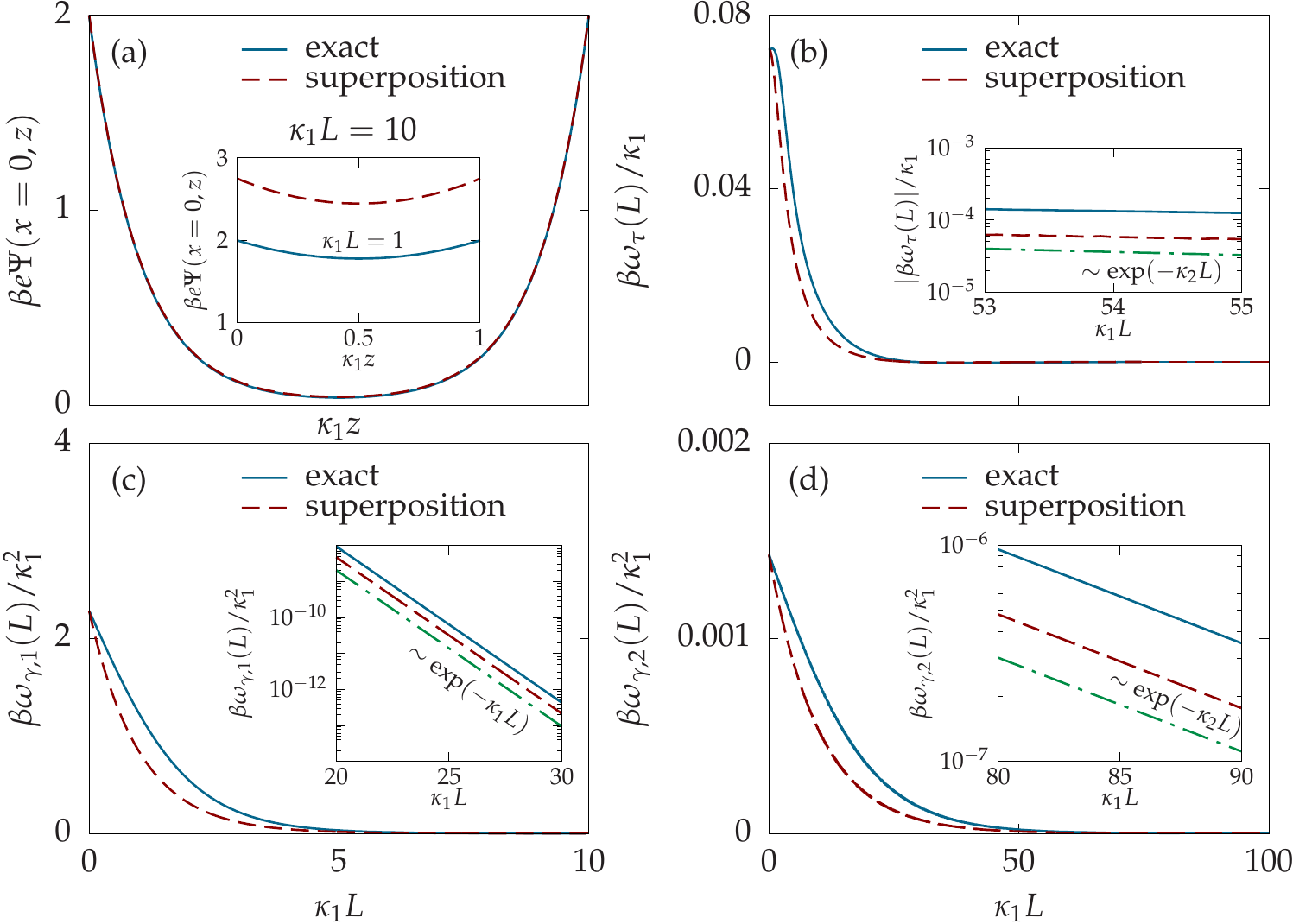}
	\caption
			[
			Comparison between the electrostatic potential, the surface interaction energy densities and 
			line interaction energy density between exact and superposition approximation for $\beta e \PsiP = 2$
			]
			{
			Variation of the surface potential $\beta e \PsiP = 2$,
			other parameters are given by their standard configuration specified in Tab.~\ref{tab:parameter}. 
			(a) Electrostatic potential $\Psi(x,z)$ within the exact and superposition calculations expressed 
			in the units of $1/\beta e$, as functions of $\kone z$.
			The separation length between the two walls is given by $\kone L = 10$ and $\kone L = 1$ in the inset. 
			Clearly, the superposition approximation fails to predict the electrostatic potential correctly for small separations.
			Upon increasing the separation, as expected, the differences between exact and superposition calculations decrease. 
			(b) Line interaction energy density $\omega_\tau(L)$ within the exact and superposition calculations
			expressed in the units of $\kone/\beta$, as functions of the separation $\kone L$.
			As the plot shows, the superposition approximation fails to capture the correct behavior. 
			In the limit of vanishing separations $\omega_\tau^e(L)$ and $\omega_\tau^s(L)$ reach the same finite value. 
			Additionally, the superposition expression predicts the exponential correctly, but underestimates the 
			line interaction always by a factor of 2 as seen by the offset in the inset. 
			(c)(d) Surface interaction energy density $\omega_{\gamma,i}(L)$ in medium $i \in \qty{1,2}$ within both 
			the exact and superposition calculations expressed in the units of $\kone^2/\beta$, as functions of 
			the separation $\kone L$.
			As the plots show, the superposition approximation fails to predict the correct behavior properly. 
			However, in the limit of vanishing separations the exact calculation and superposition approximation 
			predict the same surface interaction. 
			Although the superposition approximation predicts the monotonic exponential decay correctly, 
			it underestimates the surface interaction by a factor of 2 as seen in the offset between 
			exact and superposition calculations in the inset.
			}
	\label{fig:ES_V_PSIP2}
\end{figure}

\begin{figure}[htbp]
	\centering
	\includegraphics[width=\linewidth]{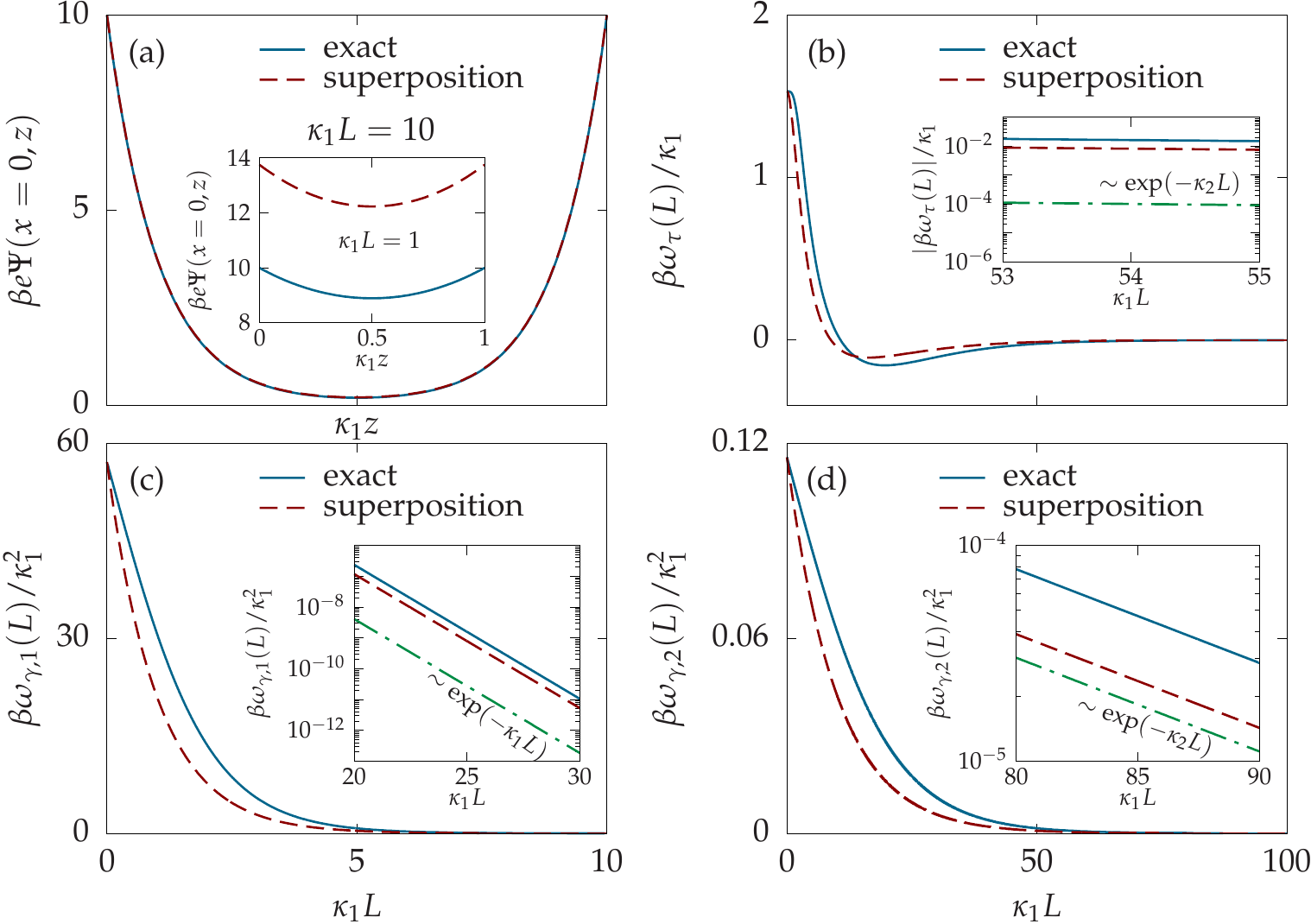}
	\caption
			[
			Comparison between the electrostatic potential, the surface interaction energy densities and 
			line interaction energy density between exact and superposition approximation for $\beta e \PsiP = 10$
			]
			{
			Variation of the surface potential $\beta e \PsiP = 10$,
			other parameters are given by their standard configuration specified in Tab.~\ref{tab:parameter}.
			(a) Electrostatic potential $\Psi(x,z)$ within the exact and superposition calculations expressed 
			in the units of $1/\beta e$, as functions of $\kone z$.
			The separation length between the two walls is given by $\kone L = 10$ and $\kone L = 1$ in the inset. 
			Clearly, the superposition approximation fails to predict the electrostatic potential correctly for small separations. 
			Upon increasing the separation, as expected, the differences between exact and superposition calculations decrease. 
			(b) Line interaction energy density $\omega_\tau(L)$ within the exact and superposition calculations 
			expressed in the units of $\kone/\beta$, as functions of the separation $\kone L$.
			As the plot shows, the superposition approximation fails to capture the correct behavior,
			but captures the non-monotonic decay. 
			In the limit of vanishing separations 
			$\omega_\tau^e(L)$ and $\omega_\tau^s(L)$ reach the same finite value.
			Additionally, the superposition expression predicts the exponential decay for large correctly,
			but underestimates the line interaction always by a factor of 2 as seen by the offset in the inset. 
			(c)(d) Surface interaction energy density $\omega_{\gamma,i}(L)$ in medium $i \in \qty{1,2}$ 
			within the exact and superposition calculations expressed in units of $\kone^2/\beta$, 
			as functions of the separation $\kone L$. 
			As the plots show, the superposition approximation fails to predict the correct behavior properly. 
			However, in the limit of vanishing separations the exact calculation and superposition 
			approximation both predict the same surface interaction. 
			Although, the superposition approximation predicts the exponential decay correctly, 
			it underestimates the surface interaction by a factor of 2 as seen in the offset between 
			exact and superposition calculations in the inset.
			}
	\label{fig:ES_V_PSIP10}
\end{figure}

This section provides comparisons between the exact calculation and superposition approximation of the electrostatic potential, 
the surface interaction energy densities and the line interaction energy density for parameters different from
the standard configuration as discussed in Section~\ref{sec:VAR}.
The plots are shown in Figs.~\ref{fig:ES_V_E2_6} to~\ref{fig:ES_V_PSIP10}.

\null
\vfill

\vfill
\null
\vfill

\vfill
\null
\vfill

\vfill
\null
\vfill

\vfill
\null
\vfill

\vfill
\null
\vfill

\vfill
\null
\vfill

\vfill
\null
\vfill

\vfill

%% file: chapters/06_Conclusion.tex

\chapter{Conclusion} 

In summary, this thesis analyzes the electrostatic interaction for a simplified model system 
by using a classical density functional theory approach.
This simplified model, consisting of two parallel walls with constant surface potential and in contact with two immiscible fluids,
is expected to mimic metallic colloidal particles trapped at a fluid-fluid interface.
Within the framework of a mean-field like linearized Poisson-Boltzmann theory, analytical expressions for the
electrostatic potentials are obtained within exact calculations as well as the superposition approximation.
Subsequently, the potential distributions inside the system are used to obtain analytical expressions
for the line and surface interaction energy densities.
The surface interaction energy densities $\omega_{\gamma,i}(L)$ for medium $i \in \qty{1,2}$ are monotonic
and decay exponentially with $\sim e^{-\kappa_i L}$ for separation length $L$ between the two walls. 
As it turns out, the superposition approximation fails to capture the behavior correctly and, even at larger separations, 
it underestimates the surface interaction energy density of the exact calculation by a factor of 2.
The line interaction energy density $\omega_\tau(L)$ shows monotonic or non-monotonic behavior
(existence of a minimum) depending on the system parameters.
Again, the superposition approximation fails to predict the behavior properly and
even at larger separation underestimates the line interaction energy density within exact calculation by a factor of 2.
This inaccuracy of the superposition approximation, especially in the limit of smaller separation,
could have a significant impact during applications depending on the experimental setup.
In addition to the surface and line interaction energy densities, analytical expressions for the surface tensions,
line tensions and interfacial tension are provided for exact and
superposition calculations.
All of these quantities are independent of the separation length between the walls, however, they differ within the exact and superposition calculations.

For future research, it would be interesting to study the interactions by using a non-linear Poisson-Boltzmann theory approach
(possibly numerically) and compare the results with those obtained within the linearized framework.
Moreover, studies could be further extended to include the impact of variations in the contact angle, particle geometry, and curvature.

%% file: backmatter/acknowledgements.tex

{\begingroup
\cleardoublepage
\thispagestyle{backmatter}
\section*{Acknowledgements}
I would like to thank Priv.-Doz.~Dr.~Markus Bier for the admission at the Max Planck Institute, 
his guidance, helpful discussions and the support in all official manners.

\noindent
Special thanks should be given to Dr.~Arghya Majee, my supervisor, for his guidance and valuable support 
as well as his generous help in the writing process.

\noindent
I would also like to thank the Department Dietrich for the nice and friendly work environment.
\endgroup}

%% file: backmatter/declaration.tex

{\begingroup
\begin{otherlanguage}{ngerman}
\cleardoublepage
\thispagestyle{backmatter}
\section*{Ehrenwörtliche Erklärung}
Ich erkläre,
\begin{itemize}
	\item 
		dass ich die vorliegende Bachelorarbeit selbstständig verfasst habe,
	\item 
		dass ich keine anderen als die angegebenen Quellen benutzt und alle wörtlich 
		oder sinngemäß aus anderen Werken übernommenen Aussagen als solche gekennzeichnet habe,
	\item 
		dass die eingereichte Arbeit weder vollständig noch in wesentlichen Teilen 
		Gegenstand eines anderen Prüfungsverfahrens gewesen ist,
	\item 
		dass ich die Arbeit weder vollständig noch in Teilen bereits veröffentlicht habe
	\item 
		dass der Inhalt des elektronischen Exemplars mit dem des Druckexemplars übereinstimmt.
\end{itemize}
\vspace*{2cm}

\noindent
Stuttgart, den 18.~August 2018 \hfill Rick Bebon
\end{otherlanguage}
\endgroup}